\documentclass[iop,twocolumn]{emulateapj}
\usepackage[table,usenames,dvipsnames, svgnames]{xcolor}
\usepackage{amsmath}
\usepackage{amssymb}
\usepackage{subfigure}
\usepackage[hang,flushmargin]{footmisc}
\usepackage[backref,breaklinks,colorlinks]{hyperref}
\usepackage{natbib}
\usepackage{graphicx}
\usepackage{multirow}
\usepackage{enumitem}
\let\tablenum\relax
\usepackage{siunitx}
\hypersetup{
    colorlinks=true,
    citecolor=Blue, % citations in text
    linkcolor=OrangeRed, % in references
    filecolor=teal,
    urlcolor=DeepPink, % \url
}

\newcommand{\planck}{{\textit{Planck}}}

\usepackage{natbib}
\def\apjl{ApJL}

\def\aap{A\&A}

\def\apjs{ApJS}

\def\clstar{\ell \left( \ell + 1 \right) C_\ell / 2 \pi}

\def\dd{\Delta_\mathrm{d}}
\def\ddp{\Delta'_\mathrm{d}}

\def\As{A_\mathrm{sync}}

\def\Bd{\beta_\mathrm{d}}
\def\Bs{\beta_\mathrm{s}}

\def\ad{\alpha_\mathrm{d}}
\def\as{\alpha_\mathrm{s}}

\newif\ifmodern
\newif\ifmanu
\def\tempfootmark#1{}

\def\correspondingauthor#1{{
\renewcommand\thefootnote{\hskip-12pt}
\footnote{Corresponding author: #1\ifmodern\vrule depth 5pt
width 0pt\relax\fi}}}

\let\saveemail\email
\def\email#1{{\let\ltx\tempfootmark
\saveemail{}}
{\renewcommand\thefootnote{\hskip-12.1pt}
\footnote{\href{mailto: #1}{#1}\ifmodern\vrule depth 7pt width
0pt\relax\else\ifmanu\vskip-4pt\else\vrule depth 7pt width 0pt\fi\fi}}}

\newcommand{\hyperfootnote}[1][]{\def\ArgI\hyperfootnoteRelay}
\newcommand\hyperfootnoteRelay[2][]{\href{#1#2}{\ArgI}\footnote{\href{#1#2}{#2}}}
\newbox\tablebox    \newdimen\tablewidth
\def\leaderfil{\leaders\hbox to 5pt{\hss.\hss}\hfil}
\def\endCMBSfourtable{\tablewidth=\wd\tablebox
    $$\hss\copy\tablebox\hss$$
    \vskip-\lastskip\vskip -2pt}
\def\tablenote#1 #2\par{\begingroup \parindent=0.8em
    \abovedisplayshortskip=0pt\belowdisplayshortskip=0pt
    \noindent
    $$\hss\vbox{\hsize\tablewidth \hangindent=\parindent \hangafter=1 \noindent
    \hbox to \parindent{$^#1$\hss}\strut#2\strut\par}\hss$$
    \endgroup}
\def\doubleline{\vskip 3pt\hrule \vskip 1.5pt \hrule \vskip 5pt}
\begin{document}
\submitted{Submitted to ApJ -- draft version August 26, 2020}
\title{CMB-S4: Forecasting Constraints on Primordial Gravitational Waves}

\author{% This file is automatically generated by a script, so edits may be overwritten. 
% author.py --optin BB_forecast/Membership_downloaded_by_ATC_20200810.csv BB_forecast/additional_authors.csv --optout BB_forecast/BB_forecast_optout.csv --rulesfile CMB-S4_rules.pkl --prefix BB_forecast/

\newcounter{CMBSFourAffiliationCount}
\newcommand{\CMBSFourAffiliation}[1]{\refstepcounter{CMBSFourAffiliationCount}\label{#1}}
\CMBSFourAffiliation{UCIrvine}
\CMBSFourAffiliation{JohnsHopkinsUniversity}
\CMBSFourAffiliation{UniversityofIllinoisatUrbana-Champaign}
\CMBSFourAffiliation{SLAC}
\CMBSFourAffiliation{UCBerkeley}
\CMBSFourAffiliation{StanfordUniversity}
\CMBSFourAffiliation{OxfordUniversity}
\CMBSFourAffiliation{LawrenceBerkeleyNationalLaboratory}
\CMBSFourAffiliation{RiceUniversity}
\CMBSFourAffiliation{Fermilab}
\CMBSFourAffiliation{UCSanDiego}
\CMBSFourAffiliation{SISSA}
\CMBSFourAffiliation{IFPU}
\CMBSFourAffiliation{INFN}
\CMBSFourAffiliation{HarvardUniversityCFA}
\CMBSFourAffiliation{UniversityofNewMexico}
\CMBSFourAffiliation{ArgonneNationalLaboratory}
\CMBSFourAffiliation{AstroParticleandCosmologyLaboratory}
\CMBSFourAffiliation{Caltech}
\CMBSFourAffiliation{CornellUniversity}
\CMBSFourAffiliation{KICP}
\CMBSFourAffiliation{UniversityofChicago}
\CMBSFourAffiliation{UniversityofMelbourne}
\CMBSFourAffiliation{UniversityofCincinnati}
\CMBSFourAffiliation{JPL}
\CMBSFourAffiliation{LMUMunich}
\CMBSFourAffiliation{UTAustin}
\CMBSFourAffiliation{CITA}
\CMBSFourAffiliation{InstitutdAstrophysiquedeParis}
\CMBSFourAffiliation{StonyBrookUniversity}
\CMBSFourAffiliation{UniversityofManchester}
\CMBSFourAffiliation{ArizonaStateUniversity}
\CMBSFourAffiliation{FloridaStateUniversity}
\CMBSFourAffiliation{CardiffUniversity}
\CMBSFourAffiliation{DartmouthCollege}
\CMBSFourAffiliation{UniversityofGeneva}
\CMBSFourAffiliation{UniversityofSussex}
\CMBSFourAffiliation{UniversityofCambridge}
\CMBSFourAffiliation{KavliInstituteforCosmologyCambridge}
\CMBSFourAffiliation{IoA}
\CMBSFourAffiliation{UniversityofTokyo}
\CMBSFourAffiliation{UniversityofGroningen}
\CMBSFourAffiliation{UniversityofToronto}
\CMBSFourAffiliation{Dunlap}
\CMBSFourAffiliation{KEK}
\CMBSFourAffiliation{CEASaclay}
\CMBSFourAffiliation{USTCH}
\CMBSFourAffiliation{UniversityofPennsylvania}
\CMBSFourAffiliation{McGillUniversity}
\CMBSFourAffiliation{NIST}
\CMBSFourAffiliation{PrincetonUniversity}
\CMBSFourAffiliation{HarvardUniversity}
\CMBSFourAffiliation{NASAGoddardSpaceFlightCenter}
\CMBSFourAffiliation{SimonFraserUniversity}
\CMBSFourAffiliation{UniversityofSouthernCalifornia}
\CMBSFourAffiliation{HaverfordCollege}
\CMBSFourAffiliation{StockholmUniversity}
\CMBSFourAffiliation{CaseWesternReserveUniversity}
\CMBSFourAffiliation{UniversityofBritishColumbia}
\CMBSFourAffiliation{UniversityofColoradoBoulder}
\CMBSFourAffiliation{UniversityofMinnesota}
\CMBSFourAffiliation{UniversityofMichigan}
\CMBSFourAffiliation{FlatironInstitute}
\CMBSFourAffiliation{Columbia}
\CMBSFourAffiliation{UniversityofVirginia}
\CMBSFourAffiliation{KavliIPMU}
\CMBSFourAffiliation{UCDavis}
\CMBSFourAffiliation{YaleUniversity}
\CMBSFourAffiliation{UniversityofPittsburgh}
\CMBSFourAffiliation{BenGurionUniversity}
\CMBSFourAffiliation{AaltoUniversity}
\CMBSFourAffiliation{UCSantaBarbara}
\CMBSFourAffiliation{PerimiterInstitue}
\CMBSFourAffiliation{SouthernMethodistUniversity}
\CMBSFourAffiliation{BerkeleyCenterforCosmologicalPhysics}
\CMBSFourAffiliation{EuropeanSouthernObservatory}
\CMBSFourAffiliation{GRAPPA}
\CMBSFourAffiliation{WashingtonUniversitySt.Louis}
\CMBSFourAffiliation{McDonnellCenterfortheSpaceSciencesWashingtonUniversity}
\CMBSFourAffiliation{UniversityofMilano-Bicocca}
\CMBSFourAffiliation{CentredePhysiquedesParticulesdeMarseille}
\CMBSFourAffiliation{KIPAC}
\CMBSFourAffiliation{InstituteforAdvancedStudy}
\CMBSFourAffiliation{PennsylvaniaStateUniversity}
\CMBSFourAffiliation{BrookhavenNationalLaboratory}
\CMBSFourAffiliation{KyotoUniversity}
\CMBSFourAffiliation{UniversityofWisconsinMadison}
\CMBSFourAffiliation{UniversitedegliStudidiMilan}
\CMBSFourAffiliation{ICJLab}
\CMBSFourAffiliation{BrownUniversity}
\CMBSFourAffiliation{SyracuseUniversity}
\CMBSFourAffiliation{UCLosAngeles}
\CMBSFourAffiliation{MSU}
The CMB-S4 Collaboration:
Kevork Abazajian,\textsuperscript{\ref{UCIrvine}}
Graeme~E.\ Addison,\textsuperscript{\ref{JohnsHopkinsUniversity}}
Peter Adshead,\textsuperscript{\ref{UniversityofIllinoisatUrbana-Champaign}}
Zeeshan Ahmed,\textsuperscript{\ref{SLAC}}
Daniel Akerib,\textsuperscript{\ref{SLAC}}
Aamir Ali,\textsuperscript{\ref{UCBerkeley}}
Steven~W.\ Allen,\textsuperscript{\ref{StanfordUniversity}}
David Alonso,\textsuperscript{\ref{OxfordUniversity}}
Marcelo Alvarez,\textsuperscript{\ref{UCBerkeley},\ref{LawrenceBerkeleyNationalLaboratory}}
Mustafa A.~Amin,\textsuperscript{\ref{RiceUniversity}}
Adam Anderson,\textsuperscript{\ref{Fermilab}}
Kam~S.\ Arnold,\textsuperscript{\ref{UCSanDiego}}
Peter Ashton,\textsuperscript{\ref{UCBerkeley}}
Carlo Baccigalupi,\textsuperscript{\ref{SISSA},\ref{IFPU},\ref{INFN}}
Debbie Bard,\textsuperscript{\ref{LawrenceBerkeleyNationalLaboratory}}
Denis Barkats,\textsuperscript{\ref{HarvardUniversityCFA}}
Darcy Barron,\textsuperscript{\ref{UniversityofNewMexico}}
Peter S.~Barry,\textsuperscript{\ref{ArgonneNationalLaboratory}}
James~G.\ Bartlett,\textsuperscript{\ref{AstroParticleandCosmologyLaboratory}}
Ritoban Basu Thakur,\textsuperscript{\ref{Caltech}}
Nicholas Battaglia,\textsuperscript{\ref{CornellUniversity}}
Rachel Bean,\textsuperscript{\ref{CornellUniversity}}
Chris Bebek,\textsuperscript{\ref{LawrenceBerkeleyNationalLaboratory}}
Amy~N.\ Bender,\textsuperscript{\ref{ArgonneNationalLaboratory},\ref{KICP}}
Bradford~A.\ Benson,\textsuperscript{\ref{Fermilab},\ref{UniversityofChicago},\ref{KICP}}
Federico  Bianchini,\textsuperscript{\ref{UniversityofMelbourne}}
Colin~A.\ Bischoff,\textsuperscript{\ref{UniversityofCincinnati}}
Lindsey Bleem,\textsuperscript{\ref{ArgonneNationalLaboratory},\ref{KICP}}
James~J.\ Bock,\textsuperscript{\ref{Caltech},\ref{JPL}}
Sebastian Bocquet,\textsuperscript{\ref{LMUMunich}}
Kimberly K.~Boddy,\textsuperscript{\ref{JohnsHopkinsUniversity},\ref{UTAustin}}
J.~Richard Bond,\textsuperscript{\ref{CITA}}
Julian Borrill,\textsuperscript{\ref{LawrenceBerkeleyNationalLaboratory},\ref{UCBerkeley}}
Fran\c{c}ois~R.\ Bouchet,\textsuperscript{\ref{InstitutdAstrophysiquedeParis}}
Thejs Brinckmann,\textsuperscript{\ref{StonyBrookUniversity}}
Michael~L.\ Brown,\textsuperscript{\ref{UniversityofManchester}}
Sean Bryan,\textsuperscript{\ref{ArizonaStateUniversity}}
Victor Buza,\textsuperscript{\ref{KICP},\ref{HarvardUniversity}}
Karen Byrum,\textsuperscript{\ref{ArgonneNationalLaboratory}}
Carlos Hervias  Caimapo,\textsuperscript{\ref{FloridaStateUniversity}}
Erminia Calabrese,\textsuperscript{\ref{CardiffUniversity}}
Victoria  Calafut,\textsuperscript{\ref{CornellUniversity}}
Robert Caldwell,\textsuperscript{\ref{DartmouthCollege}}
John~E.\ Carlstrom,\textsuperscript{\ref{UniversityofChicago},\ref{ArgonneNationalLaboratory},\ref{KICP}}
Julien Carron,\textsuperscript{\ref{UniversityofGeneva},\ref{UniversityofSussex}}
Thomas Cecil,\textsuperscript{\ref{ArgonneNationalLaboratory}}
Anthony Challinor,\textsuperscript{\ref{UniversityofCambridge},\ref{KavliInstituteforCosmologyCambridge},\ref{IoA}}
Clarence~L.\ Chang,\textsuperscript{\ref{ArgonneNationalLaboratory},\ref{KICP},\ref{UniversityofChicago}}
Yuji Chinone,\textsuperscript{\ref{UniversityofTokyo}}
Hsiao-Mei Sherry Cho,\textsuperscript{\ref{SLAC}}
Asantha Cooray,\textsuperscript{\ref{UCIrvine}}
Will Coulton,\textsuperscript{\ref{UniversityofGroningen}}
Thomas~M. Crawford,\textsuperscript{\ref{UniversityofChicago},\ref{KICP}}
Abigail Crites,\textsuperscript{\ref{UniversityofToronto},\ref{Dunlap},\ref{Caltech}}
Ari  Cukierman,\textsuperscript{\ref{SLAC},\ref{StanfordUniversity}}
Francis-Yan Cyr-Racine,\textsuperscript{\ref{UniversityofNewMexico}}
Tijmen de Haan,\textsuperscript{\ref{KEK}}
Jacques Delabrouille,\textsuperscript{\ref{AstroParticleandCosmologyLaboratory},\ref{CEASaclay},\ref{USTCH}}
Mark Devlin,\textsuperscript{\ref{UniversityofPennsylvania}}
Eleonora Di Valentino,\textsuperscript{\ref{UniversityofManchester}}
Marion Dierickx,\textsuperscript{\ref{HarvardUniversityCFA}}
Matt Dobbs,\textsuperscript{\ref{McGillUniversity}}
Shannon Duff,\textsuperscript{\ref{NIST}}
Jo Dunkley,\textsuperscript{\ref{PrincetonUniversity}}
Cora Dvorkin,\textsuperscript{\ref{HarvardUniversity}}
Joseph Eimer,\textsuperscript{\ref{JohnsHopkinsUniversity}}
Tucker Elleflot,\textsuperscript{\ref{LawrenceBerkeleyNationalLaboratory}}
Josquin Errard,\textsuperscript{\ref{AstroParticleandCosmologyLaboratory}}
Thomas Essinger-Hileman,\textsuperscript{\ref{NASAGoddardSpaceFlightCenter}}
Giulio Fabbian,\textsuperscript{\ref{UniversityofSussex}}
Chang Feng,\textsuperscript{\ref{UniversityofIllinoisatUrbana-Champaign}}
Simone Ferraro,\textsuperscript{\ref{LawrenceBerkeleyNationalLaboratory}}
Jeffrey P.~Filippini,\textsuperscript{\ref{UniversityofIllinoisatUrbana-Champaign}}
Raphael Flauger,\textsuperscript{\ref{UCSanDiego}}
Brenna Flaugher,\textsuperscript{\ref{Fermilab}}
Aurelien A.~Fraisse,\textsuperscript{\ref{PrincetonUniversity}}
Andrei Frolov,\textsuperscript{\ref{SimonFraserUniversity}}
Nicholas Galitzki,\textsuperscript{\ref{UCSanDiego}}
Patricio A.~Gallardo,\textsuperscript{\ref{CornellUniversity}}
Silvia Galli,\textsuperscript{\ref{InstitutdAstrophysiquedeParis}}
Ken Ganga,\textsuperscript{\ref{AstroParticleandCosmologyLaboratory}}
Martina Gerbino,\textsuperscript{\ref{INFN}}
Vera Gluscevic,\textsuperscript{\ref{UniversityofSouthernCalifornia},\ref{PrincetonUniversity}}
Neil Goeckner-Wald,\textsuperscript{\ref{StanfordUniversity}}
Daniel Green,\textsuperscript{\ref{UCSanDiego}}
Daniel Grin,\textsuperscript{\ref{HaverfordCollege}}
Evan Grohs,\textsuperscript{\ref{UCBerkeley}}
Riccardo Gualtieri,\textsuperscript{\ref{ArgonneNationalLaboratory}}
Jon E.~Gudmundsson,\textsuperscript{\ref{StockholmUniversity}}
Ian Gullett,\textsuperscript{\ref{CaseWesternReserveUniversity}}
Nikhel  Gupta,\textsuperscript{\ref{UniversityofMelbourne}}
Salman Habib,\textsuperscript{\ref{ArgonneNationalLaboratory}}
Mark Halpern,\textsuperscript{\ref{UniversityofBritishColumbia}}
Nils W.~Halverson,\textsuperscript{\ref{UniversityofColoradoBoulder}}
Shaul Hanany,\textsuperscript{\ref{UniversityofMinnesota}}
Kathleen Harrington,\textsuperscript{\ref{UniversityofMichigan}}
Masaya Hasegawa,\textsuperscript{\ref{KEK}}
Matthew Hasselfield,\textsuperscript{\ref{FlatironInstitute}}
Masashi Hazumi,\textsuperscript{\ref{KEK}}
Katrin Heitmann,\textsuperscript{\ref{ArgonneNationalLaboratory}}
Shawn Henderson,\textsuperscript{\ref{SLAC}}
Brandon Hensley,\textsuperscript{\ref{PrincetonUniversity}}
Charles Hill,\textsuperscript{\ref{UCBerkeley}}
J. Colin Hill,\textsuperscript{\ref{Columbia}}
Ren\'{e}e Hlo\v{z}ek,\textsuperscript{\ref{UniversityofToronto},\ref{Dunlap}}
Shuay-Pwu Patty Ho,\textsuperscript{\ref{PrincetonUniversity}}
Thuong Hoang,\textsuperscript{\ref{CornellUniversity}}
Gil Holder,\textsuperscript{\ref{UniversityofIllinoisatUrbana-Champaign}}
William Holzapfel,\textsuperscript{\ref{UCBerkeley}}
John Hood,\textsuperscript{\ref{UniversityofChicago},\ref{KICP}}
Johannes Hubmayr,\textsuperscript{\ref{NIST}}
Kevin~M.\ Huffenberger,\textsuperscript{\ref{FloridaStateUniversity}}
Howard Hui,\textsuperscript{\ref{Caltech}}
Kent Irwin,\textsuperscript{\ref{StanfordUniversity}}
Oliver Jeong,\textsuperscript{\ref{UCBerkeley}}
Bradley R.~Johnson,\textsuperscript{\ref{UniversityofVirginia}}
William C.~Jones,\textsuperscript{\ref{PrincetonUniversity}}
Jae Hwan  Kang,\textsuperscript{\ref{StanfordUniversity}}
Kirit S.~Karkare,\textsuperscript{\ref{UniversityofChicago},\ref{KICP}}
Nobuhiko Katayama,\textsuperscript{\ref{KavliIPMU}}
Reijo Keskitalo,\textsuperscript{\ref{LawrenceBerkeleyNationalLaboratory},\ref{UCBerkeley}}
Theodore Kisner,\textsuperscript{\ref{LawrenceBerkeleyNationalLaboratory},\ref{UCBerkeley}}
Lloyd Knox,\textsuperscript{\ref{UCDavis}}
Brian J.~Koopman,\textsuperscript{\ref{YaleUniversity}}
Arthur Kosowsky,\textsuperscript{\ref{UniversityofPittsburgh}}
John Kovac,\textsuperscript{\ref{HarvardUniversity},\ref{HarvardUniversityCFA}}
Ely D.~Kovetz,\textsuperscript{\ref{BenGurionUniversity}}
Steve Kuhlmann,\textsuperscript{\ref{ArgonneNationalLaboratory}}
Chao-lin Kuo,\textsuperscript{\ref{StanfordUniversity}}
Akito Kusaka,\textsuperscript{\ref{LawrenceBerkeleyNationalLaboratory}}
Anne L\"ahteenm\"aki,\textsuperscript{\ref{AaltoUniversity}}
Charles R.~Lawrence,\textsuperscript{\ref{JPL}}
Adrian T.~Lee,\textsuperscript{\ref{UCBerkeley},\ref{LawrenceBerkeleyNationalLaboratory}}
Antony Lewis,\textsuperscript{\ref{UniversityofSussex}}
Dale Li,\textsuperscript{\ref{SLAC}}
Eric Linder,\textsuperscript{\ref{LawrenceBerkeleyNationalLaboratory}}
Marilena Loverde,\textsuperscript{\ref{StonyBrookUniversity}}
Amy Lowitz,\textsuperscript{\ref{UniversityofChicago},\ref{KICP}}
Phil Lubin,\textsuperscript{\ref{UCSantaBarbara}}
Mathew S.~Madhavacheril,\textsuperscript{\ref{PerimiterInstitue}}
Adam Mantz,\textsuperscript{\ref{StanfordUniversity}}
Gabriela Marques,\textsuperscript{\ref{FloridaStateUniversity}}
Frederick Matsuda,\textsuperscript{\ref{KavliIPMU}}
Philip Mauskopf,\textsuperscript{\ref{ArizonaStateUniversity}}
Heather McCarrick,\textsuperscript{\ref{PrincetonUniversity}}
Jeffrey McMahon,\textsuperscript{\ref{UniversityofChicago},\ref{KICP}}
P.~Daniel Meerburg,\textsuperscript{\ref{UniversityofGroningen}}
Jean-Baptiste Melin,\textsuperscript{\ref{CEASaclay}}
Felipe Menanteau,\textsuperscript{\ref{UniversityofIllinoisatUrbana-Champaign}}
Joel Meyers,\textsuperscript{\ref{SouthernMethodistUniversity}}
Marius Millea,\textsuperscript{\ref{BerkeleyCenterforCosmologicalPhysics}}
Joseph Mohr,\textsuperscript{\ref{LMUMunich}}
Lorenzo Moncelsi,\textsuperscript{\ref{Caltech}}
Maria Monzani,\textsuperscript{\ref{SLAC}}
Tony Mroczkowski,\textsuperscript{\ref{EuropeanSouthernObservatory}}
Suvodip Mukherjee,\textsuperscript{\ref{InstitutdAstrophysiquedeParis},\ref{GRAPPA}}
Johanna Nagy,\textsuperscript{\ref{WashingtonUniversitySt.Louis},\ref{McDonnellCenterfortheSpaceSciencesWashingtonUniversity}}
Toshiya  Namikawa,\textsuperscript{\ref{UniversityofCambridge}}
Federico Nati,\textsuperscript{\ref{UniversityofMilano-Bicocca}}
Tyler Natoli,\textsuperscript{\ref{UniversityofChicago},\ref{KICP}}
Laura Newburgh,\textsuperscript{\ref{YaleUniversity}}
Michael D.~Niemack,\textsuperscript{\ref{CornellUniversity}}
Haruki Nishino,\textsuperscript{\ref{KEK}}
Brian Nord,\textsuperscript{\ref{Fermilab}}
Valentine Novosad,\textsuperscript{\ref{ArgonneNationalLaboratory}}
Roger O'Brient,\textsuperscript{\ref{JPL},\ref{Caltech}}
Stephen Padin,\textsuperscript{\ref{Caltech}}
Steven Palladino,\textsuperscript{\ref{UniversityofCincinnati}}
Bruce Partridge,\textsuperscript{\ref{HaverfordCollege}}
Don Petravick,\textsuperscript{\ref{UniversityofIllinoisatUrbana-Champaign}}
Elena Pierpaoli,\textsuperscript{\ref{UniversityofSouthernCalifornia}}
Levon Pogosian,\textsuperscript{\ref{SimonFraserUniversity}}
Karthik Prabhu,\textsuperscript{\ref{UCDavis}}
Clement Pryke,\textsuperscript{\ref{UniversityofMinnesota}}
Giuseppe Puglisi,\textsuperscript{\ref{UCBerkeley},\ref{LawrenceBerkeleyNationalLaboratory}}
Benjamin Racine,\textsuperscript{\ref{HarvardUniversityCFA},\ref{CentredePhysiquedesParticulesdeMarseille}}
Alexandra Rahlin,\textsuperscript{\ref{Fermilab}}
Mayuri Sathyanarayana Rao,\textsuperscript{\ref{UCBerkeley}}
Marco Raveri,\textsuperscript{\ref{UniversityofPennsylvania}}
Christian L.~Reichardt,\textsuperscript{\ref{UniversityofMelbourne}}
Mathieu Remazeilles,\textsuperscript{\ref{UniversityofManchester}}
Graca Rocha,\textsuperscript{\ref{JPL}}
Natalie A.~Roe,\textsuperscript{\ref{LawrenceBerkeleyNationalLaboratory}}
Anirban Roy,\textsuperscript{\ref{CornellUniversity}}
John E.~Ruhl,\textsuperscript{\ref{CaseWesternReserveUniversity}}
Maria Salatino,\textsuperscript{\ref{StanfordUniversity},\ref{KIPAC}}
Benjamin Saliwanchik,\textsuperscript{\ref{YaleUniversity}}
Emmanuel Schaan,\textsuperscript{\ref{LawrenceBerkeleyNationalLaboratory}}
Alessandro Schillaci,\textsuperscript{\ref{Caltech}}
Benjamin Schmitt,\textsuperscript{\ref{HarvardUniversityCFA}}
Marcel M.~Schmittfull,\textsuperscript{\ref{InstituteforAdvancedStudy}}
Douglas Scott,\textsuperscript{\ref{UniversityofBritishColumbia}}
Neelima Sehgal,\textsuperscript{\ref{StonyBrookUniversity}}
Sarah Shandera,\textsuperscript{\ref{PennsylvaniaStateUniversity}}
Blake~D.\ Sherwin,\textsuperscript{\ref{UniversityofCambridge}}
Erik Shirokoff,\textsuperscript{\ref{UniversityofChicago},\ref{KICP}}
Sara M.~Simon,\textsuperscript{\ref{Fermilab}}
An\v{z}e Slosar,\textsuperscript{\ref{BrookhavenNationalLaboratory}}
David Spergel,\textsuperscript{\ref{PrincetonUniversity}}
Tyler~St.\ Germaine,\textsuperscript{\ref{HarvardUniversity}}
Suzanne T.~Staggs,\textsuperscript{\ref{PrincetonUniversity}}
Antony Stark,\textsuperscript{\ref{HarvardUniversityCFA}}
Glenn D.~Starkman,\textsuperscript{\ref{CaseWesternReserveUniversity}}
Radek Stompor,\textsuperscript{\ref{AstroParticleandCosmologyLaboratory}}
Chris Stoughton,\textsuperscript{\ref{Fermilab}}
Aritoki Suzuki,\textsuperscript{\ref{LawrenceBerkeleyNationalLaboratory}}
Osamu Tajima,\textsuperscript{\ref{KyotoUniversity}}
Grant P.~Teply,\textsuperscript{\ref{UCSanDiego}}
Keith Thompson,\textsuperscript{\ref{StanfordUniversity}}
Ben Thorne,\textsuperscript{\ref{UCDavis}}
Peter Timbie,\textsuperscript{\ref{UniversityofWisconsinMadison}}
Maurizio Tomasi,\textsuperscript{\ref{UniversitedegliStudidiMilan}}
Matthieu Tristram,\textsuperscript{\ref{ICJLab}}
Gregory Tucker,\textsuperscript{\ref{BrownUniversity}}
Caterina Umilt\`{a},\textsuperscript{\ref{UniversityofIllinoisatUrbana-Champaign}}
Alexander van Engelen,\textsuperscript{\ref{ArizonaStateUniversity}}
Eve M.~Vavagiakis,\textsuperscript{\ref{CornellUniversity}}
Joaquin D.~Vieira,\textsuperscript{\ref{UniversityofIllinoisatUrbana-Champaign}}
Abigail G.~Vieregg,\textsuperscript{\ref{UniversityofChicago},\ref{KICP}}
Kasey Wagoner,\textsuperscript{\ref{PrincetonUniversity}}
Benjamin Wallisch,\textsuperscript{\ref{UCSanDiego},\ref{InstituteforAdvancedStudy}}
Gensheng Wang,\textsuperscript{\ref{ArgonneNationalLaboratory}}
Scott Watson,\textsuperscript{\ref{SyracuseUniversity}}
Ben Westbrook,\textsuperscript{\ref{UCBerkeley}}
Nathan Whitehorn,\textsuperscript{\ref{UCLosAngeles},\ref{MSU}}
Edward J.\ Wollack,\textsuperscript{\ref{NASAGoddardSpaceFlightCenter}}
W.~L.~Kimmy Wu,\textsuperscript{\ref{KICP}}
Zhilei Xu,\textsuperscript{\ref{UniversityofPennsylvania}}
H. Y. Eric Yang,\textsuperscript{\ref{StanfordUniversity}}
Siavash Yasini,\textsuperscript{\ref{UniversityofSouthernCalifornia}}
Volodymyr G.~Yefremenko,\textsuperscript{\ref{ArgonneNationalLaboratory}}
Ki Won Yoon,\textsuperscript{\ref{StanfordUniversity}}
Edward Young,\textsuperscript{\ref{SLAC}}
Cyndia Yu,\textsuperscript{\ref{StanfordUniversity}}
Andrea Zonca\textsuperscript{\ref{UCSanDiego}}

 \twocolumngrid
\scriptsize
\setlength{\parskip}{2pt}
\flushing % left adjust affils
%%%%%%%%%%%%%%%%%%%%%%%%%%%%%%%%%%%%%%%%%%%%%%%%%%%%%%%%%%%%
%                                                          %
% Institutional aliases file.                              %
% Originally used in CV DM whitepaper,                     %
% successfully stolen for CV 21cm roadmap whitepaper,      %
% now to live a life of its own.                           %
%                                                          %
% When editing, please respect alphabetical ordering       %
% and avoid duplication (that is the entire point).        %
%                                                          %
%                                                          %
% A summer river being crossed                             %
% how pleasing                                             %
% with sandals in my hands!                                %
%       Yosa Buson (1716-1784)                             %
%                                                          %
% 08/2020: Edits to make US, UK, and Canadian addresses uniform.
%%%%%%%%%%%%%%%%%%%%%%%%%%%%%%%%%%%%%%%%%%%%%%%%%%%%%%%%%%%%

\newcommand{\Aalto}{Aalto University, FIN-00076, Aalto, Finland}
\newcommand{\Amherst}{University of Massachusetts, Amherst, MA 01003, USA}
\newcommand{\AmsterdamAstro}{Anton Pannekoek Institute for Astronomy, University of Amsterdam, 1098~XH Amsterdam, The Netherlands}
\newcommand{\ANL}{Argonne National Laboratory, Lemont, IL 60439, USA}
\newcommand{\ANLHEP}{HEP Division, Argonne National Laboratory, Lemont, IL 60439, USA}
\newcommand{\APC}{Laboratoire Astroparticule et Cosmologie (APC), CNRS/IN2P3, Universit\'e Paris Diderot, 75205 Paris Cedex 13, France}
\newcommand{\APONMSU}{Apache Point Observatory and New Mexico State University, Sunspot, NM 88349, USA}
\newcommand{\ASU}{Arizona State University, Tempe, AZ 85287, USA}
\newcommand{\BCCP}{Berkeley Center for Cosmological Physics, University of California, Berkeley, CA 94720, USA}
\newcommand{\BenGurion}{Department of Physics, Ben-Gurion University, Be'er Sheva 84105, Israel}
\newcommand{\BNL}{Brookhaven National Laboratory, Upton, NY~11973, USA}
\newcommand{\Brown}{Brown University, Providence, RI 02912, USA}
\newcommand{\BU}{Boston University, Boston, MA 02215, USA}
\newcommand{\Buffalo}{Department of Physics, University at Buffalo, SUNY Buffalo, NY 14260 USA}
\newcommand{\Caltech}{California Institute of Technology, Pasadena, CA 91125, USA}
\newcommand{\Cardiff}{School of Physics and Astronomy, Cardiff University, The Parade, Cardiff, CF24 3AA, UK}
\newcommand{\Carleton}{Carleton University, Ottawa, ON K1S 5B6, Canada}
\newcommand{\Carnegie}{The Observatories of the Carnegie Institution for Science, 813 Santa Barbara St., Pasadena, CA 91101, USA}
\newcommand{\Cavendish}{Astrophysics Group, Cavendish Laboratory, J.J.Thomson Avenue, Cambridge, CB3 0HE, UK}
\newcommand{\CCA}{Center for Computational Astrophysics, Flatiron Institute, New York, NY~10010, USA}
\newcommand{\CCAPP}{Center for Cosmology and AstroParticle Physics, The Ohio State University, Columbus, OH 43212}
\newcommand{\CCPP}{Center for Cosmology and Particle Physics, Department of Physics, New York University, 726 Broadway, Room 1005, New York, NY 10003, USA}
\newcommand{\CPPM}{Aix Marseille Univ, CNRS/IN2P3, CPPM, Marseille, France}
\newcommand{\CEADAP}{D\'epartement d'Astrophysique, CEA Saclay DSM/Irfu, 91191 Gif-sur-Yvette, France}
\newcommand{\CERN}{CERN, Geneva, Switzerland}
\newcommand{\CfA}{Harvard-Smithsonian Center for Astrophysics, Cambridge, MA 02138, USA}
\newcommand{\CFT}{Center for Theoretical Physics, Polish Academy of Sciences, 02-668 Warsaw, Poland}
\newcommand{\Cincinnati}{University of Cincinnati, Cincinnati, OH 45221, USA}
\newcommand{\CITA}{Canadian Institute for Theoretical Astrophysics, University of Toronto, Toronto, ON M5S 3H8, Canada}
\newcommand{\CNRSA}{CNRS, Laboratoire d'Annecy-le-Vieux de Physique Th\'{e}orique, Annecy-le-Vieux, France}
\newcommand{\CNYang}{C.N.\ Yang Institute for Theoretical Physics, State University of New York Stony Brook, NY~11794, USA}
\newcommand{\CNCT}{Consejo Nacional de Ciencia y Tecnolog\'ia, Av. Insurgentes Sur 1582. Colonia Cr\'edito Constructor, Del. Benito Ju\'arez, C.P. 03940, M\'exico D.F. M\'exico}
\newcommand{\CMUCosmo}{Department of Physics, McWilliams Center for Cosmology, Carnegie Mellon University, Pittsburgh, PA 15213, USA }
\newcommand{\Columbia}{Columbia University, New York, NY 10027, USA}
\newcommand{\Cornell}{Cornell University, Ithaca, NY 14853, USA}
\newcommand{\CPthree}{CP3-Origins, 5230 Odense, Denmark}
\newcommand{\CUBoulder}{Center for Astrophysics and Space Astronomy, Department of Astrophysical and Planetary Science, University of Colorado, Boulder, CO 80309, USA}
\newcommand{\CWRU}{Case Western Reserve University, Cleveland, OH 44106, USA}
\newcommand{\daa}{David A. Dunlap Department of Astronomy and Astrophysics, University of Toronto, ON, M5S 3H4, Canada}
\newcommand{\damtp}{Department of Applied Mathematics and Theoretical Physics, University of Cambridge, Cambridge CB3 0WA, UK}
\newcommand{\Dartmouth}{Department of Physics \& Astronomy, Dartmouth College, Hanover, NH 03755, USA}
\newcommand{\DESY}{DESY, 22607 Hamburg, Germany}
\newcommand{\dfa}{Departamento de F\'{\i}sica e Astronomia, Faculdade de Ci\^{e}ncias, Universidade do Porto, Porto, Portugal}
\newcommand{\DFI}{Departamento de F\'isica, FCFM, Universidad de Chile, Blanco Encalada 2008, Santiago, Chile}
\newcommand{\DOE}{US. Department of Energy, Germantown, MD 20874, USA}
\newcommand{\drexel}{Drexel University, Philadelphia, PA 19104, USA}
\newcommand{\Duke}{Duke University and Triangle Universities Nuclear Laboratory, Durham, NC 27708, USA}
\newcommand{\DukePhys}{Department of Physics, Duke University, Durham, NC 27708, USA}
\newcommand{\dunlap}{Dunlap Institute for Astronomy and Astrophysics, University of Toronto, ON M5S 3H4, Canada}
\newcommand{\Durham}{Department of Physics, Durham University, Durham DH1 3LE, UK}
\newcommand{\ED}{University of Edinburgh, Edinburgh EH8 9YL, UK}
\newcommand{\EPFL}{Institute of Physics, Laboratory of Astrophysics, Ecole Polytechnique F\'ed\'erale de Lausanne (EPFL), Observatoire de Sauverny, 1290 Versoix, Switzerland}
\newcommand{\ESO}{European Southern Observatory, Karl-Schwarzschild-Str. 2, DE-85748 Garching b. Munchen, Germany}
\newcommand{\ETH}{ETH Zurich, Institute for Particle Physics, 8093 Zurich, Switzerland}
\newcommand{\FNAL}{Fermi National Accelerator Laboratory, Batavia, IL 60510, USA}
\newcommand{\FQAUB}{Dept. de F\' isica Qu\` antica i Astrof\' isica, Universitat de Barcelona, Mart\' i i Franqu\` es 1, E08028 Barcelona, Spain}
\newcommand{\FSU}{Florida State University, Tallahassee, FL 32306, USA}
\newcommand{\Glasgow}{University of Glasgow, Glasgow G12 8QQ, UK}
\newcommand{\GRAPPA}{GRAPPA Institute, University of Amsterdam, 1098 XH Amsterdam, The Netherlands}
\newcommand{\GSFC}{Goddard Space Flight Center, Greenbelt, MD 20771, USA}
\newcommand{\GWU}{George Washington University, Washington, DC 20052, USA}
\newcommand{\Hampton}{Hampton University, Hampton, VA 23668, USA}
\newcommand{\HarvardPhys}{Department of Physics, Harvard University, Cambridge, MA 02138, USA}
\newcommand{\Haverford}{Haverford College, Haverford, PA 19041, USA}
\newcommand{\Hawaii}{University of Hawaii, Honolulu, HI 96822, USA}
\newcommand{\HKUST}{The Hong Kong University of Science and Technology, Hong Kong SAR, China}
\newcommand{\houston}{University of Houston, Houston, TX 77204, USA}
\newcommand{\IA}{Instituto de Astrof\'{\i}sica e Ci\^encias do Espa\c{c}o (IA), Porto, Portugal}
\newcommand{\IAC}{Instituto de Astrof\'{\i}sica de Canarias, 38200 La Laguna, Tenerife, Spain}
\newcommand{\IAP}{Institut d'Astrophysique de Paris (IAP), CNRS \& Sorbonne University, Paris, France}
\newcommand{\IAS}{Institute for Advanced Study, Princeton, NJ 08540, USA}
\newcommand{\IBS}{Institute for Basic Science (IBS), Daejeon 34051, Korea}
\newcommand{\ICC}{ICC, University of Barcelona, IEEC-UB, Mart\' i i Franqu\` es, 1, E08028 Barcelona, Spain}
\newcommand{\ICCD}{Institute for Computational Cosmology, Department of Physics, Durham University, Durham DH1 3LE, UK}
\newcommand{\ICE}{Institute of Space Sciences (ICE, CSIC), Campus UAB, Carrer de Can Magrans, s/n, 08193 Barcelona, Spain}
\newcommand{\ICFUNAM}{ICFUNAM - Instituto de Ciencias F\'{i}sicas, Universidad Nacional Aut\'onoma de M\'exico,  62210 Cuernavaca, Mor., M\'exico}
\newcommand{\ICJLab}{Universit\'e Paris-Saclay, CNRS/IN2P3, IJCLab, 91405 Orsay, France}
\newcommand{\ICRR}{Institute for Cosmic Ray Resaerch, The University of Tokyo, 456 Higashi-Mozumi, Kamioka, Hida, Gifu 506-1205, Japan}
\newcommand{\ICTP}{International Centre for Theoretical Physics, Strada Costiera, 11, I-34151 Trieste, Italy}
\newcommand{\IFAE}{Institut de Fisica d'Altes Energies, The Barcelona Institute of Science and Technology, Campus UAB, 08193 Bellaterra (Barcelona), Spain}
\newcommand{\IFPU}{IFPU - Institute for Fundamental Physics of the Universe, Via Beirut 2, 34014 Trieste, Italy}
\newcommand{\IFT}{Instituto de Fisica Teorica UAM/CSIC, Universidad Autonoma de Madrid, 28049 Madrid, Spain}
\newcommand{\IFUNAM}{IFUNAM - Instituto de F\'{i}sica, Universidad Nacional Aut\'onoma de M\'exico, 04510 CDMX, M\'exico}
\newcommand{\IHEP}{Institute of High Energy Physics, Austrian Academy of Sciences, 1050 Vienna, Austria}
\newcommand{\ILL}{Institut Laue-Langevin, 71 Avenue des Martyrs, 38000 Grenoble, France}
\newcommand{\Imperial}{Theoretical Physics, Blackett Laboratory, Imperial College, London SW7 2AZ, UK}
\newcommand{\Indiana}{Indiana University, Bloomington, IN 47405, USA}
\newcommand{\INAFOATs}{INAF - Osservatorio Astronomico di Trieste, Via G.B. Tiepolo 11, 34143 Trieste, Italy}
\newcommand{\INAFOAS}{INAF - Osservatorio di Astrofisica e Scienza dello Spazio di Bologna, via Piero Gobetti 93/3, I-40129 Bologna, Italy}
\newcommand{\INFNCag}{Istituto Nazionale di Fisica Nucleare, Sezione di Cagliari,  09126 Cagliari, Italy}
\newcommand{\INFNCat}{Istituto Nazionale di Fisica Nucleare, Sezione di Catania, 95125 Catania, Italy}
\newcommand{\INFNG}{Istituto Nazionale di Fisica Nucleare, Sezione di Genova, 16146 Genova, Italy}
\newcommand{\INFN}{INFN - National Institute for Nuclear Physics, Via Valerio 2, I-34127 Trieste, Italy}
\newcommand{\INFNFE}{Istituto Nazionale di Fisica Nucleare, Sezione di Ferrara, 40122, Italy }
\newcommand{\INFNLNF}{Istituto Nazionale di Fisica Nucleare, Laboratori Nazionali di Frascati, 00044 Frascati, Italy}
\newcommand{\INFNLNS}{Istituto Nazionale di Fisica Nucleare, Laboratori Nazionali del Sud, 95125 Catania, Italy}
\newcommand{\INFNN}{Istituto Nazionale di Fisica Nucleare, Sezione di Napoli, 80125 Napoli, Italy }
\newcommand{\INFNPD}{Istituto Nazionale di Fisica Nucleare, Sezione di Padova, 35131 Padova, Italy}
\newcommand{\INFNRM}{Istituto Nazionale di Fisica Nucleare, Sezione di Roma, 00185 Roma, Italy}
\newcommand{\INFNT}{Istituto Nazionale di Fisica Nucleare, Sezione di Torino, 10125, Italy }
\newcommand{\IoA}{Institute of Astronomy, University of Cambridge, Cambridge CB3 0HA, UK}
\newcommand{\IPP}{Institute for Particle Physics, Victoria, BC V8W 3P6, Canada}
\newcommand{\IPMU}{Kavli Institute for the Physics and Mathematics of the Universe, University of Tokyo, Kashiwa, Japan}
\newcommand{\IPNL}{Universit\'e de Lyon, F-69622, Lyon, France; Universit\'e de Lyon 1, Villeurbanne; CNRS/IN2P3, Institut de Physique Nucl\'eaire de Lyon}
\newcommand{\IRFU}{IRFU, CEA, Universit\'e Paris-Saclay, F-91191 Gif-sur-Yvette, France}
\newcommand{\ITFA}{Institute for Theoretical Physics, University of Amsterdam, Science Park 904, 1098 XH Amsterdam, The Netherlands}
\newcommand{\IUCAA}{The Inter-University Centre for Astronomy and Astrophysics, Pune, 411007, India}
\newcommand{\Jerusalem}{Hebrew University of Jerusalem, 91904 Jerusalem, Israel}
\newcommand{\JHU}{Johns Hopkins University, Baltimore, MD 21218, USA}
\newcommand{\JLAB}{Thomas Jefferson National Laboratory, Newport News, VA 23606, USA}
\newcommand{\JPL}{Jet Propulsion Laboratory, California Institute of Technology, Pasadena, CA 91011, USA}
\newcommand{\KASSI}{Korea Astronomy and Space Science Institute, Daejeon 34055, Korea}
\newcommand{\kavli}{Kavli Institute for Cosmology, University of Cambridge, Cambridge CB3 0HA, UK}
\newcommand{\KEK}{High Energy Accelerator Research Organization (KEK), Tsukuba, Ibaraki 305-0801, Japan}
\newcommand{\KIAS}{School of Physics, Korea Institute for Advanced Study, Dongdaemun-gu, Seoul 130-722, Korea}
\newcommand{\KICP}{Kavli Institute for Cosmological Physics, University of Chicago, Chicago, IL 60637, USA}
\newcommand{\KIPAC}{Kavli Institute for Particle Astrophysics and Cosmology, Stanford, CA 94305, USA}
\newcommand{\KINGS}{King's College London, London WC2R 2LS, UK}
\newcommand{\Kobe}{Kobe University, 657-8501 Kobe, Japan}
\newcommand{\KPH}{Johannes Gutenberg University, 55128 Mainz, Germany}
\newcommand{\KPMU}{University of Tokyo, 277-8583  Kashiwa, Japan}
\newcommand{\KSU}{Kansas State University, Manhattan, KS 66506, USA}
\newcommand{\KwaZuluNatal}{Astrophysics and Cosmology Research Unit, School of Chemistry and Physics, University of KwaZulu-Natal, Durban 4000, South Africa}
\newcommand{\Kyoto}{Department of Physics, Kyoto University, Kyoto 606-8502, Japan}
\newcommand{\Lafayette}{Lafayette College, Easton, PA 18042, USA}
\newcommand{\LANL}{Los Alamos National Laboratory, Los Alamos, NM 87545, USA}
\newcommand{\LBL}{Lawrence Berkeley National Laboratory, Berkeley, CA 94720, USA}
\newcommand{\Leiden}{Lorentz Institute, Leiden University, Niels Bohrweg 2,Leiden, NL 2333 CA, The Netherlands}
\newcommand{\Liverpool}{University of Liverpool, Liverpool L69 7ZE, UK}
\newcommand{\LJMU}{Liverpool John Moores University, Liverpool L3 5RF, UK}
\newcommand{\LLNL}{Lawrence Livermore National Laboratory, Livermore, CA, 94550, USA}
\newcommand{\LMU}{Ludwig-Maximilians-Universit\"at, 81679 Munich, Germany}
\newcommand{\LPC}{Universit\'e Clermont Auvergne, CNRS/IN2P3, Laboratoire de Physique de Clermont, F-63000 Clermont-Ferrand, France}
\newcommand{\LPNHE}{Sorbonne Universit\'e, Universit\'e Paris Diderot, CNRS/IN2P3, Laboratoire de Physique Nucl\'eaire et de Hautes Energies, LPNHE, F-75252 Paris, France}
\newcommand{\McGill}{McGill University, Montreal, QC H3A 2T8, Canada}
\newcommand{\Melbourne}{School of Physics, The University of Melbourne, Parkville, VIC 3010, Australia}
\newcommand{\MSU}{Michigan State University, East Lansing, MI 48824, USA}
\newcommand{\Mines}{Colorado School of Mines, Golden, CO 80401, USA}
\newcommand{\MIT}{Massachusetts Institute of Technology, Cambridge, MA 02139, USA}
\newcommand{\MilanoBicocca}{Department of Physics, University of Milano - Bicocca, Piazza della Scienza 3, I-20126 Milano, Italy}
\newcommand{\MPE}{Max-Planck-Institut f\"{u}r extraterrestrische Physik (MPE), Giessenbachstrasse 1, D-85748 Garching bei M\"unchen, Germany}
\newcommand{\MPIA}{Max-Planck-Institut f\"{u}r Astrophysik, Karl-Schwarzschild-Str. 1, 85741 Garching, Germany}
\newcommand{\MPP}{Max-Planck-Institut f\"{u}r Physik (Werner-Heisenberg-Institut), F\"ohringer Ring 6, D-80805 M\"unchen, Germany}
\newcommand{\LAM}{Aix Marseille Univ, CNRS, CNES, LAM, Marseille, France}
\newcommand{\LUPM}{Laboratoire Univers et Particules de Montpellier, Univ. Montpellier and CNRS, 34090 Montpellier, France}
\newcommand{\NAOC}{National Astronomical Observatories, Chinese Academy of Sciences, PR China}
\newcommand{\NCBJ}{National Center for Nuclear Research, Ul.Pasteura 7,Warsaw, Poland}
\newcommand{\NCU}{National Central University, Taoyuan City 32001, Taiwan (R.O.C.)}
\newcommand{\NCSU}{Physics Department, North Carolina State University, Raleigh, NC 27607, USA}
\newcommand{\ND}{University of Notre Dame, Notre Dame, IN 46556, USA}
\newcommand{\NIST}{National Institute of Standards and Technology, Boulder, CO 80305, USA}
\newcommand{\NIU}{Northern Illinois University, DeKalb, Illinois 60115, USA}
\newcommand{\NMSU}{New Mexico State University, Las Cruces, NM 88003, USA}
\newcommand{\NOAO}{National Optical Astronomy Observatory, Tucson, AZ 85719 USA}
\newcommand{\Northwestern}{Northwestern University, Evanston, IL 60201, USA}
\newcommand{\Nottingham}{University of Nottingham, Nottingham NG7 2RD, UK}
\newcommand{\NPPSFAmes}{NASA Postdoctoral Program Senior Fellow, NASA Ames Research Center, Moffett Field, CA 94035, USA}
\newcommand{\NWU}{Northwestern University, Evanston, IL 60208, USA}
\newcommand{\NYU}{New York University, New York, NY 10003, USA}
\newcommand{\OIRLab}{NSF's National Optical-Infrared Astronomy Research Laboratory, Tucson, AZ 85719, USA}
\newcommand{\OK}{ University of Oklahoma, Norman, OK 73019, USA}
\newcommand{\ORNL}{Oak Ridge National Laboratory, Oak Ridge, TN 37831, USA}
\newcommand{\OSU}{The Ohio State University, Columbus, OH 43212, USA}
\newcommand{\OU}{Department of Physics and Astronomy, Ohio University, Clippinger Labs, Athens, OH 45701, USA}
\newcommand{\OskarKlein}{Oskar Klein Centre for Cosmoparticle Physics, Stockholm University, AlbaNova, Stockholm SE-106 91, Sweden}
\newcommand{\Oxford}{University of Oxford, Oxford OX1~3RH, UK}
\newcommand{\Oxy}{Occidental College, Los Angeles, CA 90041, USA}
\newcommand{\ParisSud}{Universit\'{e} Paris-Sud, LAL, UMR 8607, F-91898 Orsay Cedex, France \& CNRS/IN2P3, F-91405 Orsay, France}
\newcommand{\PI}{Perimeter Institute, Waterloo, ON N2L 2Y5, Canada}
\newcommand{\Pitt}{University of Pittsburgh and PITT PACC, Pittsburgh, PA 15260, USA}
\newcommand{\PNNL}{Pacific Northwest National Laboratory, Richland, WA 99352, USA}
\newcommand{\PNPI}{Petersburg Nuclear Physics Institute, 188300 Gatchina, Russia}
\newcommand{\Port}{Institute of Cosmology \& Gravitation, University of Portsmouth, Portsmouth PO1 3FX, UK}
\newcommand{\Princeton}{Princeton University, Princeton, NJ 08544, USA}
\newcommand{\PSU}{The Pennsylvania State University, University Park, PA 16802, USA}
\newcommand{\PSUAstro}{Department of Astronomy and Astrophysics, The Pennsylvania State University, University Park, PA 16802, USA}
\newcommand{\PSUIGC}{Institute for Gravitation and the Cosmos, The Pennsylvania State University, University Park, PA 16802, USA}
\newcommand{\Purdue}{Purdue University, West Lafayette, IN 47907, USA}
\newcommand{\PW}{Participation Worldscope, Sedona, Arizona and Hong Kong, SAR PRC}
\newcommand{\Queens}{Queen's University, Kingston, ON K7L 3N6, Canada}
\newcommand{\Queensland}{The University of Queensland, School of Mathematics and Physics, QLD 4072, Australia}
\newcommand{\QMUL}{Queen Mary University of London, London E1~4NS, UK}
\newcommand{\RAL}{Radio Astronomy Laboratory, University of California Berkeley, Berkeley, CA 94720, USA}
\newcommand{\Rice}{Department of Physics \& Astronomy, Rice University, Houston, Texas 77005, USA}
\newcommand{\RIT}{Rochester Institute of Technology, Rochester, NY 14623, USA}
\newcommand{\RomaS}{Dipartimento di Fisica, Universit\`{a} La Sapienza, P. le A. Moro 2, Roma, Italy}
\newcommand{\RUG}{Kapteyn Astronomical Institute, University of Groningen, 9700 AV Groningen, The Netherlands}
\newcommand{\Rutgers}{Department of Physics and Astronomy, Rutgers, the State University of New Jersey, Piscataway, NJ 08854, USA}
\newcommand{\SAAO}{South African Astronomical Observatory, PO Box 9, Observatory 7935, Cape Town, South Africa}
\newcommand{\SAIMSU}{Sternberg Astronomical Institute, Moscow State University, Moscow, 119992, Russia}
\newcommand{\Sanford}{Sanford Underground Research Facility, Lead, SD 57754, USA}
\newcommand{\Sassari}{Universit\`a di Sassari, 07100 Sassari,  Italy}
\newcommand{\SCIPP}{University of California, Santa Cruz, Santa Cruz, CA 95064, USA}
\newcommand{\Sejong}{Department of Physics and Astronomy, Sejong University, Seoul, 143-747, Korea}
\newcommand{\Sheffield}{University of Sheffield, Sheffield S3 7RH, UK}
\newcommand{\SHAO}{Shanghai Astronomical Observatory (SHAO), Nandan Road 80, Shanghai 200030, China}
\newcommand{\Siena}{Siena College, Loudonville, NY 12211, USA}
\newcommand{\SISSA}{SISSA - International School for Advanced Studies, Via Bonomea 265, 34136 Trieste, Italy}
\newcommand{\SimonFraser}{Department of Physics, Simon Fraser University, Burnaby, BC  V5A 1S6, Canada}
\newcommand{\SLAC}{SLAC National Accelerator Laboratory, Menlo Park, CA 94025, USA}
\newcommand{\SMU}{Southern Methodist University, Dallas, TX 75275, USA}
\newcommand{\SNOLAB}{SNOLAB, Lively, ON P3Y 1N2, Canada}
\newcommand{\SoCal}{University of Southern California, CA 90089, USA}
\newcommand{\StAndrews}{School of Physics and Astronomy, University of St Andrews, North Haugh, St Andrews KY16 9SS, UK}
\newcommand{\Stanford}{Stanford University, Stanford, CA 94305, USA}
\newcommand{\StonyBrook}{Stony Brook University, Stony Brook, NY 11794, USA}
\newcommand{\STSCI}{Space Telescope Science Institute, Baltimore, MD 21218, USA}
\newcommand{\SUNYA}{University at Albany SUNY, Albany, NY 12222, USA}
\newcommand{\SUOT}{Swinburne University of Technology, Centre for Astrophysics and Supercomputing, Melbourne, VIC 3122, Australia}
\newcommand{\SussexAstronomy}{Astronomy Centre, School of Mathematical and Physical Sciences, University of Sussex, Brighton BN1 9QH, UK}
\newcommand{\Syracuse}{Syracuse University, Syracuse, NY 13244, USA}
\newcommand{\Tamu}{Texas A\&M University, College Station, TX 77843, USA }
\newcommand{\TCU}{Department of Physics \& Astronomy, Texas Christian University, Fort Worth, TX 76129, USA}
\newcommand{\Techsource}{Techsource Incorporated, Los Alamos, NM 87544, USA}
\newcommand{\TelAviv}{Tel-Aviv University,  69978 Tel-Aviv, Israel}
\newcommand{\Temple}{Temple University, Philadelphia, PA 19122, USA}
\newcommand{\TIFR}{Tata Institute of Fundamental Research, Homi Bhabha Road, Mumbai 400005 India}
\newcommand{\Tsinghua}{Department of Physics and Tsinghua Center for Astrophysics, Tsinghua University, Beijing 100084, P R China}
\newcommand{\TUM}{Technical University of Munich, 80333 Munich, Germany}
\newcommand{\UA}{University of Alabama, Tuscaloosa, AL 35487, USA}
\newcommand{\UAS}{Department of Astronomy/Steward Observatory, University of Arizona, Tucson, AZ  85721, USA}
\newcommand{\UAM}{Universidad Aut\'onoma de Madrid, 28049, Madrid, Spain}
\newcommand{\UBC}{University of British Columbia, Vancouver, BC V6T 1Z1, Canada}
\newcommand{\UCB}{Department of Astronomy, University of California, Berkeley, CA 94720, USA}
\newcommand{\UCBP}{Department of Physics, University of California, Berkeley, CA 94720, USA}
\newcommand{\UCBSSL}{Space Sciences Laboratory, University of California, Berkeley, CA 94720, USA}
\newcommand{\UCD}{University of California, Davis, CA 95616, USA}
\newcommand{\UChicago}{University of Chicago, Chicago, IL 60637, USA}
\newcommand{\UCI}{University of California, Irvine, CA 92697, USA}
\newcommand{\UCLA}{University of California, Los Angeles, Los Angeles, CA 90095, USA}
\newcommand{\UCL}{University College London, London WC1E 6BT, UK}
\newcommand{\UCR}{University of California, Riverside, CA 92521, USA}
\newcommand{\UCSB}{University of California, Santa Barbara, CA 93106, USA}
\newcommand{\UCSC}{University of California, Santa Cruz, CA 95064, USA}
\newcommand{\UCSD}{University of California, San Diego, La Jolla, CA 92093, USA}
\newcommand{\UCT}{Department of Astronomy, University of Cape Town, Rondebosch 7701, South Africa}
\newcommand{\UFL}{University of Florida, Gainesville, FL 32611, USA}
\newcommand{\Unige}{D\'epartement de Physique Th\'eorique et CAP, Universit\'e de Gen\`eve, CH-1211 Gen\`eve 4, Switzerland}
\newcommand{\UFN}{Universit\`a Federico II di Napoli, 80125 Napoli, Italy}
\newcommand{\UGTO}{Divisi\'on de Ciencias e Ingenier\'ias, Universidad de Guanajuato, Le\'on 37150, M\'exico}
\newcommand{\UKY}{University of Kentucky, Lexington, KY 40506, USA}
\newcommand{\UMD}{University of Maryland, College Park, MD 20742, USA}
\newcommand{\UMiami}{University of Miami, Coral Gables, FL 33124, USA}
\newcommand{\UMich}{University of Michigan, Ann Arbor, MI 48109, USA}
\newcommand{\UMN}{University of Minnesota, Minneapolis, MN 55455, USA}
\newcommand{\UnB}{Instituto de F\'{i}sica, Universidade de Bras\'{i}lia, 70919-970, Bras\'{i}lia, DF, Brazil}
\newcommand{\UNC}{University of North Carolina at Chapel Hill, Chapel Hill, NC 27599, USA}
\newcommand{\UNH}{University of New Hampshire, Durham, NH 03824, USA}
\newcommand{\UNIMI}{Dipartimento di Fisica ``Aldo Pontremoli'', Universit\`a{} degli Studi di Milano, via Celoria 16, 20133 Milano, Italy}
\newcommand{\UNIPD}{Dipartimento di Fisica e Astronomia ``G. Galilei'', Universit\`a degli Studi di Padova, 35131~Padova, Italy}
\newcommand{\UNM}{Department of Physics and Astronomy, University of New Mexico, Albuquerque, NM 87131, USA}
\newcommand{\UNV}{University of Nevada, Reno, NV 89557, USA}
\newcommand{\UoM}{Jodrell Bank Center for Astrophysics, School of Physics and Astronomy, University of Manchester, Manchester M13 9PL, UK}
\newcommand{\UPenn}{Department of Physics and Astronomy, University of Pennsylvania, Philadelphia, PA 19104, USA}
\newcommand{\UR}{Department of Physics and Astronomy, University of Rochester, Rochester, NY 14627, USA}
\newcommand{\UrbanaC}{Illinois Center for Advanced Studies of the Universe \& Department of Physics, University of Illinois at Urbana-Champaign, Urbana, IL 61801, USA}
\newcommand{\USC}{The University of South Carolina, Columbia, SC 29208, USA}
\newcommand{\USD}{The University of South Dakota, Vermillion, SD 57069, USA}
\newcommand{\UTD}{University of Texas at Dallas, Texas 75080, USA}
\newcommand{\UTAustinth}{Theory Group, Department of Physics, University of Texas at Austin, Austin, TX 78712, USA}
\newcommand{\Utenn}{The University of Tennessee, Knoxville, TN 37996, USA}
\newcommand{\Utah}{ Department of Physics and Astronomy, University of Utah, Salt Lake City, UT 84112, USA}
\newcommand{\UVA}{University of Virginia, Charlottesville, VA 22903, USA}
\newcommand{\Uvic}{University of Victoria, Victoria, BC V8P 5C2, Canada}
\newcommand{\UWaterloo}{Department of Physics and Astronomy, University of Waterloo, Waterloo, ON N2L 3G1, Canada}
\newcommand{\UWisc}{Department of Astronomy, University of Wisconsin-Madison, Madison, WI 53706, USA}
\newcommand{\UWMadison}{Department of Physics, University of Wisconsin-Madison, Madison, WI 53706, USA}
\newcommand{\UW}{University of Washington, Seattle, WA 98195, USA}
\newcommand{\UWC}{Department of Physics \& Astronomy, University of the Western Cape, Cape Town 7535, South Africa}
\newcommand{\USTCH}{Department of Astronomy, School of Physical Sciences, University of Science and Technology of China, Hefei, Anhui 230026, China}
\newcommand{\Vanderbilt}{Physics \& Astronomy Department, Vanderbilt University, Nashville, TN 37235, USA}
\newcommand{\VSI}{Van Swinderen Institute for Particle Physics and Gravity, University of Groningen, 9747~AG~Groningen, The~Netherlands}
\newcommand{\VT}{Virginia Tech, Blacksburg, VA 24061, USA}
\newcommand{\VUU}{Virginia Union University, Richmond, Virginia, 23220, USA}
\newcommand{\WCA}{Waterloo Centre for Astrophysics, University of Waterloo, Waterloo, ON N2L 3G1, Canada} 
\newcommand{\Weizmann}{Weizmann Institute of Science, 76100 Rehovot, Israel}
\newcommand{\Wellesley}{Wellesley College, Wellesley, MA 02481, USA}
\newcommand{\wiscIce}{University of Wisconsin, Madison, WI 53706, USA}
\newcommand{\WM}{College of William and Mary, Newport News, VA 23606, USA}
\newcommand{\WUSL}{Washington University in St Louis, St. Louis, MO 63130, USA}
\newcommand{\McDoWUSL}{McDonnell Center for the Space Sciences, Washington University, St. Louis, MO 63130, USA}
\newcommand{\WVU}{CSEE, West Virginia University, Morgantown, WV 26505, USA}
\newcommand{\WVUGWAC}{Center for Gravitational Waves and Cosmology, West Virginia University, Morgantown, WV 26505, USA}
\newcommand{\Wyoming}{Department of Physics and Astronomy, University of Wyoming, Laramie, WY 82071, USA}
\newcommand{\Yale}{Department of Physics, Yale University, New Haven, CT 06520, USA}
\newcommand{\YorkU}{Department of Physics and Astronomy, York University, Toronto, ON M3J 1P3, Canada}

% This file is automatically generated by a script, so edits may be overwritten. 
% author.py --optin BB_forecast/Membership_downloaded_by_ATC_20200810.csv BB_forecast/additional_authors.csv --optout BB_forecast/BB_forecast_optout.csv --rulesfile CMB-S4_rules.pkl --prefix BB_forecast/

\noindent\textsuperscript{\ref{UCIrvine}}\UCI

\noindent\textsuperscript{\ref{JohnsHopkinsUniversity}}\JHU

\noindent\textsuperscript{\ref{UniversityofIllinoisatUrbana-Champaign}}\UrbanaC

\noindent\textsuperscript{\ref{SLAC}}\SLAC

\noindent\textsuperscript{\ref{UCBerkeley}}\UCBP

\noindent\textsuperscript{\ref{StanfordUniversity}}\Stanford

\noindent\textsuperscript{\ref{OxfordUniversity}}\Oxford

\noindent\textsuperscript{\ref{LawrenceBerkeleyNationalLaboratory}}\LBL

\noindent\textsuperscript{\ref{RiceUniversity}}\Rice

\noindent\textsuperscript{\ref{Fermilab}}\FNAL

\noindent\textsuperscript{\ref{UCSanDiego}}\UCSD

\noindent\textsuperscript{\ref{SISSA}}\SISSA

\noindent\textsuperscript{\ref{IFPU}}\IFPU

\noindent\textsuperscript{\ref{INFN}}\INFN

\noindent\textsuperscript{\ref{HarvardUniversityCFA}}\CfA

\noindent\textsuperscript{\ref{UniversityofNewMexico}}\UNM

\noindent\textsuperscript{\ref{ArgonneNationalLaboratory}}\ANL

\noindent\textsuperscript{\ref{AstroParticleandCosmologyLaboratory}}\APC

\noindent\textsuperscript{\ref{Caltech}}\Caltech

\noindent\textsuperscript{\ref{CornellUniversity}}\Cornell

\noindent\textsuperscript{\ref{KICP}}\KICP

\noindent\textsuperscript{\ref{UniversityofChicago}}\UChicago

\noindent\textsuperscript{\ref{UniversityofMelbourne}}\Melbourne

\noindent\textsuperscript{\ref{UniversityofCincinnati}}\Cincinnati

\noindent\textsuperscript{\ref{JPL}}\JPL

\noindent\textsuperscript{\ref{LMUMunich}}\LMU

\noindent\textsuperscript{\ref{UTAustin}}\UTAustinth

\noindent\textsuperscript{\ref{CITA}}\CITA

\noindent\textsuperscript{\ref{InstitutdAstrophysiquedeParis}}\IAP

\noindent\textsuperscript{\ref{StonyBrookUniversity}}\StonyBrook

\noindent\textsuperscript{\ref{UniversityofManchester}}\UoM

\noindent\textsuperscript{\ref{ArizonaStateUniversity}}\ASU

\noindent\textsuperscript{\ref{FloridaStateUniversity}}\FSU

\noindent\textsuperscript{\ref{CardiffUniversity}}\Cardiff

\noindent\textsuperscript{\ref{DartmouthCollege}}\Dartmouth

\noindent\textsuperscript{\ref{UniversityofGeneva}}\Unige

\noindent\textsuperscript{\ref{UniversityofSussex}}\SussexAstronomy

\noindent\textsuperscript{\ref{UniversityofCambridge}}\damtp

\noindent\textsuperscript{\ref{KavliInstituteforCosmologyCambridge}}\kavli

\noindent\textsuperscript{\ref{IoA}}\IoA

\noindent\textsuperscript{\ref{UniversityofTokyo}}\KPMU

\noindent\textsuperscript{\ref{UniversityofGroningen}}\VSI

\noindent\textsuperscript{\ref{UniversityofToronto}}\daa

\noindent\textsuperscript{\ref{Dunlap}}\dunlap

\noindent\textsuperscript{\ref{KEK}}\KEK

\noindent\textsuperscript{\ref{CEASaclay}}\CEADAP

\noindent\textsuperscript{\ref{USTCH}}\USTCH

\noindent\textsuperscript{\ref{UniversityofPennsylvania}}\UPenn

\noindent\textsuperscript{\ref{McGillUniversity}}\McGill

\noindent\textsuperscript{\ref{NIST}}\NIST

\noindent\textsuperscript{\ref{PrincetonUniversity}}\Princeton

\noindent\textsuperscript{\ref{HarvardUniversity}}\HarvardPhys

\noindent\textsuperscript{\ref{NASAGoddardSpaceFlightCenter}}\GSFC

\noindent\textsuperscript{\ref{SimonFraserUniversity}}\SimonFraser

\noindent\textsuperscript{\ref{UniversityofSouthernCalifornia}}\SoCal

\noindent\textsuperscript{\ref{HaverfordCollege}}\Haverford

\noindent\textsuperscript{\ref{StockholmUniversity}}\OskarKlein

\noindent\textsuperscript{\ref{CaseWesternReserveUniversity}}\CWRU

\noindent\textsuperscript{\ref{UniversityofBritishColumbia}}\UBC

\noindent\textsuperscript{\ref{UniversityofColoradoBoulder}}\CUBoulder

\noindent\textsuperscript{\ref{UniversityofMinnesota}}\UMN

\noindent\textsuperscript{\ref{UniversityofMichigan}}\UMich

\noindent\textsuperscript{\ref{FlatironInstitute}}\CCA

\noindent\textsuperscript{\ref{Columbia}}\Columbia

\noindent\textsuperscript{\ref{UniversityofVirginia}}\UVA

\noindent\textsuperscript{\ref{KavliIPMU}}\IPMU

\noindent\textsuperscript{\ref{UCDavis}}\UCD

\noindent\textsuperscript{\ref{YaleUniversity}}\Yale

\noindent\textsuperscript{\ref{UniversityofPittsburgh}}\Pitt

\noindent\textsuperscript{\ref{BenGurionUniversity}}\BenGurion

\noindent\textsuperscript{\ref{AaltoUniversity}}\Aalto

\noindent\textsuperscript{\ref{UCSantaBarbara}}\UCSB

\noindent\textsuperscript{\ref{PerimiterInstitue}}\PI

\noindent\textsuperscript{\ref{SouthernMethodistUniversity}}\SMU

\noindent\textsuperscript{\ref{BerkeleyCenterforCosmologicalPhysics}}\BCCP

\noindent\textsuperscript{\ref{EuropeanSouthernObservatory}}\ESO

\noindent\textsuperscript{\ref{GRAPPA}}\GRAPPA

\noindent\textsuperscript{\ref{WashingtonUniversitySt.Louis}}\WUSL

\noindent\textsuperscript{\ref{McDonnellCenterfortheSpaceSciencesWashingtonUniversity}}\McDoWUSL

\noindent\textsuperscript{\ref{UniversityofMilano-Bicocca}}\MilanoBicocca

\noindent\textsuperscript{\ref{CentredePhysiquedesParticulesdeMarseille}}\CPPM

\noindent\textsuperscript{\ref{KIPAC}}\KIPAC

\noindent\textsuperscript{\ref{InstituteforAdvancedStudy}}\IAS

\noindent\textsuperscript{\ref{PennsylvaniaStateUniversity}}\PSU

\noindent\textsuperscript{\ref{BrookhavenNationalLaboratory}}\BNL

\noindent\textsuperscript{\ref{KyotoUniversity}}\Kyoto

\noindent\textsuperscript{\ref{UniversityofWisconsinMadison}}\UWMadison

\noindent\textsuperscript{\ref{UniversitedegliStudidiMilan}}\UNIMI

\noindent\textsuperscript{\ref{ICJLab}}\ICJLab

\noindent\textsuperscript{\ref{BrownUniversity}}\Brown

\noindent\textsuperscript{\ref{SyracuseUniversity}}\Syracuse

\noindent\textsuperscript{\ref{UCLosAngeles}}\UCLA

\noindent\textsuperscript{\ref{MSU}}\MSU

\onecolumngrid}

\cleardoublepage
%------ ABSTRACT ------
\begin{abstract}
CMB-S4---the next-generation ground-based cosmic microwave background (CMB) experiment---is set to
significantly advance the sensitivity of CMB measurements and enhance 
our understanding of the origin and evolution of the Universe, 
from the highest energies at the dawn of time through the growth of structure to the present day. 
Among the %many compelling 
science cases pursued with CMB-S4, 
the quest for detecting primordial gravitational waves is a central driver of the experimental design. 
This work details the development of a forecasting framework that includes
a power-spectrum-based semi-analytic projection tool, 
targeted explicitly towards optimizing constraints on the tensor-to-scalar ratio, $r$, 
in the presence of Galactic foregrounds and gravitational lensing of the CMB. 
This framework is unique in its direct use of information from the achieved performance 
of current Stage 2--3 CMB experiments to robustly forecast the science reach 
of upcoming CMB-polarization endeavors. 
The methodology allows for rapid iteration over experimental configurations 
and offers a flexible way to optimize the design of future experiments 
given a desired scientific goal. 
To form a closed-loop process, we couple this semi-analytic tool 
with map-based validation studies, which allow for the injection of additional complexity and verification of our forecasts with several independent analysis methods. 
We document multiple rounds of forecasts for CMB-S4 using this process and
the resulting establishment of the current reference design of the primordial gravitational-wave component of the Stage-4 experiment, optimized to achieve our science goals of detecting primordial gravitational waves for $r > 0.003$ at greater than $5\sigma$, or, in the absence of a detection, of reaching an upper limit of $r < 0.001$ at $95\%$ CL.\\
\end{abstract}


\correspondingauthor{Victor Buza}
\email{vbuza@kicp.uchicago.edu}

\maketitle
%------ SECTION : INTRODUCTION ------
\section{Introduction}
\begin{figure*}[htbp]
\begin{center}
\includegraphics[height=4.0in]{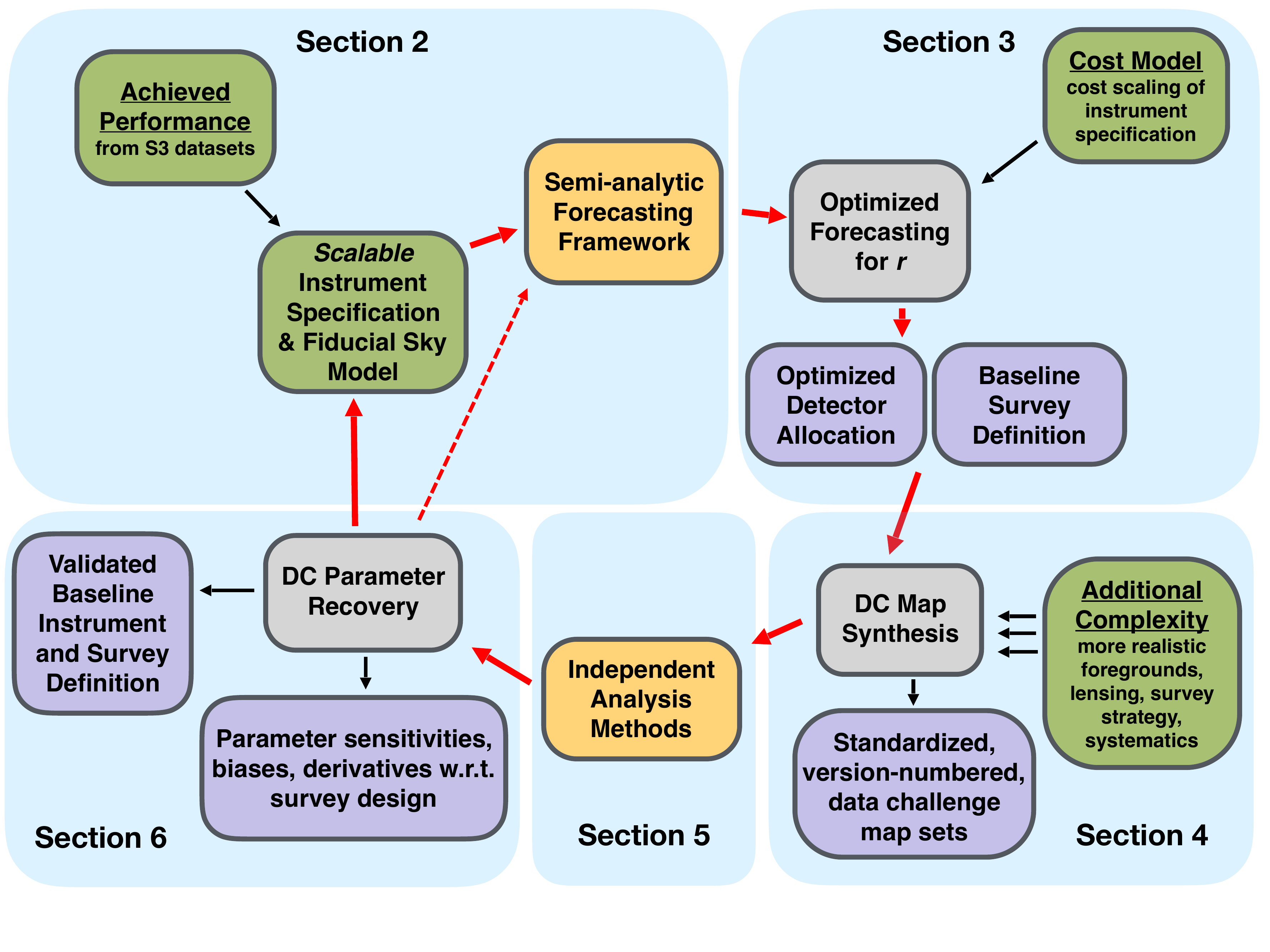}
\caption{Schematic representation of the CMB-S4 \textit{r} forecasting loop. Green boxes represent inputs, purple boxes represent outputs, yellow boxes represent large code frameworks, and gray boxes represent iterable code modules. For each stage of the loop we identify the sections of this paper in which more detail is available. We start with the achieved performance of Stage 3 datasets, in the form of full covariance matrices and noise spectra, and a set of scalable instrument specifications, as well as a fiducial sky model. These are fed as inputs to the semi-analytic optimization framework, yielding an optimized detector allocation and a baseline survey definition. Based on this definition, we develop standard data challenge (DC) noise maps, as well as a suite of signal maps with various degrees of complexity. We proceed by analyzing these maps with multiple independent component-separation analysis methods, and check for parameter recovery and the presence of biases. If the results suggest a necessary change in survey definition, or additional studies are required, the process is iterated as needed.}
\label{fig:forecasting_loop}
\end{center}
\end{figure*}

Determining the origin of structure in the Universe is one of the most important open problems in cosmology. CMB anisotropies sourced by early-Universe density perturbations are currently the most powerful observational probe of the earliest mechanisms of structure formation. It is possible that the same processes that produced the density perturbations also sourced tensor perturbations, or primordial gravitational waves (PGWs). If this is the case, detecting a PGW signal would yield insight into physics far earlier than the epoch of recombination, and allow us to build an unprecedented understanding of the earliest moments of time.

PGWs leave imprints on the polarization of the CMB~\citep{Seljak:1996gy,kamionkowski97,Hu:1997pp}. In particular, the sensitivity of CMB measurements to gravitational waves arises from the generation of polarization at the surface of last scattering: to first order, scalar perturbations produce only even-parity \textit{E}-mode polarization, while tensor perturbations produce odd-parity \textit{B}-mode polarization as well. Thus, a measurement of primordial \textit{B}-mode polarization in the CMB, parametrized by the tensor-to-scalar ratio $r$, is a direct measurement of the amplitude of tensor perturbations. Detecting $r$ has profound implications for high-energy physics and the quantum nature of gravity~\citep{krauss13}, and the potential to shed light on the mechanism that produced these primordial perturbations. 

Cosmic inflation is our current leading paradigm for what occurred in the very early Universe. It was first put forward to explain the lack of observed magnetic monopoles and to solve the flatness and horizon problems~\citep{Starobinsky:1980te, Kazanas:1980, Sato:1981, Guth, Linde, Linde:1983gd, Albrecht} and has since been an active field of research. The theory describes a period of exponential expansion in which quantum fluctuations are magnified to cosmic size and become the seeds for all structure in the Universe~\citep{Mukhanov:1981xt, Mukhanov:1982, Hawking:1982, Guth:1982, Starobinsky:1982, Bardeen:1985tr, Mukhanov:1985}. In addition to the production of PGWs~\citep[for a recent review, see][]{Kamionkowski:2015yta}, inflation makes several predictions, most of which---superhorizon fluctuations, Gaussian perturbations, adiabatic fluctuations, spatial flatness, and a nearly scale invariant scalar spectral tilt---have been confirmed, most recently by the \planck\ collaboration~\citep{planck2018X}. 

There are currently a number of ground-based experiments measuring the CMB polarization to high precision on a range of scales, and attempting to constrain the tensor-to-scalar ratio, including ACT~\citep{Aiola:2020}, BICEP/Keck~\citep{Array:2018x}, CLASS~\citep{CLASSSPIE2016}, POLARBEAR/Simons Array~\citep{Suzuki:2016, Hasegawa:2018}, and SPT~\citep{Bender:2018, Sayre:2019}, with Simons Observatory to follow soon~\citep{SOforecast}. Additionally, there are current and future balloon and satellite missions such as SPIDER~\citep{Gualtieri:2018} and LiteBIRD~\citep{Hazumi:2019}, which we expect to complement ground-based measurements. The current best constraints are $r(k_\mathrm{s}=0.05\,\text{Mpc}^{-1}, n_\mathrm{t}=0)<0.06$ at $95\%$ confidence~\citep{Array:2018x}, where $k_\mathrm{s}$ is the scalar pivot scale and $n_\mathrm{t}$ is the spectral index of the tensor modes. Ongoing efforts in both the Atacama desert 
and at the South Pole, between now and the start of CMB-S4, will steadily improve these constraints while continuing to prove the methodologies on which CMB-S4 will rely.

CMB-S4, anticipated to start observations in 2027, is intended to be the definitive ground-based CMB polarization experiment. It is designed to cross critical thresholds in constraining the \textit{B}-mode polarization signature of primordial gravitational waves and in sensitivity to new light relics, while also improving our understanding of the nature of dark energy and General Relativity on large scales~\citep{CMBS4-scibook}.
To achieve these goals requires a significant increase in sensitivity, from $2$--$4\times10^4$ detectors in Stage 3 experiments to roughly $5\times10^5$ detectors, and an unparalleled control over other sources of signal (e.g., Galactic foregrounds, gravitational lensing, etc.) and of systematics. Therefore, CMB-S4 will require telescopes at multiple frequencies, each with a maximally outfitted focal plane of pixels utilizing superconducting, photon-noise-limited detectors, and likely novel analysis techniques. To understand the optimal design for achieving the desired science goals, in particular focusing on primordial gravitational waves, we present the development of the \textit{CMB-S4 r forecasting framework} and its application towards determining the CMB-S4 baseline \textit{r} survey.

The sensitivity achieved by a CMB experiment, which observes for a given number of years with a given number of detectors, is subject to a number of efficiency factors. These include the fraction of detectors that are actually functional, the achieved sensitivity per detector versus model prediction, the fraction of days per year spent observing, the fraction of observing time spent ``on field'', and the fraction of data passing weather and other cuts. As an illustrative example, if each of the above efficiency factors were 0.65 then the product is $0.65^5\sim0.1$.  
Forecasts for the sensitivity to $r$, where excess low frequency noise or 
systematic contamination can lead to additional (potentially large) sensitivity loss, are particularly challenging.  
Most previous forecasts of the sensitivity of CMB experiments to $r$ have been 
ab initio---requiring assumptions as to the value of all efficiency factors. 
Possibly due to the human tendency towards optimism, many of these forecasts have not in practice been achieved once the data have been taken and analyzed: %~\citep{2009CMBpolworkshop}
ABS ~\citep{Hileman:2011, Kusaka:2018}; BICEP/Keck ~\citep{Hivon:2002, Yoon:2006, Nguyen:2008, Array:2015xqh}; EBEX ~\citep{Reichborn:2010, Abitbol:2018}; \textit{Planck} ~\citep{Planck:2006aa, planck2018X}; POLARBEAR ~\citep{Lee:2008, Kermish:2012eh, Adachi:2019}; QUIET ~\citep{Lawrence:2004, Araujo:2012}; SPIDER ~\citep{Montroy:2006, Fraisse:2013, Gualtieri:2018}; SPTpol ~\citep{McMahon:2009, Austermann:2012ga, Sayre:2019}.
Therefore for CMB-S4 we take an alternate approach, scaling from the overall 
performance achieved in the best available existing experimental results. 
We scale directly from published \textit{B}-mode noise spectra and bandpower covariance matrices 
derived from multi-year maps that have passed systematics null tests. 
This bypasses the need for an unbiased, individual accounting of the various 
efficiency factors, and naturally incorporates all effects which impact real-world experiments. This also differentiates our current approach from forecasts other groups have made for CMB-S4 prior to the formal existence of our collaboration~\citep{Wu:2014hta, Errard:2015cxa, Barron:2018}. 

\subsection{The CMB-S4 r Forecasting Loop}
The CMB-S4 baseline \textit{r} survey, described below, has been continuously evolving by translating science requirements to measurement and instrument requirements, based on our understanding of the impact of
astrophysical foregrounds, instrumental systematics, delensing
non-idealities, and analysis methodology.
To maintain forecasting realism as complexity increases, 
our general approach has been an iterative one. 
We rely on a closed forecasting loop, presented in
Figure \ref{fig:forecasting_loop}, to tie the semi-analytic tools (which allow for fast optimizations) to map-based studies (which can include multiple layers of additional complexity).
To establish our measurement requirements and the baseline experiment
configurations that can achieve them, we perform multiple passes
through this loop. In the figure, the section number accompanying each stage of the loop indicates the paper section in which that stage is described in detail.

The main steps describing this process are as follows.
\begin{enumerate}
       \item Develop a semi-analytic power-spectrum-level forecast, assuming noise performance that is scaled from analyses of real experiments. 
       \item Use this forecasting tool to optimize the allocation of detector effort across observing frequencies, determining certain ``checkpoints'' in survey 
       definition space.
       \item Use the checkpoint configurations to create standardized, version-numbered map-based data challenges (DCs) for validation.
       \item Estimate science parameters from the DC maps with independent component-separation analysis methods.
       \item Check that independent analyses show recovery of science parameters from these challenge maps that match analytic forecasts, either in terms of variance or bias. If they do not, we revise the forecasts accordingly.
       \item Iterate steps 1--5, injecting increasing realism in the form of: (a) sky model complexity informed by the latest data and modeling efforts; (b) survey coverage based on proven observing strategies; (c) systematics whose form, parameterization, and likely amplitude are likewise guided by real-world experience; and (d) treatment of lensing.
\end{enumerate}
\vspace{0.2in}

\subsection{CMB-S4 r Forecasting Work-flow and Evolution}
Developing the forecasting machinery for CMB-S4, and increasing the
robustness and realism of its results, has been an ongoing,
ever-evolving, group effort, performed under the auspices of the
CMB-S4 \textit{r} Forecasting Working Group. 
This work has undergone three major iterations, with results presented
in the CMB-S4 Science Book~\citep{CMBS4-scibook}, the CMB-S4 Concept Definition Task force (CDT) Report~\citep{CMBS4-CDT}, and most recently in the CMB-S4 Reference
Design Report~\citep{CMBS4-DSR}.

Each stage in the evolution of this framework has been collectively
considered, and carefully documented in the CMB-S4 Simulation and Forecasting Logbook.\footnote{\url{https://cmb-s4.org/wiki/index.php/Simulation\_and\_Forecasting\_Logbook}}
The resulting baseline survey definitions have been translated to
publicly available, version-numbered map-based Data Challenges (DCs).\footnote{\url{https://cmb-s4.org/wiki/index.php/Data\_Challenges}}
%The forecasting group has welcomed participation from members of the collaboration towards testing the strengths and shortcomings of different analysis methods on these simulations, as well as improving the simulations themselves.
Thus far, two independent groups have participated in testing the strengths and shortcomings of different analysis methods on these simulations, as well as improving the simulations themselves, as described in Sections~\ref{sec:analysis_methods} and
\ref{sec:map_results}.

Given the difficulty in describing a continuously-evolving process, for this paper we present several snapshots of our forecasting work and describe the relevant details. Sections~\ref{sec:fopt} to \ref{sec:map_results} represent a full pass through the forecasting loop, as presented in Figure \ref{fig:forecasting_loop}. In Sections~\ref{sec:fopt} and \ref{sec:optimization} we present the full semi-analytic forecasting framework and optimization process, as used in the CMB-S4 CDT Report. In Section~\ref{sec:cdt_sims} we discuss the creation of the DC4 simulation suite, corresponding to the baseline presented in that report. Sections~\ref{sec:analysis_methods} and \ref{sec:map_results} describe two independent analysis methods and the results obtained from applying these methods to DC4. In Section~\ref{sec:DSR} we describe several recent updates to the framework and the resulting findings presented in the CMB-S4 Reference Design Report. We conclude this paper with a discussion of the path forward. 

%------ SECTION : Performance-based Forecasts ------
\section{Semi-analytic Forecasting Framework}
\label{sec:fopt}
For the CMB-S4 Science Book~\citep{CMBS4-scibook}, we developed a semi-analytic forecasting framework specifically targeted towards optimizing sensitivity to the tensor-to-scalar ratio, $r$, in the presence of Galactic foregrounds and gravitational lensing of the CMB. Currently, only small aperture telescopes (SATs) have reached the level of systematics control and noise performance necessary to pursue a ground-based, high-precision measurement of \textit{B}-mode polarization down to low multipoles ($\ell\simeq 30$), targeting the $\ell \simeq 80$ peak from the polarization signature generated by PGWs at the epoch of recombination. Therefore, to forecast the performance of next-generation SATs, this machinery is based on scaling the bandpower covariance matrices (BPCMs) and noise power spectra ($N_{\ell}$) of current published SAT analyses, such as those from BICEP/Keck \citep{Array:2015xqh, Array:2018x}. This automatically builds into the forecast all real-world inefficiencies including (but not limited to): imperfect detector yield; non-uniform detector performance; read-out noise; observing inefficiency; losses due to timestream filtering; beam smoothing; and non-uniform sky coverage.

At its core, this code is based on the BICEP/Keck parametric power-spectrum-based likelihood analysis. Such types of parametric analyses have also been extensively used by the \textit{Planck} collaboration for $\ell \ge 50$~\citep{Planck:2013kta, Aghanim:2015xee, 2019arXiv190712875P}. We validate this approach using map-based simulations as part of the forecasting loop and present detailed results in Section~\ref{sec:map_results}. Our confidence in the projections is grounded in the connection to achieved performance and published results. %In particular, we emphasize the importance of using map-level signal and noise simulations as a starting point. These simulations are a good description of the maps because they pass jackknife tests derived from them.  

\begin{figure}[htbp]
\begin{center}
\includegraphics[trim={2cm 0.6cm 0.4cm 1cm},clip=true,width=\columnwidth]{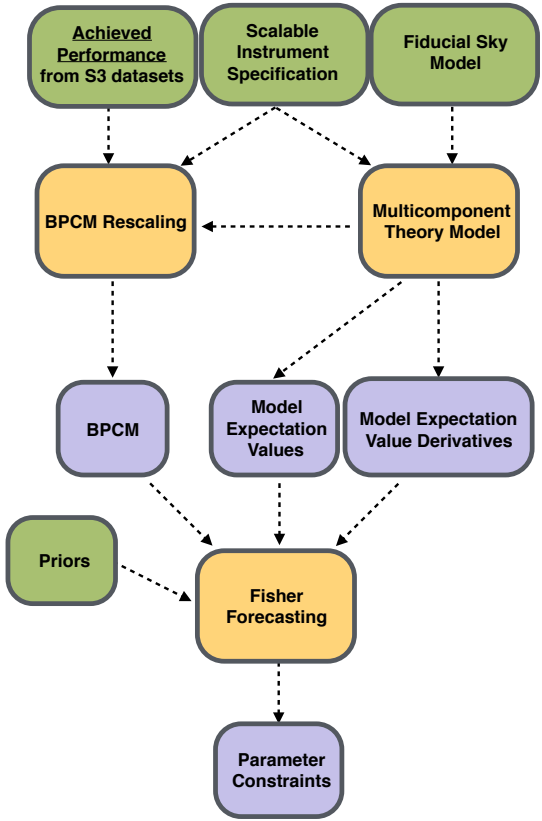}
\caption{Schematic representation of our semi-analytic forecasting framework. Green boxes represent user inputs, yellow boxes represent code modules, and purple boxes represent outputs. BPCM = bandpower covariance matrix.}
\label{fig:fisher}
\end{center}
\end{figure}

Figure~\ref{fig:fisher} presents a schematic representation of the semi-analytic forecasting framework, identifying the user inputs, code modules, and outputs. The subsections that follow describe this framework in detail.

%\vspace{-0.2in}
\subsection{Fisher Formalism}
\label{sec:fisher_formalism}
%\noindent \emph{Fisher Forecast}\\\\
Given a likelihood function of the form
\begin{equation}
L(\theta;d)\propto\frac{\exp{\left[-\frac{1}{2}(d-\mu(\theta))^{\sf{T}}\Sigma(\theta)^{-1}(d-\mu(\theta))\right]}}{\sqrt{\mathrm{det}(\Sigma(\theta))}},
\label{eq:gauss_lik}
\end{equation}
where $d$ are the data bandpowers, $\theta$ are the theory parameters, and $\mu(\theta)$ and $\Sigma(\theta)$ are the bandpower expectation values and the bandpower covariance matrix given the parameters, we can calculate the expectation value of the log-likelihood curvature, evaluated at the position of the best fit model: 
\begin{equation}
F_{ij}=-\left\langle\frac{\partial^2\log{L(\theta;d)}}{\partial\theta_i\partial\theta_j}\right\rangle.
\label{eq:FIM}
\end{equation}
This quantity, called the Fisher information matrix, measures how steeply the likelihood falls as we move away from the best-fit model, and $F^{-1}$ can be thought of
as the best possible covariance matrix for the measurement errors on the parameters $\theta_i$. It can be shown that $\sqrt{(F^{-1})_{ii}}$ is the minimum obtainable standard deviation on the desired parameters~\citep[e.g.,][]{Cramer:1946, Kendall:1979, Tegmark:1996a}.

Inserting Equation \ref{eq:gauss_lik} into Equation \ref{eq:FIM} yields
\begin{equation}F_{ij}=\frac{\partial\mu^{\sf{T}}}{\partial\theta_i}\Sigma^{-1}\frac{\partial\mu}{\partial\theta_j}+\frac{1}{2}\mathrm{Tr}(\Sigma^{-1}\frac{\partial\Sigma}{\partial\theta_i}\Sigma^{-1}\frac{\partial\Sigma}{\partial\theta_j}).
\label{eq:fisher_eq}
\end{equation}
We then calculate our parameter constraints as
\begin{equation}
\sigma_{i}=\sqrt{(F^{-1})_{ii}}.
\end{equation}
In all the projections below, for each fiducial model considered, we choose to fix the covariance matrix with respect to the theory parameters, i.e., $\Sigma(\theta)=\Sigma$, making the second term of Equation~\ref{eq:fisher_eq} identically zero. Equation~\ref{eq:fisher_eq} provides a clear picture of how the construction of the covariance matrix $\Sigma$ impacts the final constraints, and how its misestimation could lead to constraints that are far too optimistic. It is with this in mind that we have decided to compute our BPCMs by directly scaling the values in achieved covariance matrices (see Section~\ref{sec:bpcm}).

\subsection{Forecasting Inputs}
In this section we briefly describe the set of inputs to the forecasting code, shown as green boxes in Figure~\ref{fig:forecasting_loop}. We discuss the key inputs in more detail in Sections~\ref{sec:Exptspec} and ~\ref{sec:multicomp}.

\vspace{1em}
\textit{Achieved Performance}: The code takes signal and noise simulations of the BICEP/Keck datasets~\citep{Array:2015xqh, Array:2018x} and derives the bandpower covariance matrix and the ensemble-averaged signal and noise bandpowers. These inputs contain information about the actual on-sky map noise achieved from multiple receivers, over multiple years, at 95, 150, and \SI{220}{\GHz}, including all real-world penalties. %the real penalties for detector yield, distribution of detector performance, weather, and observing efficiency. These inputs also fold in the incomplete mode coverage due to sky coverage, scan strategy, beam smoothing, and filtering during data analysis. 
A more detailed description of these simulations is available in Appendix H of \citet{Array:2018x}. Similar information from another experiment could easily be substituted. To project the performance of CMB-S4 channels, we assume that we can scale down the achieved noise based on increased detector count and integration time and that we can apply beam-size and noise-equivalent temperature (NET) rescalings to account for the differences in experimental design. 

%We operate under the assumption that detector noise is uncorrelated and we do not suffer heavily from common-mode atmospheric noise.

\vspace{1em}
\textit{Scalable Instrument Specification}: To specify the forecast instrument we start by selecting a set of observing frequency bands, with bandpass functions describing the response of each band. Then, for each observing band, we must provide the beam size, number of detectors, and ideal per-detector NET. The last two items can be used to make an idealized calculation of the instrument sensitivity in each band. We emphasize that we use these ideal performance numbers \textit{only} for scaling between frequency bands, by comparing to similarly-calculated ideal sensitivities of BICEP/Keck. The scaling factors are ultimately applied to the achieved sensitivities, as described in Section \ref{sec:bpcm}, to obtain performance-based sensitivities in our desired bands. We note that for the frequency bands in which we do not currently have existing data, we extrapolate from the closest frequency for which we do.

\vspace{1em}
\textit{Fiducial Sky Model}: A fiducial parametric model describing the foreground and CMB signal. Our standard model, which has 14 parameters, is discussed in Section \ref{sec:multicomp} and in Appendix A.
%For the projections included in this paper, we choose the foreground amplitudes for dust and synchrotron to the best-fit values in the BICEP/Keck patch ($A_{dust}=4.25 \mu K^2_{CMB}$, $A_{sync}<3.8 \mu K^2_{CMB}$ at $95\%$). We consider two options for tensors, r=0 or r=0.003.

\vspace{1em}
\textit{Priors}: If we have external prior knowledge for a given parameter $\theta_i$, we can introduce this information by adding $P_i=1/\sigma_i^2$ to the diagonal of the Fisher matrix, where $\sigma_i$ is the width of the prior. 

\subsection{CMB-S4 Scalable Instrument Specification}
\label{sec:Exptspec}
\begin{figure*}[htbp]
\begin{center}
\includegraphics[height=3.2in]{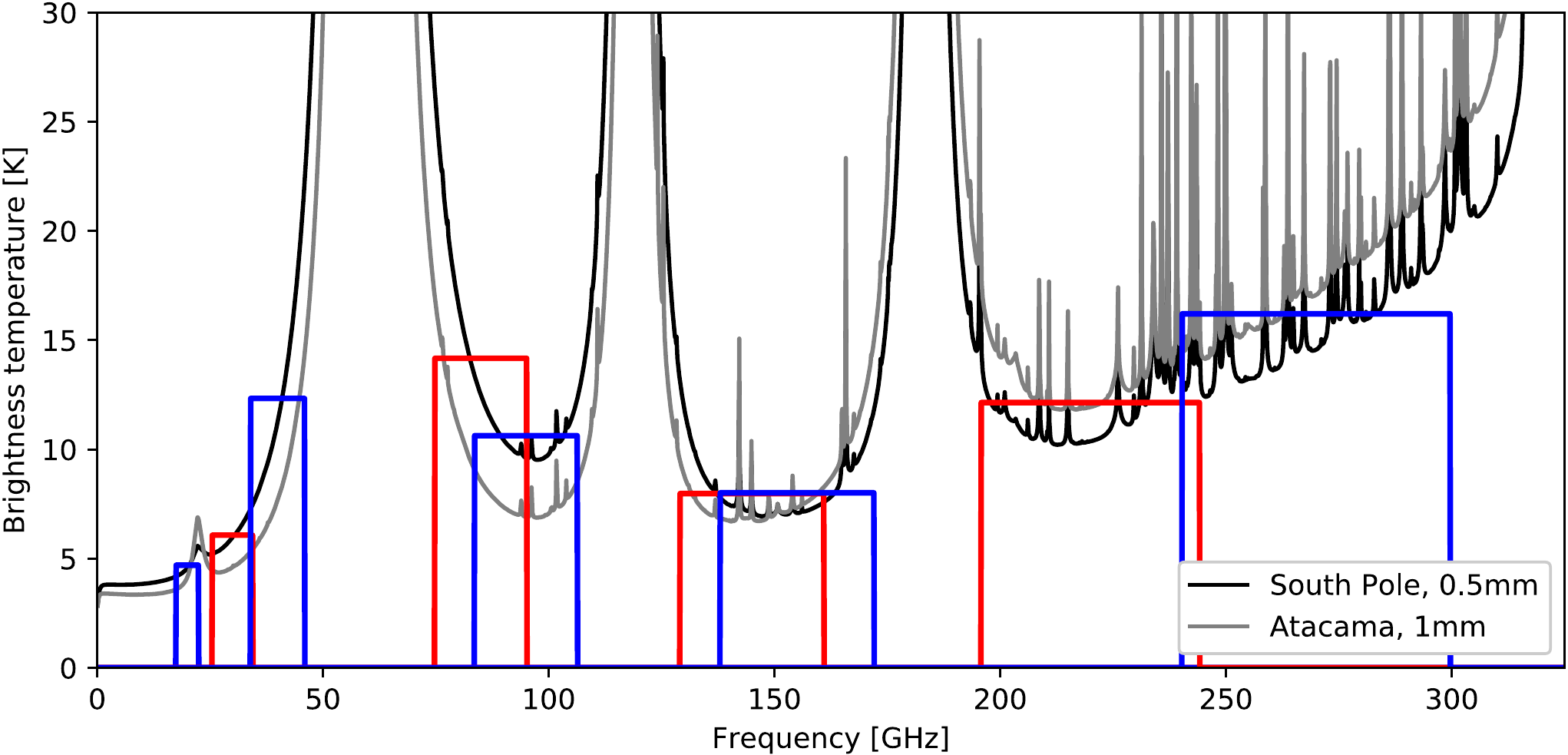}
\caption{Calculated atmospheric brightness temperature spectra (at zenith) for the South Pole at \SI{0.5}{\mm} precipitable water vapor (PWV) and Atacama at \SI{1.0}{\mm} PWV (both are near the median values). Atmospheric spectra are generated using the \texttt{am} Atmospheric Model~\citep{Paine:2017am}. The top-hat bands, in red and blue, are plotted on top of these spectra, with the height of each rectangle equal to the band-averaged brightness temperature using the South Pole spectrum. Details about the bands, such as fractional bandwidths, are presented in Table 3-1 of \cite{CMBS4-DSR}.}
\label{fig:bpass}
\end{center}
\end{figure*}

To span the four atmospheric windows available to ground-based experiments (Figure~\ref{fig:bpass}) and have enough channels to mitigate against complex foregrounds, we assume eight channels at 30, 40, 85, 95, 145, 155, 220, and \SI{270}{\GHz}, which are placed on small-aperture telescopes. 
For these low-resolution instruments, we pick 0.52\nobreakdash-m apertures, motivated by proven SAT Stage-3 experiments~\citep{BICEP3}, yielding the beams presented in Table~\ref{tab:instrument}. In addition, we also include a 20\nobreakdash-GHz channel on a large-aperture telescope (LAT).\footnote{The inclusion of a 20\nobreakdash-GHz channel is the result of insight gained from an earlier iteration through the forecasting loop (performed for the CMB-S4 Science Book), which demonstrated that for specific foreground models, sizeable biases were present due to synchrotron residuals. To mitigate against such biases the reference design was updated to include this additional channel.
Placing this low-frequency band on a SAT would result in a very broad beam, which would dominate the noise at the relevant scales; to circumvent this, as mentioned above, we place it on a LAT. This means that while the scaling of the noise for this channel is still calculated from achieved performance, we use noise parameters (see Equation \ref{eq:Nlfit}) that are more in tune with what has been achieved by LATs at the time of this forecasting~\citep{Louis2017, Henning:2017nuy}, i.e., the $1/f$ noise component is characterized by an $\ell_{\mathrm{knee}}=200$, keeping $\gamma$ -- the slope of this component -- the same as for the small-aperture noise, and a beam of $\Theta_{\nu}=11'$ FWHM (assuming a 6\nobreakdash-m aperture).\\}

We derive the split in each window by separating the overlapping bands as far as possible while still keeping the calculated per-detector NET within 10--15\% of the NET for a detector that spans the full window. %~\cite{Barkats:2017net} 
The ideal per-detector  NETs were calculated with {\tt NETlib.py}\footnote{\url{https://cmb-s4.org/wiki/index.php/New\_NET\_Calculator\_and\_Validation}}\textsuperscript{,}\footnote {\url{https://github.com/dbarkats/NET\_forecast\_python}} at the South Pole and Chile, using the 10-year MERRA2 median atmospheric profiles~\citep{2017_merra2}.
We use the average of NET calculated for detectors at the two sites, which are 214, 177, 224, 270, 238, 309, 331, 747 and \SI{1281}{\micro\mathrm{K}_{\mathrm{CMB}}\sqrt{\mathrm{s}}} for our nine channels at 20--\SI{270}{\GHz}, respectively.\footnote{Going beyond the number of bands for the current reference design increases manufacturing complexity and costs, and decreases the per-channel NET, thereby reducing the overall sensitivity in each channel; while exploring alternative options with 5 bands (20, 30, 95, 155, 270 GHz) and 7 bands (20, 30, 95, 155, 220, 270 GHz) has shown that choosing fewer bands leads to statistically significant biases on $r$ after marginalization over foreground residuals (an increase in bias of up to $\Delta r=1.5\times10^{-4}$ when compared to the current count). Balancing these considerations, we have chosen the proposed configuration with nine frequency bands for this work. Going forward, as instrumentation choices are finalized, we anticipate a possible revision of this design.}.
These NETs are lower than similar BICEP/Keck channels (which are 287.6, 313.1 and \SI{837.7}{\micro\mathrm{K}_{\mathrm{CMB}}\sqrt{\mathrm{s}}} for 95, 150 and \SI{220}{GHz}) predominantly because they are calculated for a 100\nobreakdash-mK thermal bath, as opposed to \SI{250}{mK} (which was also used in the CMB-S4 Science Book forecasts). 
%Note that this is the only departure from achieved performance.
%We want to emphasize that these NET
%values are only used to determine the appropriate scalings 
%to the current achieved performance values, and not to calculate ab-initio sensitivities.

We also fold in information from two WMAP channels, 23 and \SI{33}{\GHz}, and seven \textit{Planck} channels, 30, 44, 70, 100, 143, 217, and \SI{353}{\GHz}, though this extra information is only relevant in the early stages of CMB-S4 observation.

Since the forecasts use scaled BICEP/Keck bandpower statistics, they also use the same bandpower window functions and uniform binning: nine multipole bins with $\Delta\ell = 35$, spanning a multipole range of $21 \leq \ell \leq 335$.

\subsubsection{Delensing Treatment}
\label{sec:delens_treatment}

One of the main challenges for detecting primordial \textit{B} modes is the lensing \textit{B}-mode contribution from the weak lensing of \textit{E} modes as the CMB photons travel to us. For instrument noise levels below 5 $\mu$K-arcmin, this lensing signal becomes an important source of contamination and its sample variance significantly worsens our constraining power on PGWs~\citep{Smith:2010gu}. Unlike Galactic foregrounds, the lensing signal is achromatic and cannot be mitigated with multi-frequency observations. However, its contribution can be reduced by knowing the cumulative gravitational lensing potential $\phi$ along the line of sight, and having a high-fidelity \textit{E}-mode map. Together, the two can be combined to form a lensing \textit{B}-mode template by lensing the \textit{E}-mode map with the $\phi$ field and subtracting this template from the measured \textit{B}-mode map. This technique is known as $delensing$~\citep{Knox:2002pe, Seljak:2003pn, Carron2017, Carron:2017b}. 

Reconstructing $\phi$ with high S/N requires high-sensitivity, high-angular resolution CMB polarization maps~\citep{Lewis:2006fu}. Therefore, in addition to the low-resolution effort, we assume a separate high-resolution large-aperture instrument dedicated to measuring the intermediate- and small-scale information necessary to delens the \textit{B}-mode map. This instrument is assumed to have 1-arcminute resolution and noise performance equivalent to the 145\nobreakdash-GHz channel from the small-aperture telescopes. These experiment specifications are revised for the CMB-S4 Reference Design Report and updated in Section~\ref{sec:DSR}. 
The translation between detector effort and map noise in the delensing instrument is based on the method used for the low-resolution instrument (as described in Section ~\ref{sec:noisim} and Equation~\ref{eq:Nlfit}), but using SPTpol achieved performance \citep{Sayre:2019}, i.e., without incurring penalties from non-idealities specific to low-resolution instruments and low-$\ell$ analysis, such as low-$\ell$ mode filtering and non-uniform coverage. Following the iterative formalism in \citet{Smith:2010gu}\footnote{Delensing estimators which are technically more optimal have been introduced in \citet{Carron2019} and \citet{Millea:2020}, and we are currently developing tools to test their feasibility in further iterations of our forecasting.}, using $\ell_{\mathrm{min}}=300$ and $\ell_{\mathrm{max}}=4000$ for the $\phi$ reconstruction and $\ell_{\mathrm{min}}=30$ for the \textit{E}-mode map, we convert the map noise in the delensing survey to a delensing efficiency, or equivalently a fractional residual in the lensed \textit{B}-mode power, specified by setting the residual lensing amplitude $A_\mathrm{L}$ to the corresponding level.

The detector effort dedicated to the delensing instrument comes out of the total detector effort budget for the $r$ survey, and the distribution of effort between the low-resolution and delensing instruments is part of the optimization process, as shown in Figures~\ref{fig:sigr} and \ref{fig:mapdepth}.

\subsection{Multicomponent Theory Model}
\label{sec:multicomp}
%\noindent \emph{Multicomponent Theory Model}:
%This module calculates the parametrized model of one or more signal fields, including how the different signals scale to the various observing frequencies.
Our model includes a CMB component parametrized by $r$ and the residual lensing amplitude, $A_\mathrm{L}$, and components of polarized dust and synchrotron emission. We assume that the synchrotron scales as a simple power law in both frequency and $\ell$. For the dust we assume a power-law scaling in $\ell$ and a modified blackbody spectral energy distribution (SED). We allow for spatially correlated synchrotron and dust, parameterized by a single correlation parameter $\epsilon$~\citep{Choi:2015s, Spass:2018}; the effective frequency scaling of this correlation depends on the relative strength of the two components. In addition, we also consider dust and synchrotron frequency decorrelation parameters, which allow their spatial pattern to change with frequency, suppressing the correlation of foreground signals between observing bands. A detailed description of the full parametric model is presented in Appendix \ref{app:Appendix A}. The current model is easily extendable to accommodate additional complexities that have not yet been captured; alternatively, other models could be substituted here as well.

\vspace{0.1in}
\noindent
The model parameters are:
\begin{itemize}[leftmargin=*,label={--}]
  \setlength\itemsep{0em}
\item{$r$, tensor-to-scalar ratio, at pivot scale $k_\mathrm{s}=0.05 \, \text{Mpc}^{-1}$}
\item[] \quad and spectral index of the tensor modes $n_\mathrm{t}=0$;
\item{$A_\mathrm{L}$, residual lensing amplitude};
\item{$A_\mathrm{dust}$, dust amplitude, in $\mu \mathrm{K}^2_\text{CMB}$, 
\item[] \qquad \quad at $353$ GHz and $\ell=80$};
\item{$\beta_\mathrm{d}$, dust spectral index};
\item{$T_\mathrm{d}$, dust temperature};
\item{$\alpha_\mathrm{d}$, dust spatial spectral index};
\item{$\Delta_\mathrm{d}$, dust frequency correlation, 
\item[] \qquad between 217 and 353 GHz, at $\ell=80$};
\item{$EE/BB_\mathrm{dust}$, power ratio for dust};
\item{$A_\mathrm{sync}$, synchrotron amplitude, in $\mu \mathrm{K}^2_\text{CMB}$, 
\item[] \qquad \quad at $23$ GHz and $\ell=80$};
\item{$\beta_\mathrm{s}$, synchrotron spectral index};
\item{$\alpha_\mathrm{s}$, synchrotron spatial spectral index};
\item{$\Delta_\mathrm{s}$, synchrotron frequency correlation,
\item[] \qquad between 23 and 33 GHz, at $\ell=80$};
\item{$EE/BB_\mathrm{sync}$, power ratio for synchroton};
\item{$\epsilon$, dust/synchrotron spatial correlation}.
\end{itemize}

For a given auto- or cross-spectrum, we step through the model components, combine the appropriate amplitude functions for the two bands contributing to the spectrum, and apply the bandpower window functions to obtain the binned expectation values. Finally, we sum over model components to find the total expectation value for that spectrum. In addition, since a Fisher forecast requires knowledge of the response of the model expectation values with respect to the model parameters, we also output the derivatives of the model expectation values.

The Fisher matrix that we consider is usually 10-dimensional. The parameters we constrain are $r$, $A_\mathrm{dust}$, $\beta_\mathrm{d}$, $\alpha_\mathrm{d}$, $\Delta_\mathrm{d}$, $A_\mathrm{sync}$, $\beta_\mathrm{s}$, $\alpha_\mathrm{s}$, $\Delta_\mathrm{s}$, $\epsilon$. We fix
$T_\mathrm{d}=$\SI{19.6}{\K} because this parameter is mostly degenerate with $A_\mathrm{dust}$ for observations below \SI{300}{\GHz}, where the SED is in the Rayleigh-Jeans limit. The parameter $A_\mathrm{L}$ is assumed to be known, but its value is adjusted to represent varying levels of delensing, as discussed in Section~\ref{sec:delens_treatment}. The
$EE/BB$ ratios are not relevant for calculations presented here because we are focusing on constraints from the $BB$ spectrum only; however, these parameters are left in for possible future forecasting. 
The fiducial model used for forecasting is centered at either $r = 0$ or $0.003$, with $A_\mathrm{dust}=4.25\,\mu \mathrm{K}^2_\mathrm{CMB}$~\citep[best-fit value from][]{Array:2015xqh} and $A_\mathrm{sync}=3.8\,\mu \mathrm{K}^2_\mathrm{CMB}$~\citep[$95\%$ upper limit from][]{Array:2015xqh}. In Sections \ref{sec:fopt}--\ref{sec:map_results} we assume no variation of these foreground amplitudes over the sky fractions relevant to this study, i.e., they are always pinned to the values listed above. We revisit this assumption in Section \ref{sec:DSR}. The spatial and  frequency spectral indices are centered at the preferred \textit{Planck} and WMAP values~\citep{planckiXXII,Fuskeland}: 
$\beta_\mathrm{d}=1.59$ (with Gaussian prior of width $0.11$); $\beta_\mathrm{s}=-3.10$ (with a Gaussian prior of width $0.30$); $\alpha_\mathrm{d}=-0.42$; $\alpha_\mathrm{s}=-0.6$; and the dust/synchrotron correlation is centered at $\epsilon=0$. Following \cite{Aghanim:2017}, the central dust correlation value is taken to be $\Delta_\mathrm{d}=0.97$ (3\% decorrelation) and 
the synchrotron correlation value is assumed to be $\Delta_\mathrm{s}=1$ (no fiducial synchrotron decorrelation). Unless otherwise stated, the parameters have flat unbounded priors.  

\subsection{Bandpower Covariance Matrix Rescaling}
%\noindent \emph{BPCM (Bandpower Covariance Matrix) Rescaling}:
\label{sec:bpcm}

\subsubsection{Signal Scaling}
%\textit{Signal Scaling}:
The output model expectation values are also useful in the construction of our bandpower covariance matrix. To construct the BPCM components, we use lensed-$\Lambda$CDM + BICEP/Keck noise simulations. %(described in Appendix H of \cite{Array:2018x}) 
However, because we have the individual signal-only, noise-only, and signal$\times$noise terms, we can record all the individual BPCM components:
\begin{itemize}[leftmargin=1em,label=]
\label{it:bpcm_cpts}
  \item{sig = signal-only terms $\mathrm{Cov}(S_i\times S_j, S_k\times S_l)$;}
  \item{noi = noise-only terms $\mathrm{Cov}(\mathcal{N}_i\times \mathcal{N}_j, \mathcal{N}_k\times \mathcal{N}_l)$;}
  \item{sn1 = signal$\times$noise terms $\mathrm{Cov}(S_i\times \mathcal{N}_j,S_k\times \mathcal{N}_l)$;}
  \item{sn2 = signal$\times$noise terms $\mathrm{Cov}(S_i\times \mathcal{N}_j, \mathcal{N}_k\times S_l)$;}
  \item{sn3 = signal$\times$noise terms $\mathrm{Cov}(\mathcal{N}_i\times S_j, S_k\times \mathcal{N}_l$);}
  \item{sn4 = signal$\times$noise terms $\mathrm{Cov}(\mathcal{N}_i\times S_j, \mathcal{N}_k\times S_l)$.}
\end{itemize}
Here $S$ are signal simulations, $\mathcal{N}$ are 
noise simulations, and the indices $i, j, k, l$ run over the experimental frequency channels. 
%($\text{CMB-S4}_{\nu=\SI{20}{\GHz}}$,  $\text{CMB-S4}_{\nu=\SI{30}{\GHz}}$, etc.). 
%fields in the analysis, i.e., all combinations of a map type $(T, E, B)$ and an experiment ($S4_{20}$,  $S4_{30}$, etc.). 
%For many combinations of indices, we set certain covariance terms to be identically zero - $Cov(N_{S4_{20}}\times N_{S4_{20}}, N_{S4_{30}}\times N_{S4_{30}})$, for example. 

While calculating the covariances from the signal and noise simulations, we also record the average signal bandpowers from the simulations. For a new signal sky model, we can calculate the new bandpower expectation values, and rescale the signal components in the bandpower covariance matrix by the appropriate power of the ratio of the recorded average signal bandpowers and the newly calculated expectation values. The full BPCM construction and rescaling procedure is presented in Appendix~\ref{app:Appendix B}. 

When we do this, we set to zero any term that has an expectation value of zero (under the assumption that signal and noise are uncorrelated, and different signals
are uncorrelated) to reduce the Monte Carlo error in the resulting covariance matrix, given the relatively modest number of 499 realizations used. We also set to zero the covariance between bandpowers that are separated by more than one bin in $\ell$, but, importantly, preserve the covariance between the auto- and cross-spectra of the different frequency bands.

It is worth noting that this procedure allows us to have different numbers of degrees of freedom per bandpower for noise than for signal, which is a byproduct of signal and noise entering differently in a real analysis. This complication is often ignored in other forecasts by setting the noise and signal degrees of freedom to be identical.

The ability to estimate a BPCM for any model means that only a single set of simulations is necessary, and one does not have to run simulations for any and all conceivable scenarios. As already mentioned, in all the projections below we choose to fix $\Sigma(\theta)=\Sigma$ and hence we only apply the rescaling step once per fiducial model considered, i.e., we do not rescale our BPCM at every step along the way.

%\vspace{1em}
\subsubsection{Noise Scaling}
\label{sec:noi_scale}
%\textit{Noise Scaling}:
In addition to scaling from one signal model to another, recording all the covariance terms allows us to rescale the noise components as well. Given a dataset for which we have simulations, the noise scaling can be performed in one of two ways. The first is to take a frequency present in the dataset and scale down the noise in the BPCM by the desired amount. In particular, each BPCM component is scaled independently by $\sqrt{N_{\ell,\mathrm{S4}}/N_{\ell,\mathrm{BK}}}$ for each factor of $\mathcal{N}$ present. %i.e., each cross term is scaled by $N_{l,S4}/N_{l,BK}$, while the noise-only term is scaled by $(N_{l,S4}/N_{l,BK})^2$. 
The second way is to add an additional frequency, for which we do not yet have data, by taking the covariance structure of an existing frequency, scaling down the noise by the amounts described above, and then expanding the BPCM by filling it in with the appropriate variance and covariance terms between the new band and all the existing ones. These tools facilitate the construction of a new data structure to explore any combination of frequency bands, with any sensitivity in each band.

%To obtain projected weights for any of the S4 channels from achieved sensitivities, we scale the achieved survey weights from BICEP/Keck at $\{95, 150, 220\}$ GHz. 

%By definition, the achieved per-detector-year survey weight can be written as follows
%\begin{equation}
%w_{BK,achieved}^{per-det-yr}=\frac{t_{obs}}{\alpha_{BK,ideal/achieved} NET^2_{BK,ideal}},
%\end{equation}
%where $\frac{t_{obs}}{\alpha_{BK,ideal/achieved}}$ is the factor that takes us from the ideal scenario to reality, and that encapsulates the less-than-ideal observing time, receiver performance, cuts, etc. We never have to actively calculate this factor because we operate with ratios of survey weights. 

To obtain $N_\ell$ values for a CMB-S4 channel by scaling the achieved $N_\ell$ values of a Stage-3 channel, we have to first scale by the ratio of their respective survey weights and then scale by the ratio of beam window functions, $B_\ell^2$. The survey weight is defined as $w=2f_\mathrm{sky}/\sigma_\mathrm{map}^2$, where $f_\mathrm{sky}$ is the effective sky area and $\sigma_\mathrm{map}$ is the $Q$/$U$ map noise level. The input simulations use measured non-Gaussian $B_\ell$ shapes, but we rescale based on Gaussian approximations that are close to the true functions. We can write the noise spectrum of a CMB-S4 channel as
\begin{equation}
N_{\ell,\mathrm{S4}}=N_{\ell,\mathrm{BK}}\frac{w_{\mathrm{BK,achieved}}}{w_{\mathrm{S4,projected}}}\frac{B^2_{\ell,\mathrm{S4}}}{B^2_{\ell,\mathrm{BK}}},
\label{eq:Nl}
\end{equation}
where $B^2_{\ell,\nu}=\exp{\frac{-\ell(\ell+1)\Theta^2_{\nu}}{8\log(2)}}$, $\Theta_{\nu}$ is the full width at half maximum (FWHM), in radians, of the Gaussian beam, and $w_{\mathrm{i, achieved}}$ is the achieved integrated survey weight of a particular instrument.

To obtain the projected achieved survey weight for any of the CMB-S4 channels, we rescale the achieved survey weights as 
\begin{equation}
\frac{w_{\mathrm{S4,projected}}}{w_{\mathrm{BK,achieved}}}=\frac{n_{\mathrm{S4}}^{\mathrm{det-yr}}}{n_{\mathrm{BK}}^{\mathrm{det-yr}}}\frac{NET^2_{\mathrm{BK,ideal}}}{NET^2_{\mathrm{S4,ideal}}},
\end{equation}
where $n_{\mathrm{S4}}^{\mathrm{det-yr}}$ is the number of detector-years assumed for CMB-S4 at any particular frequency, and $n_{\mathrm{BK}}^{\mathrm{det-yr}}$
is the number of detector-years in the BK15 dataset, with the instruments in their final state~\citep{Array:2018x}.

%$w_{BK,achieved}^{per-det-yr}=$83.3, 58.6, and 3.9~$\mu K^{-2}$ are the average BICEP/Keck per detector-year survey weights at 95, 150, and 220~GHz, calculated directly from multi-year maps and therefore including all the non-idealities described in the sections above~\citep{Array:2018x}. %The per-detector values consistent with published results are $w_{BK,achieved}^{per-det-yr}=\{83.3, 58.6, 3.9\} \mu K^{-2}$ (\cite{Array:2018x}). 
The implicit assumption in this step is that the performance of this new CMB-S4 frequency channel falls short of idealized performance by the same factor as the real map from which we are scaling. The survey weight scaling is always performed from the closest frequency for which we have available simulation inputs: 20--95\nobreakdash-GHz are scaled from BICEP/Keck 95\nobreakdash-GHz data; 145 and \SI{155}{\GHz} are scaled from BICEP/Keck 150\nobreakdash-GHz data; and 220 and \SI{270}{\GHz} are scaled from BICEP/Keck 220\nobreakdash-GHz data.

Since we are using BICEP/Keck products, which are calculated with a particular sky mask $m$ (with $f^{\mathrm{BK}}_{\mathrm{sky}}=\frac{\Omega_{\mathrm{pix}}}{4\pi}\sum_i{m_i} \simeq 1$\%, where the sum is over the pixel $i$), we must also scale these products appropriately to evaluate the effect of different sky fractions. We propagate the effects of $f_{\mathrm{sky}}$ in the noise spectra and BPCM in two ways: first, we inflate the \(N_l\) values by a factor \(\beta=f^{\mathrm{S4}}_{\mathrm{sky
}}/f^{\mathrm{BK}}_{\mathrm{sky}}\), which boosts the (signal $\times$ noise) and (noise $\times$ noise) terms of the covariance matrix by $\beta$ and $\beta^2$, to take into account the redistribution of the achieved sensitivity onto a larger patch. Note that the (signal $\times$ signal) component remains unchanged in this step. Second, we scale down the entire covariance matrix by a factor of $\beta$ to increase the number of degrees of freedom in the BPCM, accounting for the fact that we are now observing more modes. This procedure scales the signal and noise degrees of freedom independently, preserving the relative effects that filtering and non-uniform coverage have on the covariance structure. We revisit the way this scaling is performed in Section~\ref{sec:DSR}, where we separately take into account the impact of realistic observing strategies on the various components. 

With the $N_\ell$ scalings in hand, we can perform the aforementioned BPCM operations to arrive at a scaled CMB-S4 BPCM that encompasses the intricacies of realistic observing conditions.

\section{Optimized Forecasting for \it{\lowercase{r}}}
\label{sec:optimization}
In this section, we answer the following question: given a fixed amount of effort and the instrument specifications offered in the previous sections, what is the optimal distribution of effort for foreground cleaning and delensing such that the tightest constraint on $r$ is achieved? To do this, we set up an optimization process that calculates the steepest descent through the ten dimensional space (effort in the nine single-frequency low-resolution channels plus one high-resolution channel for delensing). 

We operate in discrete units of effort, with a single unit defined to be equivalent to 500 detector-years at \SI{150}{\GHz} (similar to a BICEP/Keck 150\nobreakdash-GHz receiver observing for 1 year). For other channels, the number of detector-years per unit of effort is calculated as $n_{\mathrm{det},150}\times \left(\nu/\SI{150}{\GHz}\right)^2$. We define ``effort'' in these units because it is proportional to focal plane area, which is one of the strongest drivers of the overall project cost. 

At each step of the algorithm, we allocate a unit of effort in each dimension. 
For each separate allocation, we rescale the BPCM, compute a new Fisher matrix, and calculate the resulting $\sigma(r)$. We then compare the constraints and permanently assign one half of the unit of effort to the channel that produces the largest improvement in $\sigma(r)$ and the other half to its atmospheric window counterpart (the groupings are 30/40, 85/95, 145/155, and 220/\SI{270}{\GHz}). This last step, motivated by earlier iterations through the forecasting loop, is enforcing a split in order to reduce the foreground biases obtained under various foreground models.  
Projections run to a total of 6000 units of effort, which would be equivalent to 500,000 150\nobreakdash-GHz detectors operating for six years. Stage-4 scale surveys seem likely to be in the range of 1--3$\times10^6$ detector-years, an order of magnitude increase from Stage-3 experiments. Though it is generally prohibitive to calculate the entire 10-dimensional hypercube of $\sigma(r)$, we have validated our approach with a full-grid calculation at various points in the optimization. 

\begin{figure}[htbp]
\begin{center}
\includegraphics[height=4.5in]{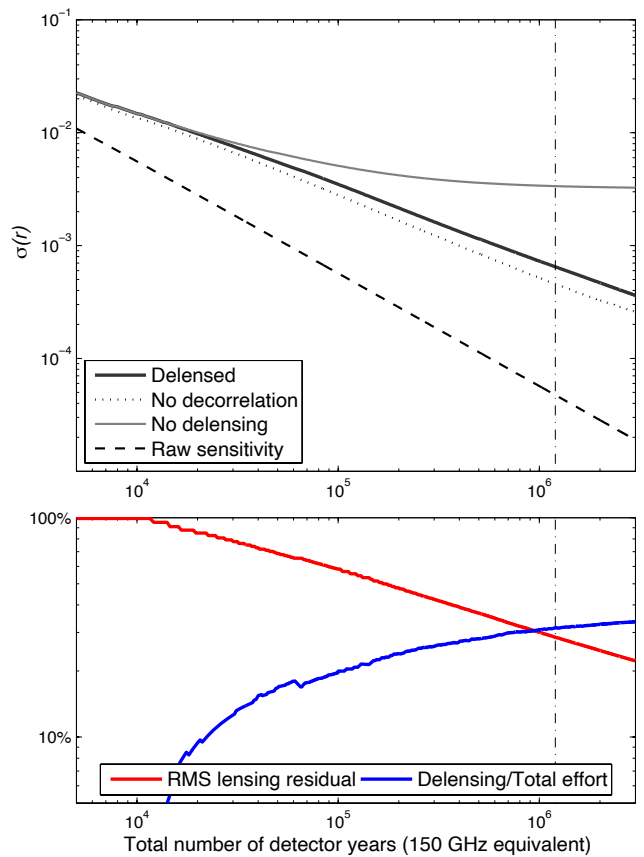}
\caption{Top: optimized constraints on $r$ for $3\%$ sky fraction as a function of total effort. We include in solid black the case with delensing, allowing for decorrelation of the foregrounds, in solid gray the case without delensing, in dotted gray the case where no decorrelation is allowed in the model (with delensing), and in dashed black the raw sensitivity in the absence of foregrounds and lensing. Bottom: for the delensed case (with decorrelation), we show the fraction of effort spent on removing the lensing sample variance and the resulting rms lensing residual.}
\label{fig:sigr}
\end{center}
\end{figure}

Figure~\ref{fig:sigr} shows the optimized constraints on $r$ as a function of total effort, as well as the fraction of effort spent on removing the lensing sample variance and the resulting map rms lensing residual, for the no-detection scenario (i.e., $r=0$). To reach the desired science goal of $\sigma(r)=5\times 10^{-4}$, for $f_{\mathrm{sky}}=3\%$ (see discussion on sky fraction below), requires $1.2\times 10^{6}$ 150\nobreakdash-GHz equivalent detector-years (or $1.8 \times 10^6$ when including marginalization over the decorrelation parameters). 

\begin{figure}[htbp]
\begin{center}
\includegraphics[height=2.7in]{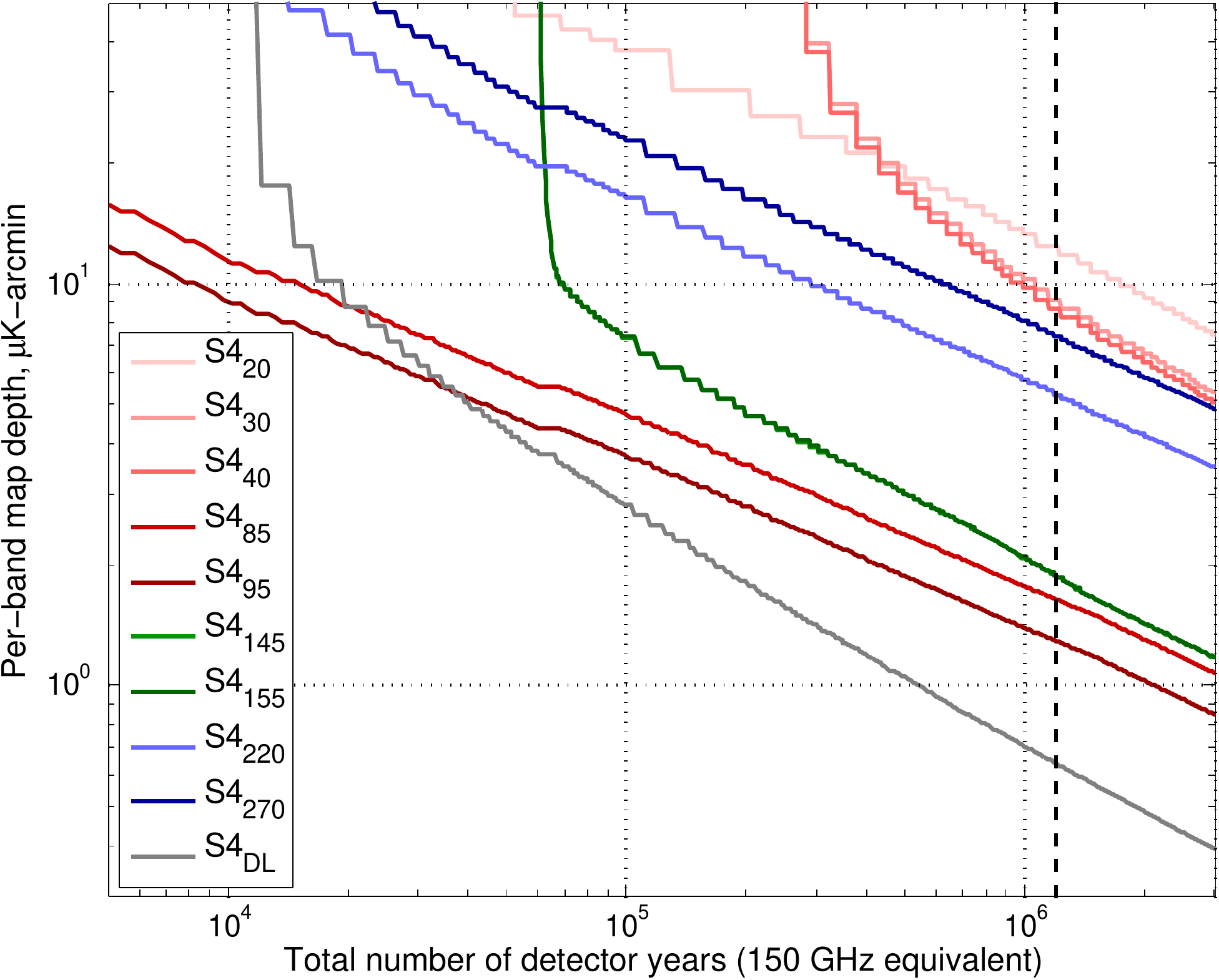}
% The y-axis of this plot would be better labelled ``per band map depth''.
\caption{Optimized map depth in each of
the small-aperture channels as well as in the delensing channel, for $f_{\mathrm{sky}}=3$\%, corresponding to the delensed (with decorrelation) case in
Figure~\ref{fig:sigr}.}
\label{fig:mapdepth}
\end{center}
\end{figure}

Upon obtaining the optimized detector count distribution, we obtain the input noise spectra according to Equation~\ref{eq:Nl}. To use these spectra to create noise simulations (discussed in Section~\ref{sec:noisim}), it is useful to distill them to a few input parameters. To that end we fit them to the formula
\begin{equation}
N_{\ell,\mathrm{fit}} = \frac{\ell(\ell+1)}{2\pi}\frac{\Omega_{\mathrm{pix}}}{B_\ell^2}\left(1+\left(\frac{\ell}{\ell_{\mathrm{knee}}}\right)^\gamma\right)\sigma_{\mathrm{map}}^2
\label{eq:Nlfit}
\end{equation}
and obtain the map depth $\sigma_{\mathrm{map}}$, slope $\gamma$, and $\ell_{\mathrm{knee}}$ values. For the small-aperture data, we find $\ell_\mathrm{knee} =$ 50--60 with $\gamma$ of $-2$ to $-3$, depending on the frequency. The optimal distribution of effort is presented in Figure~\ref{fig:mapdepth} and the configuration that achieves the science goal in 4 years (or 6 years when marginalizing over decorrelation) is summarized in Table~\ref{tab:instrument}.

\begin{figure}[htbp]
\begin{center}
\includegraphics[height=4.5in]{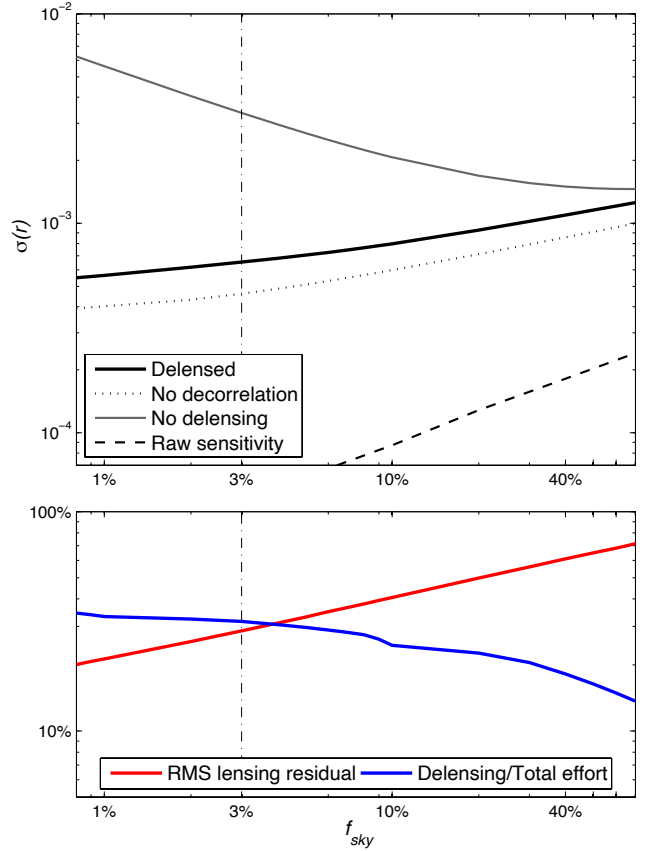}
\caption{Top: optimized constraints on $r$ as a function of sky fraction, for a fixed effort of $1.2\times 10^{6}$ 150-GHz-equivalent detector-years. We include in solid black the case with delensing, allowing for decorrelation of the foregrounds, in solid gray the case without delensing, in dotted gray the case where no decorrelation is allowed in the model (with delensing), and in dashed black the raw sensitivity in the absence of foreground and lensing. Bottom: for the delensed case (with decorrelation), we show the fraction of effort spent on removing the lensing sample variance and the resulting rms lensing residual.}
\label{fig:sigr_fsky}
\end{center}
\end{figure}

As mentioned, it is also necessary to optimize the sky fraction. The trade-off between raw sensitivity, ability to remove foregrounds, and ability to delens results is a complicated optimization problem with respect to sky coverage. Figure~\ref{fig:sigr_fsky} shows the $r$ sensitivity forecast for CMB-S4 as a function of the observed sky fraction for the case that we only
have an upper limit ($r=0$). In this case, the
optimization prefers a deep survey that targets as small an area as
possible. This conclusion, of course, depends on the forecasting assumptions; to
that end we would like to draw attention to several key factors. 

First, holding the desired constraint on $r$ fixed, the level to which we rely on delensing to decrease the sample variance increases at smaller sky fractions, as expected. For example, as shown in Figure~\ref{fig:sigr_fsky}, achieving the forecasted sensitivity on $r$ for a survey targeting 1\% of the sky will require a $> 80$\% reduction in the map rms level of the CMB lensing \textit{B}-modes. While from a sensitivity standpoint it is possible to achieve these levels, the extent to which systematic effects and small-scale foregrounds will need to be constrained may become too stringent~\citep{Carron2017, Adachi:2019}.

Second, the current optimization assumes uniform foreground behavior across the sky (with amplituide equal to that in the BICEP2/Keck region), while in reality the average amplitude, and possibly the complexity of foregrounds increase as larger sky areas are targeted. This effect would steepen the optimization curve at larger sky fractions and increase our preference for small amounts of sky. 

Third, in the case of a detection, a practical consideration for the robustness of the final $r$ result is its reproducibility across the sky. It is therefore useful
to observe multiple $1$\% patches from which we can derive and compare separate cosmological constraints. 

Finally, technical aspects of \textit{E/B} separation of CMB maps may heavily disfavor patches smaller than about $1$\% of the sky due to cut-sky effects~\citep{Bunn:2003, Smith:2006zm, Tolan:2016}. 

Balancing the forecasting results with these concerns, we have chosen $3$\% as the default sky fraction for CMB-S4 $r$ constraints (assuming a true value of
$r = 0$). This choice was made for both the CMB-S4 Science Book~\citep{CMBS4-scibook} and CMB-S4 CDT Report~\citep{CMBS4-CDT} versions of the forecasts. As mentioned in Section \ref{sec:noi_scale}, we revisit the issue of sky coverage in Section \ref{sec:DSR} with updated assumptions about our survey strategy and how we perform the $f_{\mathrm{sky}}$ scaling from the achieved products. 

\begin{table*}  
\begingroup
\newdimen\tblskip \tblskip=5pt
\vskip 3mm
\caption{Optimized instrument configuration for the PGW survey, as presented in the CMB-S4 CDT Report.}
\label{tab:instrument}
\nointerlineskip
\vskip -2mm
\footnotesize
\setbox\tablebox=\vbox{
 \newdimen\digitwidth 
 \setbox0=\hbox{\rm 0} 
 \digitwidth=\wd0 
 \catcode`*=\active 
 \def*{\kern\digitwidth}
 \newdimen\signwidth 
 \setbox0=\hbox{+} 
 \signwidth=\wd0 
 \catcode`!=\active 
 \def!{\kern\signwidth}
 \newdimen\decimalwidth 
 \setbox0=\hbox{$3$} 
 \digitwidth=\wd0 
 \catcode`|=\active 
 \def|{\kern\digitwidth}
\halign{\hbox to 0.7in{#\leaderfil}\tabskip 1em&
    #\hfil\tabskip 2em&
    \hfil#\hfil\tabskip 1em&
    \hfil#\hfil&
    \hfil#\hfil&
    \hfil#\hfil&
    \hfil#\hfil&
    \hfil#\hfil&
    \hfil#\hfil&
    \hfil#\hfil&
    \hfil#\hfil&
    \hfil#\hfil\tabskip=1.5em&
  \hfil#\hfil\tabskip=0pt\cr
\noalign{\doubleline}
\noalign{\vskip 0pt}
\omit&&\multispan9\hfil Frequency [GHz]\hfil\cr
\noalign{\vskip -2pt}
\omit&&\multispan9\hrulefill\cr
\noalign{\vskip 2pt}
\omit \hfil Science Goal\hfil&\omit\hfil Item\hfil&20&30&40&85&95&145&155&220&270&DL&Total\cr
\noalign{\vskip 4pt\hrule\vskip 4pt}
\omit&\omit\hfil{\bf }\hfil\cr
%\omit&\# detectors [150-GHz-equiv$^{\rm a}$]&\dots&6.5\,k&6.5\,k&53\,k&53\,k&20\,k&20\,k&17\,k&17\,k&193\,k\cr
$r$&No. of detectors&                                            130&260&470&17\,k&21\,k&18\,k&21\,k&34\,k&54\,k&84\,k&250\,k\cr
\omit&Angular resolution [FWHM]&11\arcmin&77\arcmin&58\arcmin&27\arcmin&24\arcmin&16\arcmin&15\arcmin&11\arcmin&8.5\arcmin&1.0\arcmin\cr
%\omit&\omit\hfil{\bf 1 x 6-m telescope}\hfil\cr
%\omit&\# detectors [150-GHz-equiv$^{\rm a}$]&7.5\,k&7.5\,k&7.5\,k&\dots&63\,k&25\,k&\dots&5\,k&2.5\,k&118\,k\cr
%\omit&\# detectors&                                          130&250&500&\dots&25\,k&25\,k&\dots&8.7\,k&8.7\,k&*68\,k\cr
%\omit&Angular resolution [FWHM]&11\arcmin&7.0\arcmin&5.2\arcmin&\dots&2.2\arcmin&1.4\arcmin&\dots&1.0\arcmin&0.8\arcmin\cr
\noalign{\vskip 2pt\hrule\vskip 3pt}}}
\endCMBSfourtable
%\tablenote {{a}} The number of 150-GHz detectors that would be required to cover the same focal-plane area as the actual detectors at the specified frequency.\par          
\endgroup
\vskip 2mm
\label{tab:distribution}
\end{table*}

%------ SECTION : Simulations ------
\section{Map-Based Simulations}
\label{sec:cdt_sims}

Using simulations to optimize the design of a CMB experiment inevitably involves a trade-off between the degree of detail that the simulations are able to capture and the computational cost of generating and analyzing them. This trade-off includes the choice of the domain in which the simulation is generated, ranging from the most detailed (but most expensive) time domain, through the map domain, to the most simplified (but most flexible) spectral domain.  Inclusion of additional detail can help validate semi-analytic results, explore their sensitivity to assumptions about foreground models, sky coverage, and instrumental noise and systematics, and in more mature stages of design can inform specific instrument and survey strategy choices.

Here we review the methods used to explore the parameter space for the PGW survey, including map-level noise simulations, sky models, and observation strategy. We also describe our approach to modeling instrumental systematics, the delensing survey, and the analysis methods.

In addition, we use these simulations to validate the spectral domain forecasts for configurations where the approaches are directly comparable, thereby closing the loop presented in Figure \ref{fig:forecasting_loop}.

\subsection{Map Noise Simulations}
\label{sec:noisim}
We use Equation~\ref{eq:Nlfit} to obtain the desired noise prescription, for a fixed total effort, and then generate 499 Gaussian noise realizations at each band. As in the Science Book, we have mostly used a circular sky area of 3\%. Small-aperture cameras have a very wide instantaneous field of view and hence the observed sky region necessarily has a large edge taper. For the nominal 3\% sky coverage simulations, we assumed a circular sky patch centred at RA=\SI{0}{\degree}, Dec=\SI{-45}{\degree} (slightly below the BICEP/Keck patch) with full coverage out to radius $\theta<\SI{12}{\degree}$ and ``relative hits'' tapering to zero with a cosine-squared shape for $\SI{12}{\degree}<\theta<\SI{27}{\degree}$. The noise realizations are divided by the square root of this coverage pattern such that the noise ``blows up around the edge'' as it does in real maps. We also assume an $\ell_{\mathrm{min}}=30$ cut-off below which we do not recover any information.

%The simulations as described above, including the extension to a 20\, GHz channel on the small-aperture telescope, we refer to here as simulations of the ``Science Book Configuration.''\footnote{Note that this configuration differs slightly from the one that appears in the 2016 edition of the Science Book, due mainly to updates in frequency band definitions, but the total assumed effort is the same.} This is to distinguish it from the strawperson configuration which has both a larger number of 150\, GHz-equivalent detector-years and a 20 GHz channel with a resolution (and $\ell _{\rm knee}$) appropriate for a 6m aperture.

For the CMB-S4 Reference Design Report (and Section \ref{sec:DSR}), we include an explicit scan strategy on the sky and produce more realistic sky coverage patterns, but for the moment we regard the above as a reasonable compromise between idealism and reality.

\subsection{Foreground Models}
\label{sec:skymod}

To make simulated sky maps, we add realizations of lensed CMB (both with and without a PGW component) to models of the Galactic foregrounds.  So far, we have run simulations with seven foreground models, which we now describe.

\begin{enumerate}
\setcounter{enumi}{-1}

\item Simple Gaussian realizations of synchrotron and dust with power-law angular power spectra at amplitudes set to match the observations in the BICEP/Keck field, and simple uniform SEDs (power law for synchrotron, modified blackbody for dust).

\item The PySM\footnote{\url{https://github.com/bthorne93/PySM\_public}} model {\tt a1d1f1s1}, where the letters refer to anomalous microwave emission (AME), dust, free-free, and synchrotron, respectively, and the numbers are the base models described in~\cite{Thorne:2016ifb}. Free-free and AME are assumed to be unpolarized in this model
and thus do not affect the analysis in this paper.

\item The PySM model {\tt a2d4f1s3}, where the models have been updated to variants that are also described in~\cite{Thorne:2016ifb}.  Note that these include 2\% polarized AME, a curvature of the synchrotron SED, and a two-temperature model for dust.

\item The PySM model {\tt a2d7f1s3}, where the dust model has been updated to a more sophisticated physical characterization of dust grains as described in~\cite{Hensley2015}. This model is interesting in that it does not necessarily conform to the modified blackbody SED.

\item The dust in Model 3 is replaced by a model of polarized dust emission that incorporates H{\sc i} column density maps as tracers of the dust intensity structures, and a phenomenological description of the Galactic magnetic field as described in \cite{Ghosh:2017}.  The model is expanded beyond what is described in that paper to produce a modest amount of decorrelation of the dust emission pattern as a function of frequency motivated by the analysis of \textit{Planck} data in~\cite{Aghanim:2017}.

\item A toy model where the strong dust decorrelation suggested in Figure~3 of~\cite{Aghanim:2017} is taken at face value ($\Delta^{217\times353}=0.85$, at $\ell=80$) and scaled to other frequencies using the functional form given in appendix B of~\cite{vansyngel16}, with a linear scaling in $\ell$.
While such a model is not ruled out by current data, it appears to be very hard to produce such strong decorrelation in physics-based models.  We also note that~\cite{Sheehy:2017} and \cite{planckiLIV} have re-analyzed the same \textit{Planck} data and, while they find that the high level of decorrelation in this model is still consistent with the data, their best fit to that same data has no decorrelation.

\item A model based on MHD simulations~\citep{Kritsuk:2017} of the Galactic magnetic field, which naturally produces non-Gaussian correlated dust and synchrotron emission.

\end{enumerate}

Models 1 to 4 use the large-scale modes of the real sky as measured above the noise in the \textit{Planck} data.  This means that these models are intrinsically ``single-realization,'' and this must be borne in mind when interpreting the results. Models 4 and 6 are not based on \textit{Planck} data, but still contain a fixed signal realization. Models 0 and 5 have different seeds for each signal map and include the (Gaussian) sample variance. The PySM models fill in the small-scale structure with power-law Gaussian extrapolations, while Models 4 and 6 naturally produce non-Gaussian small-scale structure. However, all of these models are consistent with current data, and the more complex models are not necessarily more accurate reflections of reality.

\subsection{Instrumental Systematics}

Control of instrumental systematics is a critical design consideration.  However, predicting and modeling these effects realistically is a difficult task that is dependent on actual instrument and survey design details, and furthermore, their impact on actual results comes not through the modeled effects but through unmodeled residuals.
Many existing CMB experiments have published in-depth studies that use calibration data and simulations to set upper limits on a wide variety of effects, \textit{e.g.}~\citet{Keisler:2015hfa}, ~\citet{2015_BICEP2_III},~\citet{essinger/etal:2016},~\citet{Adachi:2019},~\citet{Karkare:2019}.
There have also been several publications that examine the effect of specific classes of instrumental systematics on a generic polarization experiment, \textit{e.g.}~\citet{Hu:2002vu},~\citet{ODea:2007},~\citet{Shimon:2008},~\citet{Wallis:2014},~\citet{Duivenvoorden:2019}.

For this study, in the absence of detailed instrument and survey designs, we have taken the first steps in simulating various generic classes of additive systematic by injecting additional noise-like components into the maps and then re-analyzing them without knowledge of what was put in.  We have experimented with components that are both correlated and uncorrelated across frequency bands, and that have white, $1/\ell$, and $\hbox{white} + 1/\ell$ spectra, at varying levels compared to single-frequency map noise or, for correlated cases, combined map noise. The leading-order effects of such components can be mitigated via explicit modeling or filtering, but they may still produce map-level residuals. Examples of mechanisms in this class include bandpass mismatches, beam and pointing variations, calibration variations, cross-talk effects, half-wave-plate leakage, ground pickup, and readout irregularities.

To assess the impact of instrumental systematics on measurement requirements,
for the purpose of determining both the required survey depths and the maximum allowable levels of systematic effects in the final single-frequency survey maps, our general procedure is to feed parameterizations of various systematic effects into semi-analytic forecasts and judge at what levels classes of systematics introduce parameter biases or additional uncertainties that are significant compared to the science targets for those parameters.

Other classes of systematics could be simulated by manipulating the analysis procedure only.  Examples of such effects include uncertainties in the bandpasses, polarization angles, calibration, and beam shapes. Such examples are not presented here and are left for future work.

%Description of additive systematics--rough draft: simulating unmodeled residuals, broad range of effects
%which include unmodeled residuals after best efforts to subtract or filter multiplicative
%and additive effects including beam mismatches, bandpass mismatches, calibration variation,
%cross-talk effects, half-wave plate leakages, ground pickup, and readout irregularities.
%Requirements for control of residual additive contamination are stated as pecentage of
%single-frequency survey noise, or (in the case of correlated systematics) total combined
%noise levels.

%Multiplicative systematics--rough draft: bandpass uncertainty, polarization angle,
%calibration, and beam shape uncertainty. including citations.

\subsection{Delensing}
\label{sec:delens}

We have generated high-resolution simulated maps on which we intend to run explicit lensing reconstruction and then include that information in the analysis. While we are currently working on this analysis, this process has not yet converged, and so for the present, we approximate delensing by scaling down the $\Lambda$CDM lensing signal by the appropriate factor, as described in Section ~\ref{sec:delens_treatment}.

%------ SECTION : Analysis Methods ------
\section{Simulation Analysis Methods}
\label{sec:analysis_methods}

To make simulated maps, the noise realizations described in Section~\ref{sec:noisim} are added to the sky models described in Section~\ref{sec:skymod}.
For each realization, one then has a stack of multi-frequency $I,Q,U$ maps containing non-uniform noise, foregrounds, and signal, and the challenge is to re-analyze them to recover the parameter of interest (in this case $r$).  This can be done by different teams using different methods and could be carried out in a blind manner, although we have not done this yet.

So far, we have experimented with two methods.  The first is a spectral internal linear combination (ILC) method~\citep[e.g.,][]{Tegmark:2003ve}, which determines the linear combination of multipole coefficients that minimizes the foreground and noise power without altering the CMB contribution. This method only relies on the frequency dependence of the CMB and does not rely on assumptions about the spectral dependence of foreground components. Determination of cleaning coefficients in each bin ($\Delta \ell=31$) leads to the smallest foreground residuals, but yields foreground residuals that are difficult to model. Relying on a single cleaning coefficient across all bins leads to foreground residuals that are easy to model but large. As a consequence, we use the same cleaning coefficients in three neighboring bins, which results in residuals that are acceptable and can still be marginalized in the power spectrum likelihood analysis. For the marginalization, since foreground residuals are typically dominated by dust, we assume an $\ell$-dependence of the residuals across the three bins that share a common cleaning coefficient that is consistent with the $\ell$-dependence observed by \planck\ for dust. For models that exhibit decorrelation between different dust components, this model is no longer correct. This could be improved with better understanding of foregrounds, but no attempts were made to do so in this analysis method.

The second method is an evolution of the parametric multi-component fit to the ensemble of auto- and cross-spectra, as used for the BICEP/Keck analysis to date \citep{Ade:2015tva, Array:2015xqh, Array:2018x}.  This method fits the observed bandpowers to a model composed of the lensing expectation plus dust and synchrotron contributions and a possible $r$ component.  Dust and synchrotron each have an amplitude ($A_\mathrm{dust}$ and $A_\mathrm{sync}$), a spatial spectral parameter ($\alpha_\mathrm{d}$ and $\alpha_\mathrm{s}$),  and a frequency spectral parameter ($\beta_\mathrm{d}$ and $\beta_\mathrm{s}$).  We also allow a dust/synchrotron correlation ($\epsilon$), and decorrelation of the foreground patterns over frequency ($\Delta_\mathrm{d}$ and $\Delta_\mathrm{s}$). This model is equivalent to the one described in Section~\ref{sec:multicomp}.

Both of these analysis methods are only close to optimal when the foreground behavior is close to uniform across the observing field.  For analysis of larger fields,  algorithms that fit more complex behavior will likely be required, for example, modeling the frequency spectral indices individually in (large) pixels.

%------ SECTION : Results ------
\section{Map-Based Results}
\label{sec:map_results}
\begin{table*}[htbp]
\begingroup
\newdimen\tblskip \tblskip=5pt
\caption{Results of two analysis methods applied to map-based simulations assuming the CMB-S4 CDT Report~\citep{CMBS4-CDT} configuration and our suite of sky models (DC4). All simulations assume an instrument configuration including a (high-resolution) 20-GHz channel, a survey of 3\% of the sky with $1.2\times10^6$ 150-GHz-equivalent detector-years, and $A_\mathrm{L} = 0.1$.} %This statement was true for the CDT: Note that this configuration is not the final strawperson concept, and in particular has fewer detector-years.
\label{tab:modres}
\nointerlineskip
%\vskip -3mm
\footnotesize
\setbox\tablebox=\vbox{
 \newdimen\digitwidth
 \setbox0=\hbox{\rm 0}
 \digitwidth=\wd0
 \catcode`*=\active
 \def*{\kern\digitwidth}
 \newdimen\signwidth
 \setbox0=\hbox{+}
 \signwidth=\wd0
 \catcode`!=\active
 \def!{\kern\signwidth}
 \newdimen\decimalwidth
 \setbox0=\hbox{$3$}
 \digitwidth=\wd0
 \catcode`|=\active
 \def|{\kern\digitwidth}
\halign{\hbox to 1.0in{#\leaderfil}\tabskip 1em&
    \hfil#\hfil&
    \hfil#\hfil&
    \hfil#\hfil&
    \hfil#\hfil&
    \hfil#\hfil&
    \hfil#\hfil&
   \hfil#\hfil\tabskip=0pt\cr
\noalign{\doubleline}
\noalign{\vskip 3pt}
\omit&&\multispan2\hfil ILC\hfil&\multispan2\hfil Parametric (no decorrelation)\hfil&\multispan2\hfil Parametric (incl. decorrelation)\hfil\cr
\noalign{\vskip -2pt}
\omit&&\multispan2\hrulefill&\multispan2\hrulefill&\multispan2\hrulefill\cr
\noalign{\vskip 2pt}
\omit\hfil$r$ value\hfil&Sky model&$\sigma(r)\times10^{-4}$&$r$ bias $\times10^{-4}$&$\sigma(r)\times10^{-4}$&$r$ bias $\times10^{-4}$&$\sigma(r)\times10^{-4}$&$r$ bias $\times10^{-4}$\cr
\noalign{\vskip 3pt\hrule\vskip 5pt}
0&       0&4.4 &$-0.2$ &*4.4&!!0.2* &5.7&!0.3\cr
\omit& 1&4.6 &!0.8 &*4.7&!6.8&6.4&!5.2\cr
\omit& 2&4.7 &!0.7 &*4.8&!3.8&6.5&!1.9\cr
\omit& 3&4.6 &!1.2 &*4.7&!6.0&6.7&!0.7\cr
\omit& 4&6.5 &!4.8 &*7.9&!43&8.3&$-7.7$\cr
\omit& 5\rlap{$^{\rm a}$}&18*&!17 &**31*&!340&15&!0.2\cr
\omit& 6&4.8 &$-1.8$ &*4.8&!0.6&6.5&!1.8\cr
\noalign{\vskip 4pt}
\hline
\noalign{\vskip 4pt}
0.003& 0&6.6 &$-0.7$ &*6.2&!0.3&8.1&!0.4\cr
\omit&  1&6.9 &!0.9 &*6.5&!6.9&8.5&!5.4\cr
\omit&  2&6.5 &$-0.1$ &*6.4&!3.9&7.9&!1.9\cr
\omit&  3&7.0 &!1.4 &*6.6&!6.7&8.7&!0.9\cr
\omit&  4&11 &!7.1 &!10 &!*51*&11&$-6.2$\cr
\omit&  5\rlap{$^{\rm a}$}&23 &!17 &*34&!350&!17* &!0.4\cr
\omit&  6&7.5 &$-0.2$ &*7.1&**1.4 &8.6&!2.5\cr
\noalign{\vskip 3pt\hrule\vskip 3pt}}}
\endCMBSfourtable
\tablenote {{a}} An extreme decorrelation model---see Section~\ref{sec:skymod}. In the right column the parametric analysis includes a decorrelation parameter. No attempt is made in the ILC analysis to model the decorrelation. The middle columns shows the parametric analysis when we do not include deccorelation parameters. \par
\endgroup
\vglue 14pt
\end{table*}
%
%\noalign{\vskip 3pt\hrule\vskip 5pt}
%0&       0&*5.4&!0.3&*4.4&!0.2&*5.6&!0.3\cr
%\omit& 1&*8.6&!1.1&*4.7&!6.9&*6.4&!5.3\cr
%\omit& 2& *8.0&$-2.3$&*4.8&!4.0&*6.6&!2.0\cr
%\omit& 3& *6.8&!0.5&*4.7&!6.1&*6.7&!0.7\cr
%\omit& 4&*7.5&!5.0&*7.8&!42.0|&*8.1&$-5.8$\cr
%\omit& 5\rlap{$^{\rm a}$}&*16!&18|&33.6&*341.8!&*12.8|&$-0.3$\cr
%\omit& 6&*5.7&$*-0.4$ &*4.8&!0.6&*6.5&!1.8\cr
%\noalign{\vskip 4pt}
%\hline
%\noalign{\vskip 4pt}
%0.003& 0&*7.2&$-4.0$&*6.2&!0.3&*8.1&!0.4\cr
%\omit&  1&10!&!0.2&*6.5&!7.0&*8.5&!5.5\cr
%\omit&  2&10!&$-2.8$&*6.5&!4.1&*8.0&!2.1\cr
%\omit&  3&*7.4&$-1.4$&*6.6&!6.8&*8.7&!1.1\cr
%\omit&  4&10!&!5.8&*9.9&!49.7|&10.3&$*-4.0$|\cr
%\omit&  5\rlap{$^{\rm a}$}&*20& 20|&36.1&*351.7!&*14.1|&$-0.5$\cr
%\omit&  6&*8.0& $-0.4$&*7.1&!1.6&*8.6&!2.6\cr

Table~\ref{tab:modres} summarizes the results of the analysis for simulations of the optimized configuration obtained in Section \ref{sec:optimization}
($1.2\times10^6$ 150\nobreakdash-GHz-equivalent detector-years) and residual lensing power
$A_\mathrm{L}=0.1$. The lensing residual is expected for iterative $EB$ delensing according to~\citet{Smith:2010gu} for the sensitivity and angular resolution of the delensing survey. The results from the parametric analysis naturally depend on whether
a marginalization over decorrelation is performed, while the ILC analysis did not
attempt to capture the effects of decorrelation on the recovery of $r$ and
$\sigma(r)$. This is evidenced by the large bias for the ILC method for Model 5
when compared to the parametric analysis that directly accounts for a possible
decorrelation (last column). In general, we see that for $r=0$ the simple
Gaussian foreground Model 0 gives $\sigma(r)\approx 5\times10^{-4}$, exactly as
expected from the semi-analytic formalism.  As we progress to the more complex
foreground models, $\sigma(r)$ is generally in the range 5--8$\times10^{-4}$. 

The level of biases is generally below $1.0\sigma$ for all the models.
These simulations are sets of 499 realizations, so the statistical uncertainty
on the bias is approximatively $0.04 \sigma$.
However, the strong decorrelation in Model 5, as well as the high-significance detection of decorrelation in the parametric analysis of Model 4, do significantly increase $\sigma(r)$ and the level of bias. While the parametric method is able to account for the decorrelation, by construction information is lost, and in fact if one believed in such a scenario, a different re-optimization to concentrate the sensitivity at closer-in (less decorrelated) frequencies would be called for.

Table~\ref{tab:straw} shows results on detection significance for the CDT Report configuration for sky Model 6. For $r=0$, the $95\%$ upper limit is about $2.1\sigma(r)$. The value of the tensor-to-scalar ratio for which we expect a $5\sigma$ detection after 4 years of operation is $r=0.004$.\footnote{The reason this is larger than 5 times the quoted $\sigma(r)$ is sample variance.} For a tensor-to-scalar ratio of $r=0.003$, the median detection significance after 4~years is expected to be $4\sigma$.  If a detection were to be emerging at this point, extending the run time to 8~years would be justified in order to reach a $5\sigma$ detection.

While $\sigma(r)$ can be precisely forecast for given assumptions, the true achieved detection level for $r$ depends on the particular realization of the $B$-mode field in the observed patch of sky. Therefore we can only forecast a distribution of detection levels. For a tensor-to-scalar ratio of $r=0.003$ and 8 years of observing we expect to achieve more than a $3\sigma$ detection with a probability of 0.99, more than $4\sigma$ with a probability of 0.93, more than $5\sigma$ with a probability of 0.53, and more than $6\sigma$ with a probability of 0.14. For simplicity, we focus on $\sigma(r)$, and on median detection levels as well as median 95\% confidence upper limits to state the typical outcome.

\begin{table*}[htbp]
\begingroup
\newdimen\tblskip \tblskip=5pt
\caption{Results on detection significance for the CMB-S4 CDT report~\citep{CMBS4-CDT} configuration, using the two analysis methods. For the $r=0$ model we report the 95\% confidence level upper limit (CL UL).}  
%This statement was true for CDT report: Note that this has an increase in detector-year effort versus the configuration in Table~\ref{tab:modres}.
\label{tab:straw}
\nointerlineskip
\vskip -2mm
\footnotesize
\setbox\tablebox=\vbox{
 \newdimen\digitwidth
 \setbox0=\hbox{\rm 0}
 \digitwidth=\wd0
 \catcode`*=\active
 \def*{\kern\digitwidth}
 \newdimen\signwidth
 \setbox0=\hbox{+}
 \signwidth=\wd0
 \catcode`!=\active
 \def!{\kern\signwidth}
 \newdimen\decimalwidth
 \setbox0=\hbox{$3$}
 \digitwidth=\wd0
 \catcode`|=\active
 \def|{\kern\digitwidth}
\halign{\hbox to 1.0in{#\leaderfil}\tabskip 1em&
    \hfil#\hfil&
    \hfil#\hfil&
    \hfil#\hfil&
    \hfil#\hfil&
    \hfil#\hfil&
   \hfil#\hfil\tabskip=0pt\cr
\noalign{\doubleline}
\noalign{\vskip 3pt}
\omit&&&\multispan2\hfil ILC\hfil&\multispan2\hfil Parametric (incl. decorrelation)\hfil\cr
\noalign{\vskip -2pt}
\omit&&&\multispan2\hrulefill&\multispan2\hrulefill\cr
\noalign{\vskip 2pt}
\omit\hfil $r$ value\hfil&\omit\hfil Duration\hfil&Sky model&95\% CL UL&Detection significance&95\% CL UL&Detection significance \cr
\noalign{\vskip 3pt\hrule\vskip 5pt}
0&\omit\hfil4 years\hfil&       6&$1.0\times 10^{-3}$&\dots&$1.0\times 10^{-3}$&\dots\cr
\noalign{\vskip 4pt}
\hline
\noalign{\vskip 4pt}
0.003&\omit\hfil4 years\hfil& 6&\dots&4.0&\dots&4.2\cr
\noalign{\vskip 2pt}
\omit&\omit\hfil8 years\hfil& 6&\dots&5.1&\dots&5.6\cr
\noalign{\vskip 3pt\hrule\vskip 3pt}}}
\endCMBSfourtable
\endgroup
\vglue 15pt
\end{table*}

The numbers in Table~\ref{tab:modres} clearly show dependence on the foreground model used in the simulation. If the actual foregrounds are substantially different than any of these cases, then the biases could be larger. To obtain some understanding of how large the biases could be, and what instrument modifications might help to reduce them, we have also looked at ILC biases in the extreme case that the foreground residuals are not modeled or marginalized over, but simply absorbed into the estimated \textit{B}-mode power spectrum. Doing so with dedicated simulations based on sky Model 6 increases the magnitude of the bias on $r$ to $4.1\times 10^{-4}$. The dominant contribution to the bias comes from synchrotron residuals, which motivated placing one lower-frequency channel on the LAT (reducing the magnitude of the bias to $1.8 \times 10^{-4}$). 

\begin{table*}[htbp]
\begingroup
\newdimen\tblskip \tblskip=5pt
\caption{Map-based simulation results for dedicated simulations containing systematics (DC3). Simulations here assume the Science Book Configuration \citep{CMBS4-scibook}, i.e., an instrument configuration including a (low-resolution) 20-GHz channel, a survey of 3\% of the sky with $1.0\times10^6$ 150-GHz-equivalent detector-years, and $A_\mathrm{L} = 0.1$. We report sky Model 3 and $r=0$, with additive systematic effects in varying combinations, the amplitudes of which are specified as percentages of survey noise, for the white (A) and $1/\ell$ (B) components.}
\label{tab:sysres}
\nointerlineskip
\vskip -3mm
\footnotesize
\setbox\tablebox=\vbox{
 \newdimen\digitwidth
 \setbox0=\hbox{\rm 0}
 \digitwidth=\wd0
 \catcode`*=\active
 \def*{\kern\digitwidth}
 \newdimen\signwidth
 \setbox0=\hbox{+}
 \signwidth=\wd0
 \catcode`*=\active
 \def*{\kern\digitwidth}
 \newdimen\signwidth
 \setbox0=\hbox{+}
 \signwidth=\wd0
 \catcode`!=\active
 \def!{\kern\signwidth}
\halign{\hbox to 1.6in{#\leaderfil}\tabskip 1em&
    \hfil#\hfil\tabskip 0.5em&
    \hfil#\hfil\tabskip 2em&
    \hfil#\hfil\tabskip 0.5em&
    \hfil#\hfil\tabskip 1em&
    \hfil#\hfil\tabskip 0.5em&
    \hfil#\hfil\tabskip 0.5em&
    \hfil#\hfil\tabskip 0.5em&
   \hfil#\hfil\tabskip=0pt\cr
\noalign{\doubleline}
\noalign{\vskip 3pt}
\omit&\multispan2\hfil Uncorrelated\hfil&\multispan2\hfil Correlated\hfil&\multispan2\hfil ILC\hfil&\multispan2\hfil Parametric\hfil\cr
\noalign{\vskip -2pt}
\omit&\multispan2\hrulefill&\multispan2\hrulefill&\multispan2\hrulefill&\multispan2\hrulefill\cr
\noalign{\vskip 2pt}
\omit \hfil Systematic\hfil&A [\%]&B [\%]&A [\%]&B [\%]&$\sigma(r)\times10^{-4}$&$r$ bias $\times10^{-4}$&$\sigma(r)\times10^{-4}$&$r$ bias $\times10^{-4}$\cr
\noalign{\vskip 4pt\hrule\vskip 4pt}
None& 0& 0& 0& 0& 5.3& 0.0*& 7.2& 0.0*\cr
Uncorrelated white& 3.3& 0& 0& 0& 6.0& 0.84& 8.0& 0.63\cr
Uncorrelated $1/\ell$& 0& 6.8& 0& 0& 5.0& 0.99& 7.0& 0.85\cr
Correlated white& 0& 0& 5.8& 0& 6.3& 1.2*& 7.3& 1.4\cr
Correlated $1/\ell$& 0& 0& 0& 11& 5.2& 1.0*& 6.7& 0.97\cr
Uncorrelated white + $1/\ell$& 1.6& 3.5& 0& 0& 5.6& 0.89& 7.5& 0.76\cr
Correlated white + $1/\ell$& 0& 0& 2.9& 5.3& 5.5& 0.98& 6.9& 1.0\cr
Both, white + $1/\ell$& 0.8& 1.7& 1.5& 2.6& 5.6& 1.1*& 7.9& 0.98\cr
\noalign{\vskip 3pt\hrule\vskip 3pt}}}
\endCMBSfourtable
\endgroup

\end{table*}

Table~\ref{tab:sysres} summarizes the results of the analysis of simulations including additive systematic effects on top of foreground Model 3 (note: these simulations correspond to DC3). Different combinations of uncorrelated and correlated contamination with varying spectra are considered. The levels of systematic contamination for these simulations were chosen to predict biases on $r$ of $1 \times 10^{-4}$ in semi-analytic forecasts.  We can see that the different combinations explored increase the bias on $r$ by amounts that typically vary from 0.5 to 1.5$ \times 10^{-4}$ for the two separate analyses, over the different cases.  We find that to restrict the bias on $r$ to this level, the sum of additive contamination effects needs to be controlled to 3--7\%  of the single-frequency survey noise, or (in the case of correlated systematics) 6--11\% of the total combined noise levels.  Such percentages are consistent with the upper limits currently achieved for residual additive systematic contamination compared to survey noise
by small-aperture experiments~\citep[e.g.,][]{Array:2015xqh}.  Assuming that CMB-S4 will include a sustained effort to continue to control, understand, and model systematic effects down to levels limited by survey noise, these percentages provide reasonable benchmark requirements.

Results of simulating systematic errors in the determination of bandpasses vary by analysis method.  The construction of the ILC method makes it largely insensitive to such errors.  The parametric analysis, which includes parametrized models of the frequency spectra of different foregrounds, shows biases on $r$ at the $1 \times 10^{-4}$ level for uncorrelated random deviations in band\nobreakdash-center determination of 0.8\%, or for correlated deviations of 2\%; we adopt these as reasonable benchmark requirements to accommodate a variety of both blind and astrophysical foreground modeling approaches.

\section{Modifications Leading to the Reference Design}
\label{sec:DSR}

In this section, we describe updates to the framework and the reference design, which attempt to take into account the impact of realistic observing strategies, realistic focal\nobreakdash-plane layouts and mapping onto optics tubes, as well as a more conservative approach to our delensing forecasts. We also seek to answer the question of siting, with the South Pole and Chile being the two choices considered. The work in this section has led to the forecasts
and plans presented in the most recent CMB-S4 document---the CMB-S4 Reference Design Report \citep{CMBS4-DSR}. While this section is self-contained, we direct the reader to the aforementioned source for more details. 

In previous versions of our forecasting, we have had the ability to choose the number of detectors in each frequency band in a continuously variable manner, as shown in Figure~\ref{fig:mapdepth}. For the reference design, a mapping of detectors into dichroic optics tubes has been carried out, while seeking to maintain the band distribution as determined in the optimization calculations. In scaling the achieved performance from the existing monochromatic instruments to dichroic detectors and optics, no degradation of optical performance has been assumed at this stage, but this assumption should be verified with upcoming data. This results in the configuration described in \cite{CMBS4-DSR} with 18 SAT tubes, observing for 7 years. We use this configuration to scale the BICEP/Keck noise bandpower covariance matrix in the same way as described in Section~\ref{sec:bpcm}. A further re-scaling is applied to account for sky coverage, as explained below.

\subsection{Sky Coverage Effects}
\begin{figure}[htbp]
\begin{center}
\includegraphics[clip, trim=0cm 1cm 0cm 1cm, width=0.85\columnwidth]{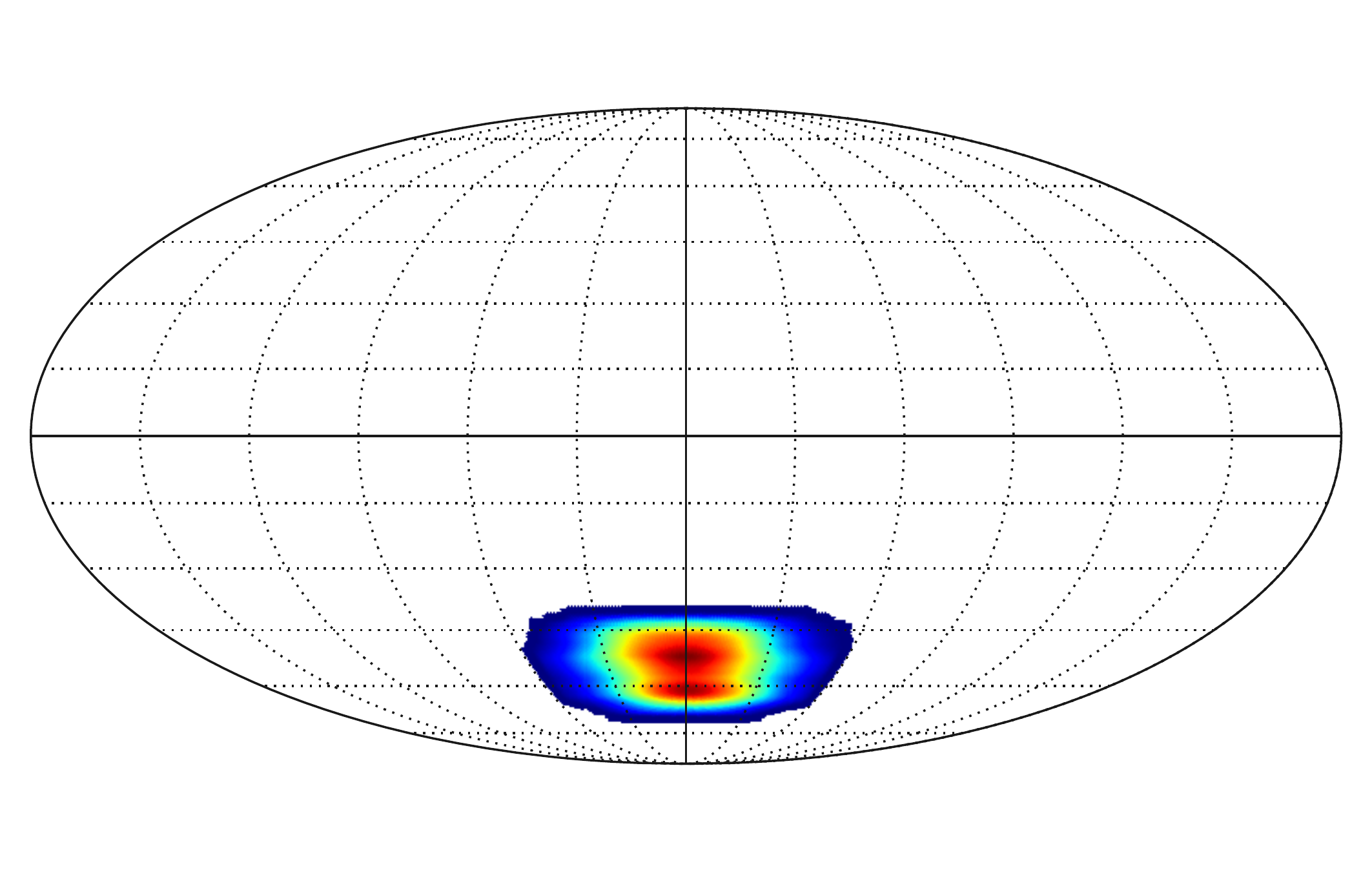} 
\includegraphics[clip, trim=0cm 1cm 0cm 1cm, width=0.85\columnwidth]{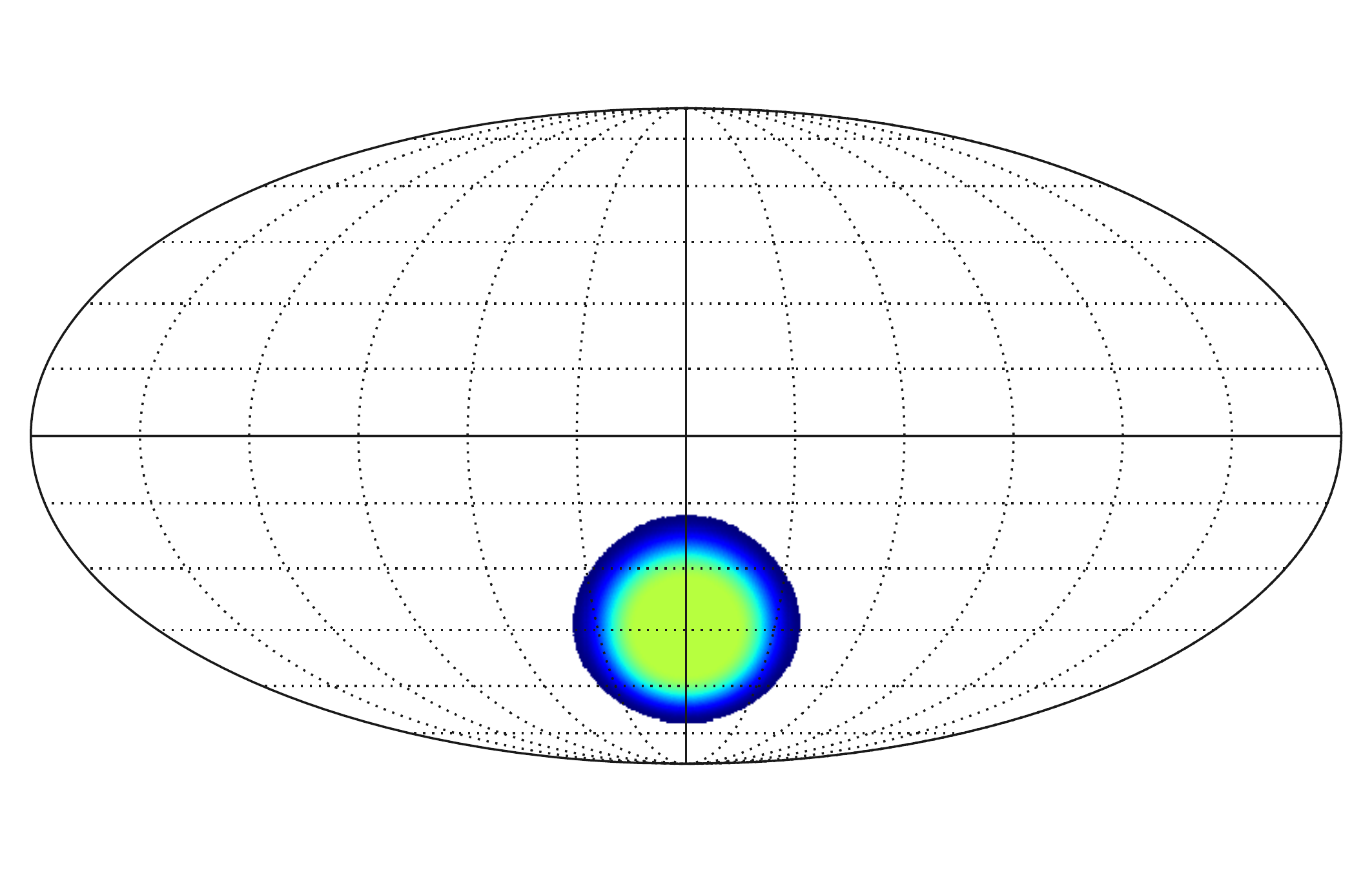}
\includegraphics[clip, trim=0cm 1cm 0cm 1cm, width=0.85\columnwidth]{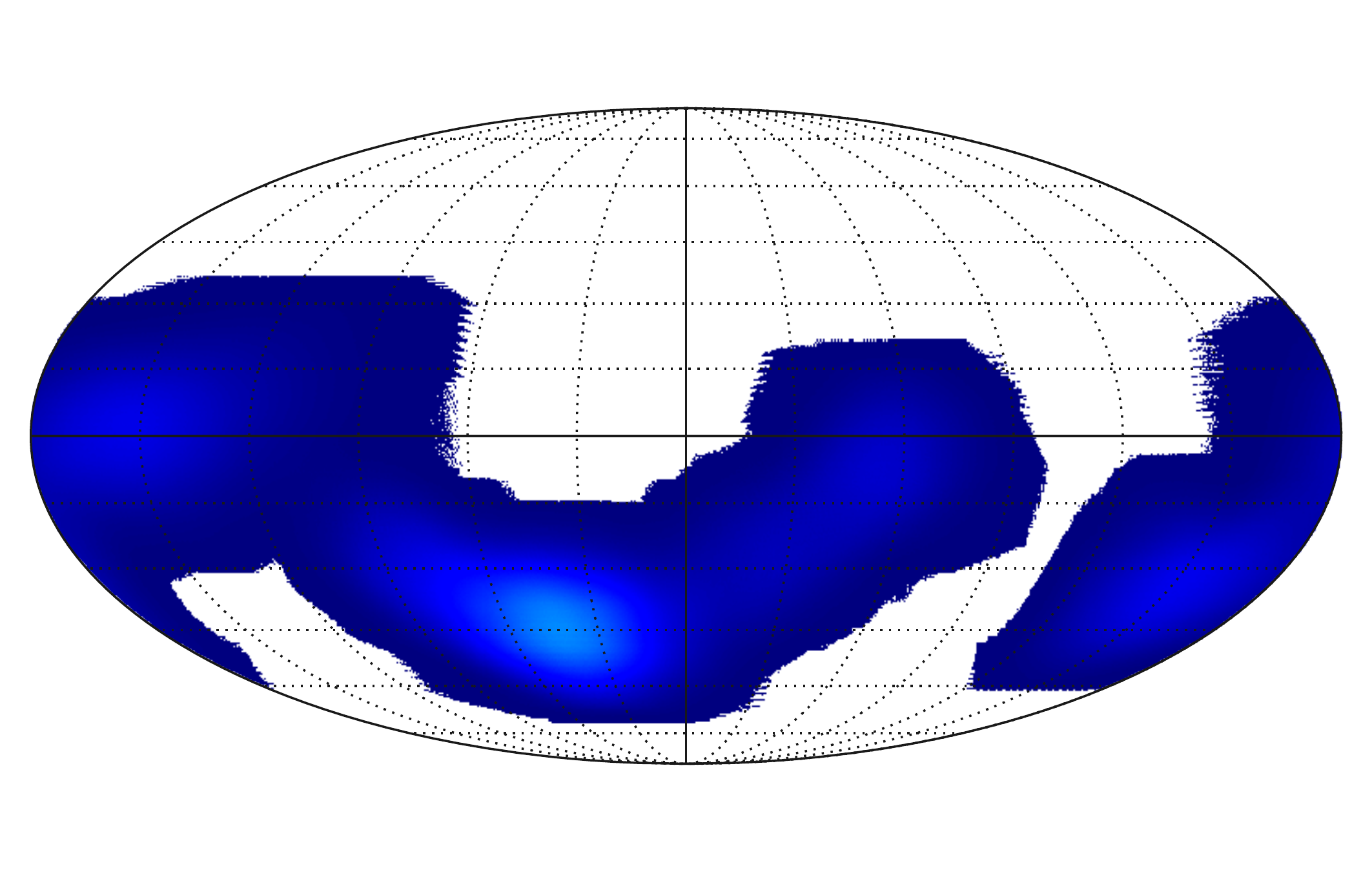}
\includegraphics[clip, trim=0cm 1cm 0cm 1cm, width=0.85\columnwidth]{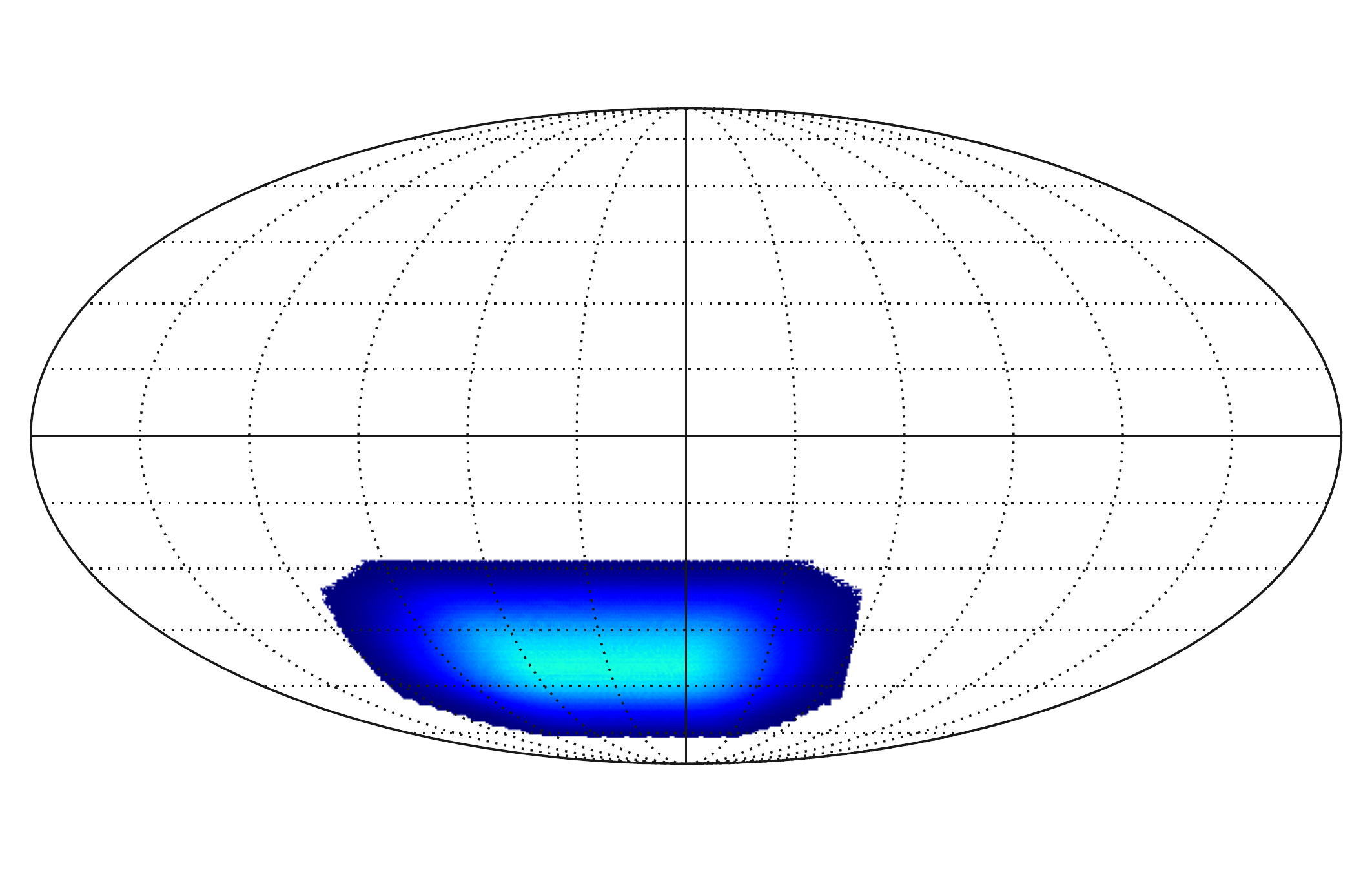}
\includegraphics[clip, trim=0cm 1cm 0cm 1cm, width=0.85\columnwidth]{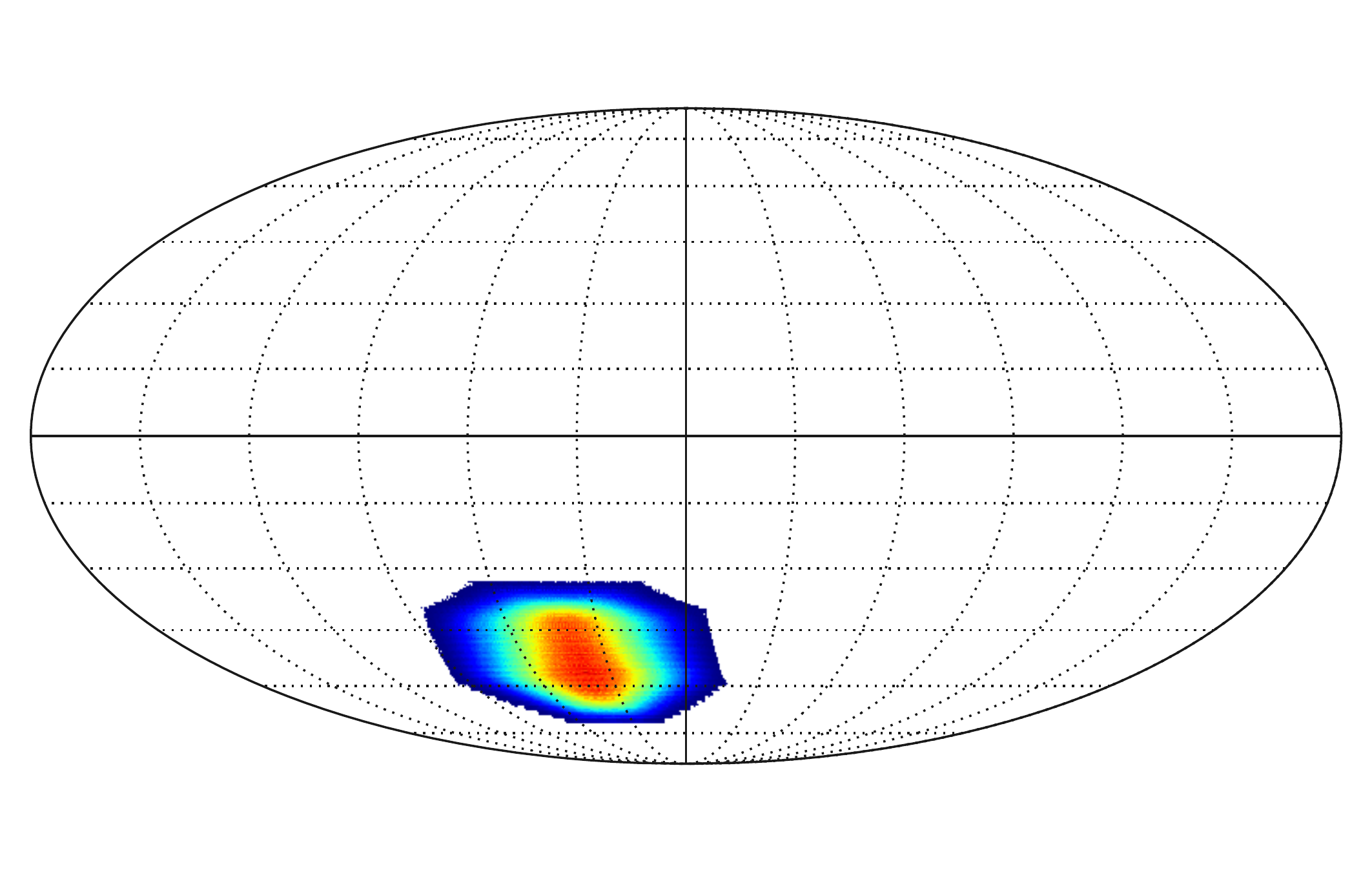}
\includegraphics[width=0.85\columnwidth]{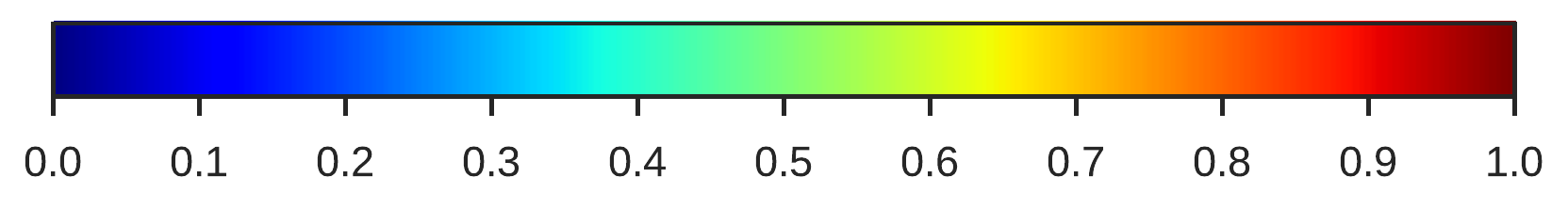}
\end{center}
\caption{Hit patterns on the sky for small aperture telescope surveys. Top panel: the actual BICEP3 2017 hit pattern (peak normalized). Second panel: idealized circular pattern as used in Section~\ref{sec:cdt_sims}. Third panel: simulated ``Chile full'' pattern, Fourth panel: simulated ``Pole wide'' pattern. Fifth panel: simulated ``Pole deep'' pattern.
Each pattern is normalized to the same hit sum as in the top panel, and
the color scales are the same. The ``Chile deep'' and ``Chile shallow'' regions referred to in the text are sub-regions of the ``Chile full'' pattern.}
\label{fig:relhits}
\end{figure}

The semi-analytic calculations of Section~\ref{sec:bpcm}
assumed a simplified re-scaling for sky area, while the map based simulations of Section~\ref{sec:map_results} assumed an idealized circular sky patch, which is not actually achievable with a practical instrument from a site at any latitude.
Figure~\ref{fig:relhits} compares our prior assumptions to more realistic hit patterns.

From the South Pole, it is possible to concentrate the coverage onto a compact region of sky, but from Chile the region that can be observed is affected by Earth's rotation, resulting in more extended coverage. The large instantaneous field of view of the SAT telescopes means that there is a minimum field size that can be achieved, and that there is always a strong edge taper in the coverage pattern.

We have performed a calculation that attempts to optimize simulated SAT observations from Chile to produce the densest possible coverage on a $3$\% patch of low-foreground sky, resulting in the overall pattern shown in Figure~\ref{fig:relhits} as ``Chile full.'' We segment this into its deepest part, which we call ``Chile deep,'' and the remainder, which we call ``Chile shallow.''

From the South Pole, one can scan the same patch at all times of the day and year at the same observing elevation, with the size of the observed patch controlled by the length of the scan throw in Right Ascension. A minimal-length scan results in the pattern shown in the figure as ``Pole deep.''
Lengthening the scan while remaining in the low foreground sky results in the pattern ``Pole wide.'' In the results below, ``Pole deep'' and ``Pole wide'' are therefore ``either/or'' options.

Because the noise increases in regions with less observing time, the effective sky area for noise is larger than the effective sky area for signal---and both of these also depend on the weighting applied when analyzing the maps.
The patterns shown in Figure~\ref{fig:relhits} have the effective sky fractions reported in Table~\ref{tab:skyfrac}, assuming inverse noise-variance weighting.

\begin{table}[htbp]
\renewcommand{\arraystretch}{1.7}
 \centering
 \caption{Effective sky fractions for signal, noise, and signal-cross-noise, as percentages, for the observation patterns shown in Figure~\ref{fig:relhits}, and the case of inverse noise-variance weighting (i.e., Equations~\ref{eq:scalsky1}--\ref{eq:scalsky3} with $w_i = h_i$). Here ``SP'' is the South Pole and ``CH'' is Chile. Note: the CH Shallow numbers appear larger than CH Full due to the effects of the weighting.
  % We also report the fraction of sky with non zero coverage, $ f_{\rm sky}^{\rm non-zero}$. 
  }
  \label{tab:skyfrac}
  \begin{tabular}{c|ccccc}
  \hline 
   & \text{SP deep} &  \text{SP wide}&  \text{CH deep}& \text{CH shallow}  & \text{CH full} \\ %\text{BK}
  \hline 
 $ f_{\rm sky}^{\rm sig} $& 1.9  & 4.3 & 2.4 & 10& 5.9\\ %1.0
 $ f_{\rm sky}^{\rm noi} $ & 2.9 & 6.5 & 3.4& 20 & 18 \\ %1.3
 $ f_{\rm sky}^{\rm cross}$  & 2.5 & 5.5 & 3.0 & 16 & 12\\ %1.1
%$ f_{\rm sky}^{\rm non-zero} $ & 5.0 & 12 & 5.0 & 47 & 52 \\ %2.4
  \hline
  \end{tabular}
  {\raggedright For comparison, the BICEP2/Keck and BICEP3 values are (1.0, 1.3, 1.1) and (1.9, 2.7, 2.3) for signal, noise, and signal-cross-noise respectively. \par}
\end{table}

We can take into account the above effects by rescaling the
BICEP/Keck BPCM in a more sophisticated manner.
First, we need to scale the noise due to distributing the effort on a patch of sky larger than the original BICEP/Keck one. The noise is scaled by the effective noise factor
\begin{equation}
 f_{\rm eff}^{\rm noise} =   \frac{\Omega_{\rm pix}}{4\pi} \frac{\sum_i  w_i^2 h_i^{-1} \sum_i  h_i}{\sum_i w_i^2},
\end{equation}
where $\Omega_{\rm pix}$ is the solid angle of a single pixel, $w_i$ are the weights for pixel $i$, and $h_i$ are the hit counts.

Second, since we observe a different number of modes, we need to scale the signal, noise and signal-cross-noise contributions of the BPCM by the factors

\begin{equation}
\label{eq:scalsky1}
 f_{\rm sky}^{\rm noise} =   \frac{\Omega_{\rm pix}}{4\pi} \frac{(\sum_i  w_i^2 h_i^{-1})^2}{\sum_i w_i^4 h_i^{-2}},
\end{equation}
\begin{equation}
\label{eq:scalsky2}
 f_{\rm sky}^{\rm signal} =   \frac{\Omega_{\rm pix}}{4\pi} \frac{(\sum_i w_i^2)^2}{\sum_i w_i^{4}},
\end{equation}
\begin{equation}
\label{eq:scalsky3}
 f_{\rm sky}^{\rm cross} =  \frac{\Omega_{\rm pix}}{4\pi} \frac{\sum_i w_i^2 \sum_i  w_i^2 h_i^{-1}}{\sum_i w_i^4 h_i^{-1}}.
\end{equation}

We also need to take out the effect of these factors from the original BPCM. In the BICEP/Keck analysis, the weights are the inverse noise variance, i.e., $w_i = h_i$.
In the case of CMB-S4, we will never be noise-dominated, either due to an actual primordial signal, or due to the lensing and foreground residuals, so here we use the inverse variance of the total signal and noise to determine the optimal weighting.

Lastly, while the scanning strategy used here has been optimized to mostly observe the cleanest available sky, some regions with high Galactic emission are also observed. Realistically, we would mask such regions when analyzing the maps. To assess this effect, we used masks based on a smoothed \planck\ \SI{353}{\GHz} polarized intensity map, keeping the cleanest 30\% or 60\% of the full sky (28\% and 58\% after apodization). We then used these masks to disregard some of the pixels, resulting in a degradation of the constraints on $r$.

\subsection{Delensing Revised}
The CMB-S4 PGW science goal can only be achieved if the majority of the lensing $B$ modes can be removed. The optimization in Section~\ref{sec:optimization} assumed a single frequency channel assigned to the higher resolution delensing
observations. The strength of polarized foregrounds at small angular scales is currently poorly constrained by data; the reference design, therefore, includes some additional coverage at higher and lower frequencies.\footnote{Though studies related to the role polarized small-scale galactic foreground residuals play in the delensing process have been performed in \citet{Fabbian:2019} and \citet{Beck2020}, a complete understanding of their impact is still missing.}

To forecast the delensing performance, we proceed in two steps. For a given LAT configuration and sky coverage, we derive the noise levels for an ILC that minimizes the variance of components with a frequency dependence that differs from that of a blackbody \citep{Tegmark:1995pn}. In this step, we assume that polarized foreground emission is dominated by Galactic synchrotron and thermal dust emission. Using the ILC noise power spectrum, we then forecast the performance expected for iterative $EB$ delensing \citep{Smith:2010gu}.

The LAT reference design, established independently from this work, and described in \cite{CMBS4-DSR}, includes 2 LATs in Chile for the wide field survey targeting science goals other than PGWs, and 1 LAT at the South Pole for the PGW delensing survey. For the Chile LATs, and a wide-area survey covering 70\% of the sky, the two-step procedure predicts that 73\% of the lensing power can be removed in the ``Chile shallow'' region after 7 years of observation. Similarly, for the single LAT at the South Pole dedicated to delensing of the approximately 3\% ``Chile deep'' and ``Pole deep'' regions, we expect to be able to remove close to 90\% of the lensing power after 7 years of observation.

%The numbers given above assume inverse noise-variance weighting. For the reference design such weighting is typically suboptimal, and in all the forecasts presented below we employ weights that account for both signal (e.g., for $r=0$ lensing residual after foreground removal) and noise. For the same survey, this leads to slightly higher noise and lensing residuals, but
%reduced $\sigma(r)$ due to reduced sample variance.

The numbers given above assume identical hits maps for the South Pole LATs and SATs, and assume inverse noise-variance weighting rather than a weighting scheme that accounts for both signal and noise. For the lensing residuals achieved by the reference design, inverse noise-variance weighting for the SAT maps is suboptimal because the signal (e.g., for $r=0$ lensing residual after foreground removal) is measured with signal-to-noise ratio above unity over a region that extends beyond the region that dominates the analysis in inverse noise-variance weighting. As a consequence, in all the forecasts presented below we employ weights that account for both signal and noise, which significantly increases the number of modes that contribute to the analysis, especially for small, deep patches. Even though, for the same survey, this leads to slightly higher noise and lensing residuals, the increase in the number of modes overall reduces the lensing sample variance contribution to $\sigma(r)$. In practice we determine the optimal weights iteratively, accounting for the effect on the SAT and LAT analyses and find rapid convergence.

\subsection{Results}
\begin{figure}[htbp]
\begin{center}
\includegraphics[width=0.9\columnwidth]{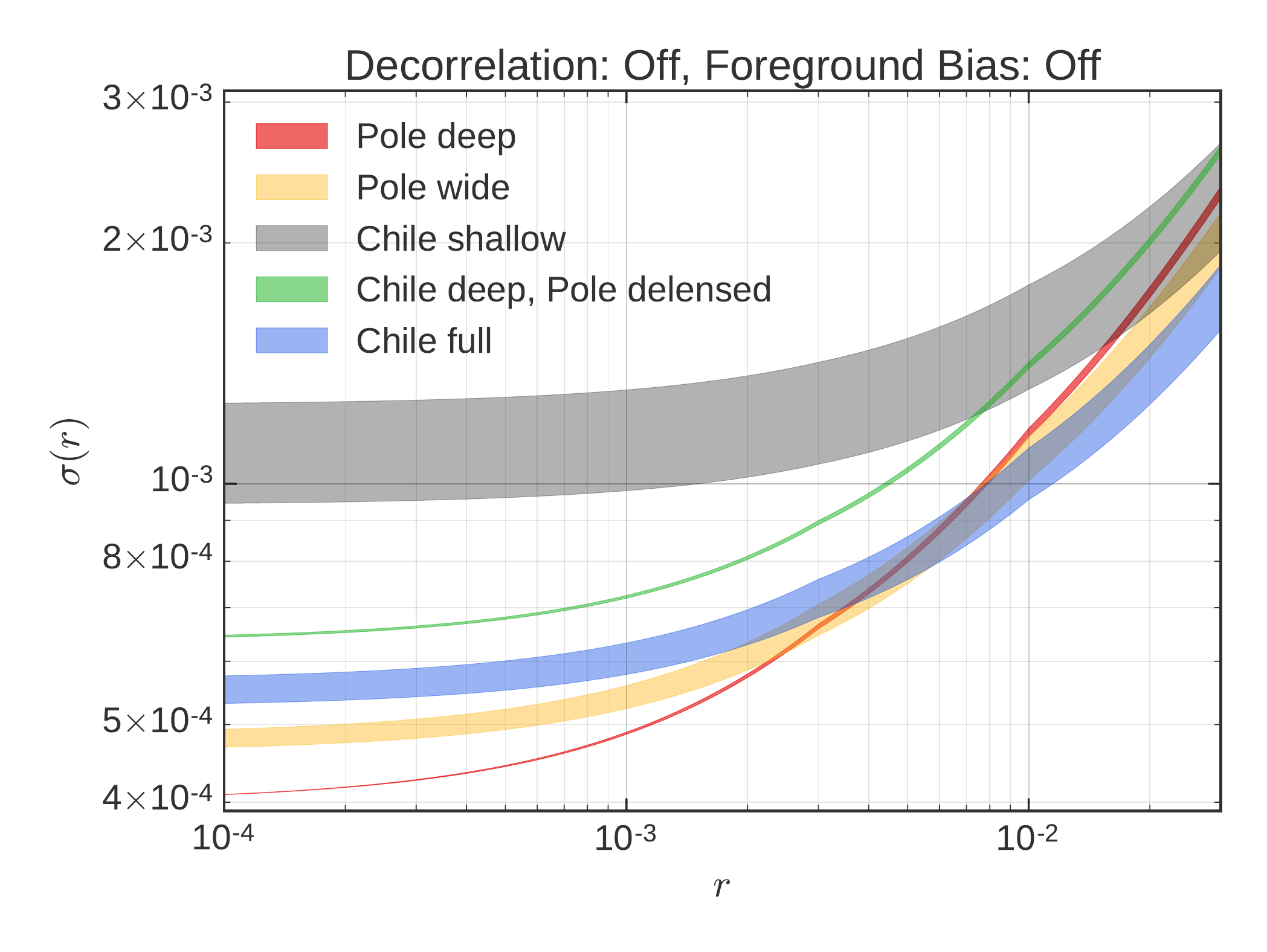}

\vspace{-0.25in}

\includegraphics[width=0.9\columnwidth]{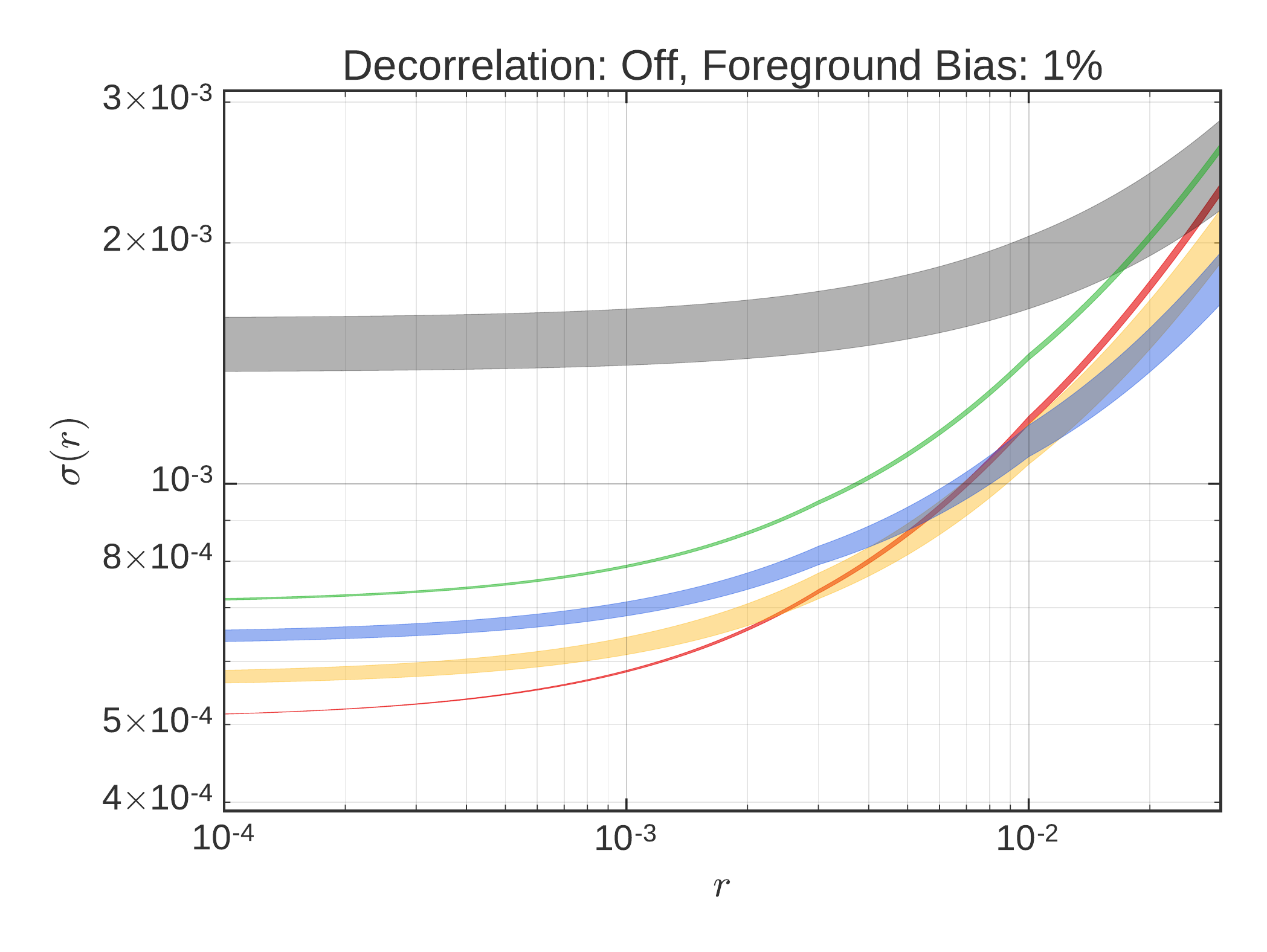}

\vspace{-0.25in}

\includegraphics[width=0.9\columnwidth]{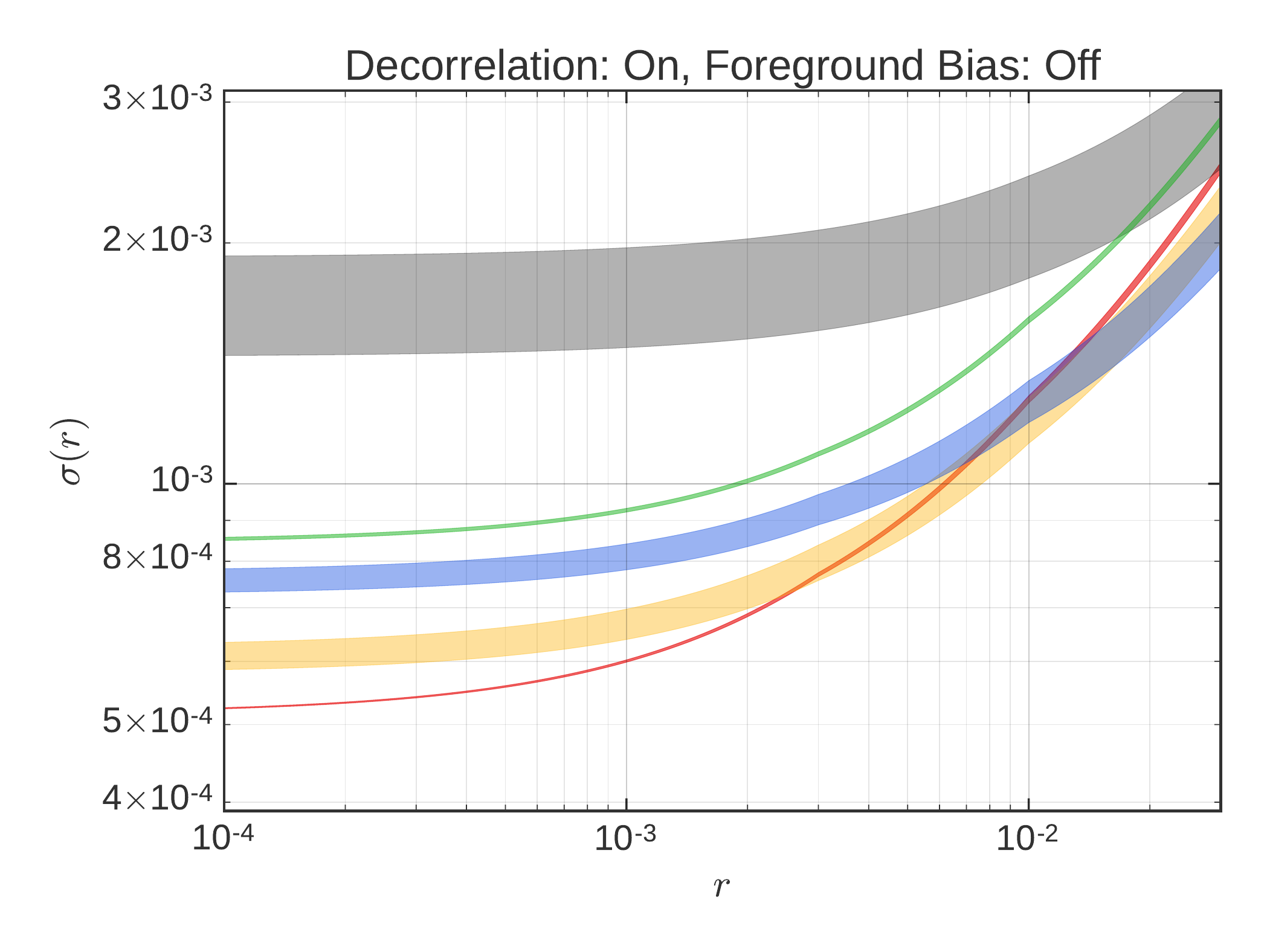}

\vspace{-0.25in}

\includegraphics[width=0.9\columnwidth]{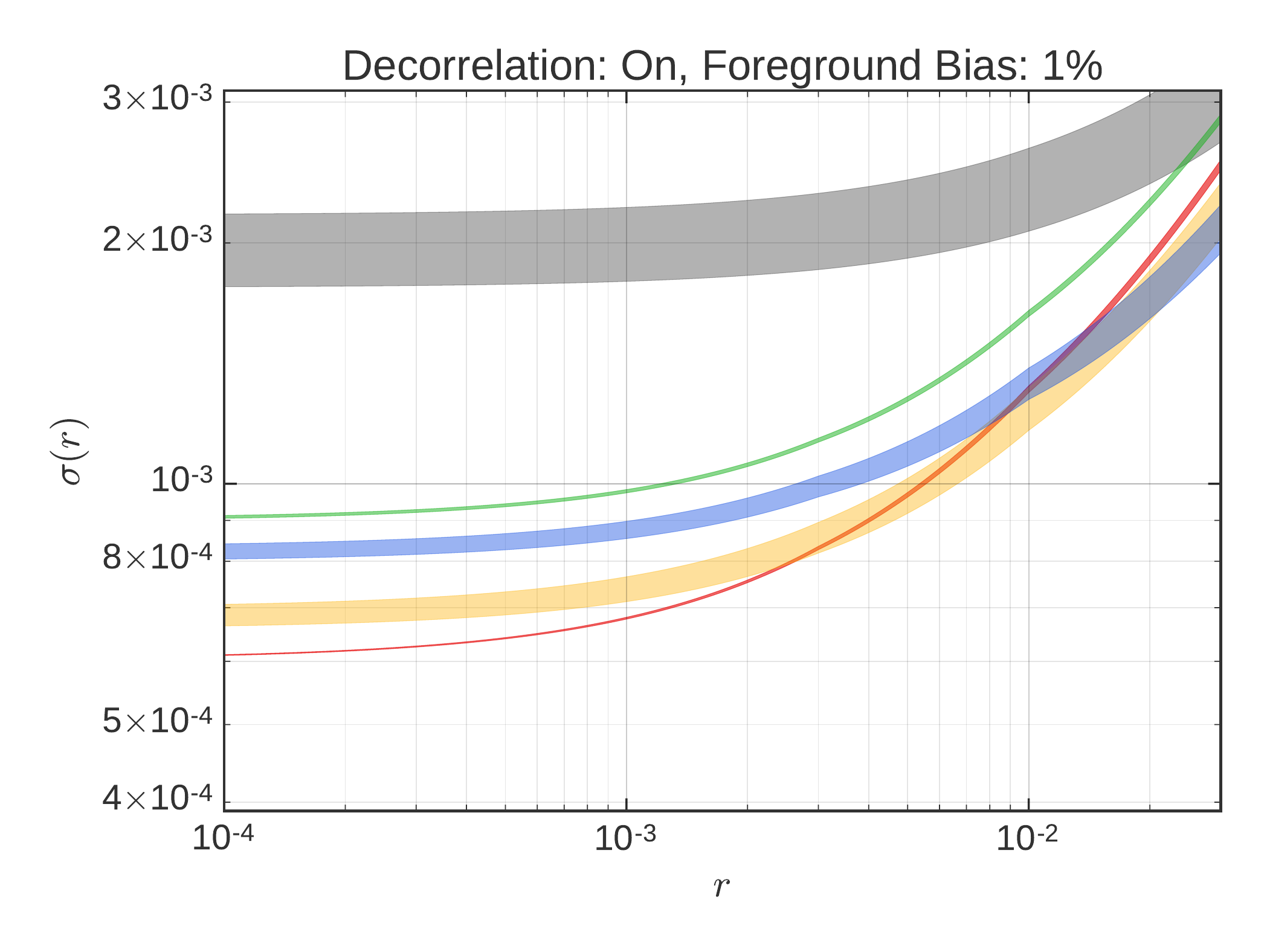}
\caption{Constraints on $r$ as a function of the value of $r$. We assume an instrument with 18 SAT optics tubes and an observation time of 7 years, with the five masks defined in Section \ref{sec:DSR}. Each band shows different Galactic cuts, based on {\it Planck\/} polarized foregrounds: the upper edge uses the cleanest 28\% of the full sky, whereas the lower edge uses the 58\% cleanest. We explore the effect of turning on foreground decorrelation in the forecasting. We also explore adding a foreground bias, in quadrature, with a 1\% value of the equivalent $r$ of the current foreground minimum of the BK15 data at $\ell=80$.
% The last sentence need explaination - but there is none in the text.
}
\label{fig:sigrvsr_4panels}
\end{center}
\end{figure}

The covariance matrices calculated as described above are used to produce the
results given in this section, where the number, siting, and coverage patterns
of the SATs are varied. In all cases, a delensing LAT at the South Pole is assumed to concentrate its coverage on a small patch of sky, while delensing over larger
sky areas is assumed to be available from the Chilean LATs.

As mentioned earlier, we split the Chilean coverage shown as ``Chile full'' in Figure~\ref{fig:relhits} into a deep patch, ``Chile deep'', which overlaps with the ``Pole deep'' region, and call the remainder ``Chile shallow.''
We then make separate forecasts for each Chilean sub-region using the appropriate delensing level for that sub-region. To obtain forecasts for the full region we add the $\sigma(r)$ results in simple inverse quadrature, thereby making the approximation of independence of the measured modes.
When we combine these with South Pole observations, we mimic a joint analysis over the overlapping region by taking the sum of the ``Pole deep'' and ``Chile deep'' coverage maps and computing the corresponding weights and lensing residuals, and add the ``Chile shallow'' results in inverse quadrature.

We also explore the possibility of unmodeled foreground residuals contributing residual power to the cleaned maps, assuming a foreground cleaning to 1\% at \SI{95}{GHz}. We do so by adding in quadrature to the $r$ forecasts a foreground bias equal to 1\% of the equivalent $r$ of the foreground minimum of the BK15 data at \SI{95}{GHz} and $\ell=80$.

Since some parts of the ``Chile shallow'' coverage lie closer to the Galactic plane, we boost the foreground level in the shallow region, and the foreground bias if it is included, by a factor of $3$ with respect to the deep patch. This scaling is based on the dust amplitude measured in the map-based simulations for surveys observed from the South Pole or from Chile, using a map with spatially varying foregrounds based on \textit{Planck} data. 

In Figure~\ref{fig:sigrvsr_4panels}, we show the dependence of $\sigma(r)$ on $r$ for the different coverage masks. We calculate these constraints for $r=0$, 0.003, 0.01, and 0.03, for the different Galactic cuts, and show the linear interpolation between these points, sampled on a high-resolution linearly-spaced grid. We find that the survey strategy from the South Pole is always favored in the limit of small $r$, with the crossover point depending on the specific assumptions. This comes as a direct result of the fact that while a larger fraction of the sky can be observed from Chile, one can concentrate the available sensitivity more deeply from the South Pole (see Figure~\ref{fig:relhits}). We note that the delensing requirements are more stringent for the smaller surveys (such as the South Pole one), highlighting the importance of periodically revisiting the assumptions made for the delensing survey of CMB-S4 as new data from Stage-3 experiments become available.

In Tables~\ref{tab:sigr_7yr_nodecorr_0}--\ref{tab:sigr_7yr_decorr_0} we present a set of $\sigma(r)$ results for seven years of observations and $r=0$, and in Tables~\ref{tab:sigdeg_7yr_nodecorr_0.003}--\ref{tab:sigdeg_7yr_decorr_0.003} we present a set of significance of detection levels for $r=0.003$, while varying the number of optics tubes at the South Pole and Chile over a wide range of options.
We show results for two different variants:
%(i) with no marginalization over the decorrelation parameters and observing efficiency in Chile equal to that at Pole for the 28\% cleanest polarized sky;
%(ii) assuming observing efficiency in Chile is half that at Pole;
%(iii) assuming 1\% unmodeled foreground residual uncertainty;
%(ii) marginalizing over the foreground decorrelation parameters. 
%and (v) keeping the 58\% cleanest sky.
%The results are moderately degraded in each of the variants. 
(i) with no marginalization over the decorrelation parameters and for the 28\% cleanest polarized sky; and (ii) the same as (i), but marginalizing over the foreground decorrelation parameters.

\renewcommand{\arraystretch}{1.5}
\begin{table}[htbp]
\setlength\tabcolsep{0pt}
 \centering
 \caption{Combined $10^4 \times \sigma(r)$ values (\textit{smaller} numbers are better), assuming $r=0$ after 7 years of observation, keeping only the 28\% cleanest part of the sky, assuming no decorrelation and an observing efficiency in Chile the same as at the South Pole. The bolded cells conserve the 18 SAT optics tube count of the reference design, while non-bolded cells explore other counts at the South Pole and in Chile.}
 \label{tab:sigr_7yr_nodecorr_0}
\begin{tabular}{c|cccccc}
\hline\hline
   \text{Chile}$\backslash$\text{Pole} & 0 & 6 & 9 & 12& 18& 30        \\
\hline
   0 & & \cellcolor[rgb]{1.000000,0.690888,0.000000}\text{ 6.3 }   & \cellcolor[rgb]{0.915033,0.957683,0.000000}\text{ 5.0 }   & \cellcolor[rgb]{0.653595,0.827477,0.000000}\text{ 4.5 }   & \cellcolor[rgb]{0.444444,0.723312,0.000000}\textbf{ 4.0 } & \cellcolor[rgb]{0.235294,0.619146,0.000000}\text{ 3.5  } \\
   6 & \cellcolor[rgb]{0.994054,0.000000,0.000000}\text{ 12 }    & \cellcolor[rgb]{1.000000,0.949250,0.000000}\text{ 5.5 }   & \cellcolor[rgb]{0.705882,0.853518,0.000000}\text{ 4.7 }   & \cellcolor[rgb]{0.549020,0.775394,0.000000}\textbf{ 4.3 } & \cellcolor[rgb]{0.392157,0.697270,0.000000}\text{ 3.9 }   & \cellcolor[rgb]{0.183007,0.593105,0.000000}\text{ 3.5  } \\
   9 & \cellcolor[rgb]{1.000000,0.392464,0.000000}\text{ 8.7 }   & \cellcolor[rgb]{0.915033,0.957683,0.000000}\text{ 5.1 }   & \cellcolor[rgb]{0.653595,0.827477,0.000000}\textbf{ 4.5 } & \cellcolor[rgb]{0.496732,0.749353,0.000000}\text{ 4.1 }   & \cellcolor[rgb]{0.339869,0.671229,0.000000}\text{ 3.8 }   & \cellcolor[rgb]{0.183007,0.593105,0.000000}\text{ 3.4 } \\
  12 & \cellcolor[rgb]{1.000000,0.568397,0.000000}\text{ 7.1 }   & \cellcolor[rgb]{0.758170,0.879559,0.000000}\textbf{ 4.8 } & \cellcolor[rgb]{0.549020,0.775394,0.000000}\text{ 4.3 }   & \cellcolor[rgb]{0.444444,0.723312,0.000000}\text{ 4.0 }   & \cellcolor[rgb]{0.287582,0.645188,0.000000}\text{ 3.7 }   & \cellcolor[rgb]{0.130719,0.567064,0.000000}\text{ 3.4   } \\
  18 & \cellcolor[rgb]{1.000000,0.875433,0.000000}\textbf{ 5.7 } & \cellcolor[rgb]{0.601307,0.801435,0.000000}\text{ 4.4 }   & \cellcolor[rgb]{0.444444,0.723312,0.000000}\text{ 4.1 }   & \cellcolor[rgb]{0.339869,0.671229,0.000000}\text{ 3.8 }   & \cellcolor[rgb]{0.235294,0.619146,0.000000}\text{ 3.6 }   & \cellcolor[rgb]{0.130719,0.567064,0.000000}\text{ 3.3  } \\
  30 & \cellcolor[rgb]{0.601307,0.801435,0.000000}\text{ 4.4 }   & \cellcolor[rgb]{0.339869,0.671229,0.000000}\text{ 3.9 }   & \cellcolor[rgb]{0.287582,0.645188,0.000000}\text{ 3.7 }   & \cellcolor[rgb]{0.235294,0.619146,0.000000}\text{ 3.6 }   & \cellcolor[rgb]{0.183007,0.593105,0.000000}\text{ 3.4 }   & \cellcolor[rgb]{0.078431,0.541023,0.000000}\text{ 3.2  } \\
\hline
\end{tabular}

 \centering
  \caption{Same as Table~\ref{tab:sigr_7yr_nodecorr_0}, but assuming additional foreground decorrelation parameters.}
  \label{tab:sigr_7yr_decorr_0}
 \begin{tabular}{c|cccccc}
\hline\hline
   \text{Chile}$\backslash$\text{Pole} & 0 & 6 & 9 & 12& 18& 30        \\
\hline
   0 & & \cellcolor[rgb]{1.000000,0.419531,0.000000}\text{ 8.4 }   & \cellcolor[rgb]{1.000000,0.608997,0.000000}\text{ 6.7 }   & \cellcolor[rgb]{1.000000,0.801615,0.000000}\text{ 6.0 }   & \cellcolor[rgb]{0.967320,0.983724,0.000000}\textbf{ 5.2 } & \cellcolor[rgb]{0.601307,0.801435,0.000000}\text{ 4.4 } \\
   6 & \cellcolor[rgb]{0.895937,0.000000,0.000000}\text{ 16 }    & \cellcolor[rgb]{1.000000,0.554864,0.000000}\text{ 7.3 }   & \cellcolor[rgb]{1.000000,0.727797,0.000000}\text{ 6.2 }   & \cellcolor[rgb]{1.000000,0.912341,0.000000}\textbf{ 5.6 } & \cellcolor[rgb]{0.862745,0.931642,0.000000}\text{ 5.0 }   & \cellcolor[rgb]{0.549020,0.775394,0.000000}\text{ 4.3 } \\
   9 & \cellcolor[rgb]{1.000000,0.027067,0.000000}\text{ 12 }    & \cellcolor[rgb]{1.000000,0.608997,0.000000}\text{ 6.8 }   & \cellcolor[rgb]{1.000000,0.801615,0.000000}\textbf{ 5.9 } & \cellcolor[rgb]{1.000000,0.949250,0.000000}\text{ 5.4 }   & \cellcolor[rgb]{0.810458,0.905600,0.000000}\text{ 4.9 }   & \cellcolor[rgb]{0.549020,0.775394,0.000000}\text{ 4.3 } \\
  12 & \cellcolor[rgb]{1.000000,0.270665,0.000000}\text{ 9.7 }   & \cellcolor[rgb]{1.000000,0.653979,0.000000}\textbf{ 6.4 } & \cellcolor[rgb]{1.000000,0.875433,0.000000}\text{ 5.7 }   & \cellcolor[rgb]{1.000000,0.986159,0.000000}\text{ 5.3 }   & \cellcolor[rgb]{0.758170,0.879559,0.000000}\text{ 4.8 }   & \cellcolor[rgb]{0.549020,0.775394,0.000000}\text{ 4.2 } \\
  18 & \cellcolor[rgb]{1.000000,0.487197,0.000000}\textbf{ 7.8 } & \cellcolor[rgb]{1.000000,0.838524,0.000000}\text{ 5.8 }   & \cellcolor[rgb]{1.000000,0.986159,0.000000}\text{ 5.3 }   & \cellcolor[rgb]{0.862745,0.931642,0.000000}\text{ 5.0 }   & \cellcolor[rgb]{0.705882,0.853518,0.000000}\text{ 4.6 }   & \cellcolor[rgb]{0.496732,0.749353,0.000000}\text{ 4.1 } \\
  30 & \cellcolor[rgb]{1.000000,0.764706,0.000000}\text{ 6.0 }   & \cellcolor[rgb]{0.915033,0.957683,0.000000}\text{ 5.1 }   & \cellcolor[rgb]{0.810458,0.905600,0.000000}\text{ 4.8 }   & \cellcolor[rgb]{0.705882,0.853518,0.000000}\text{ 4.6 }   & \cellcolor[rgb]{0.601307,0.801435,0.000000}\text{ 4.3 }   & \cellcolor[rgb]{0.392157,0.697270,0.000000}\text{ 4.0 } \\
\hline
\end{tabular}

%\end{table}
%
%\begin{table}[ht!]
%\setlength\tabcolsep{0pt}
 \centering
 \caption{Combined detection significance (\textit{larger} numbers are better) for $r=0.003$ after 7 years of observation, keeping only the 28\% cleanest part of the sky, assuming no decorrelation and an observing efficiency in Chile the same as at the South Pole. The bolded cells conserve the 18 SAT optics tube count of the reference design, while non-bolded cells explore other counts at the South Pole and in Chile.}
 \label{tab:sigdeg_7yr_nodecorr_0.003}
\begin{tabular}{c|cccccc}
\hline\hline
   \text{Chile}$\backslash$\text{Pole} & 0 & 6 & 9 & 12& 18& 30        \\
\hline
   0 & & \cellcolor[rgb]{1.000000,0.726644,0.000000}\text{ 3.7 }   & \cellcolor[rgb]{1.000000,0.989619,0.000000}\text{ 4.5 }   & \cellcolor[rgb]{0.839216,0.919923,0.000000}\text{ 4.9 }   & \cellcolor[rgb]{0.650980,0.826175,0.000000}\textbf{ 5.4 } & \cellcolor[rgb]{0.415686,0.708989,0.000000}\text{ 6.0 } \\
   6 & \cellcolor[rgb]{0.884045,0.000000,0.000000}\text{ 2.2 }   & \cellcolor[rgb]{1.000000,0.920415,0.000000}\text{ 4.3 }   & \cellcolor[rgb]{0.870588,0.935548,0.000000}\text{ 4.8 }   & \cellcolor[rgb]{0.729412,0.865236,0.000000}\textbf{ 5.2 } & \cellcolor[rgb]{0.556863,0.779300,0.000000}\text{ 5.6 }   & \cellcolor[rgb]{0.352941,0.677739,0.000000}\text{ 6.1 } \\
   9 & \cellcolor[rgb]{1.000000,0.025375,0.000000}\text{ 3.0 }   & \cellcolor[rgb]{0.949020,0.974610,0.000000}\text{ 4.6 }   & \cellcolor[rgb]{0.760784,0.880861,0.000000}\textbf{ 5.1 } & \cellcolor[rgb]{0.635294,0.818362,0.000000}\text{ 5.4 }   & \cellcolor[rgb]{0.494118,0.748051,0.000000}\text{ 5.8 }   & \cellcolor[rgb]{0.290196,0.646490,0.000000}\text{ 6.3 } \\
  12 & \cellcolor[rgb]{1.000000,0.685121,0.000000}\text{ 3.6 }   & \cellcolor[rgb]{0.839216,0.919923,0.000000}\textbf{ 4.9 } & \cellcolor[rgb]{0.666667,0.833987,0.000000}\text{ 5.3 }   & \cellcolor[rgb]{0.556863,0.779300,0.000000}\text{ 5.6 }   & \cellcolor[rgb]{0.431373,0.716801,0.000000}\text{ 5.9 }   & \cellcolor[rgb]{0.243137,0.623053,0.000000}\text{ 6.4 } \\
  18 & \cellcolor[rgb]{1.000000,0.948097,0.000000}\textbf{ 4.4 } & \cellcolor[rgb]{0.666667,0.833987,0.000000}\text{ 5.3 }   & \cellcolor[rgb]{0.541176,0.771488,0.000000}\text{ 5.7 }   & \cellcolor[rgb]{0.447059,0.724614,0.000000}\text{ 5.9 }   & \cellcolor[rgb]{0.337255,0.669927,0.000000}\text{ 6.2 }   & \cellcolor[rgb]{0.164706,0.583991,0.000000}\text{ 6.6 } \\
  30 & \cellcolor[rgb]{0.635294,0.818362,0.000000}\text{ 5.4 }   & \cellcolor[rgb]{0.400000,0.701176,0.000000}\text{ 6.0 }   & \cellcolor[rgb]{0.305882,0.654302,0.000000}\text{ 6.2 }   & \cellcolor[rgb]{0.243137,0.623053,0.000000}\text{ 6.4 }   & \cellcolor[rgb]{0.149020,0.576178,0.000000}\text{ 6.6 }   & \cellcolor[rgb]{0.023529,0.513679,0.000000}\text{ 6.9 } \\
\hline
\end{tabular}

 \centering
 \caption{Same as Table~\ref{tab:sigdeg_7yr_nodecorr_0.003}, but assuming additional foreground decorrelation paramaters.}
 \label{tab:sigdeg_7yr_decorr_0.003}
\begin{tabular}{c|cccccc}
\hline\hline
   \text{Chile}$\backslash$\text{Pole} & 0 & 6 & 9 & 12& 18& 30        \\
\hline
   0 & & \cellcolor[rgb]{1.000000,0.025375,0.000000}\text{ 3.0 }   & \cellcolor[rgb]{1.000000,0.685121,0.000000}\text{ 3.6 }   & \cellcolor[rgb]{1.000000,0.823529,0.000000}\text{ 4.0 }   & \cellcolor[rgb]{0.996078,0.998047,0.000000}\textbf{ 4.5 } & \cellcolor[rgb]{0.760784,0.880861,0.000000}\text{ 5.1 } \\
   6 & \cellcolor[rgb]{0.806741,0.000000,0.000000}\text{ 1.7 }   & \cellcolor[rgb]{1.000000,0.583622,0.000000}\text{ 3.5 }   & \cellcolor[rgb]{1.000000,0.795848,0.000000}\text{ 3.9 }   & \cellcolor[rgb]{1.000000,0.920415,0.000000}\textbf{ 4.3 } & \cellcolor[rgb]{0.933333,0.966797,0.000000}\text{ 4.7 }   & \cellcolor[rgb]{0.713725,0.857424,0.000000}\text{ 5.2 } \\
   9 & \cellcolor[rgb]{0.895937,0.000000,0.000000}\text{ 2.3 }   & \cellcolor[rgb]{1.000000,0.726644,0.000000}\text{ 3.7 }   & \cellcolor[rgb]{1.000000,0.865052,0.000000}\textbf{ 4.1 } & \cellcolor[rgb]{1.000000,0.975779,0.000000}\text{ 4.4 }   & \cellcolor[rgb]{0.870588,0.935548,0.000000}\text{ 4.8 }   & \cellcolor[rgb]{0.666667,0.833987,0.000000}\text{ 5.3 } \\
  12 & \cellcolor[rgb]{0.961348,0.000000,0.000000}\text{ 2.8 }   & \cellcolor[rgb]{1.000000,0.795848,0.000000}\textbf{ 3.9 } & \cellcolor[rgb]{1.000000,0.920415,0.000000}\text{ 4.3 }   & \cellcolor[rgb]{0.980392,0.990235,0.000000}\text{ 4.6 }   & \cellcolor[rgb]{0.823529,0.912111,0.000000}\text{ 4.9 }   & \cellcolor[rgb]{0.635294,0.818362,0.000000}\text{ 5.4 } \\
  18 & \cellcolor[rgb]{1.000000,0.482122,0.000000}\textbf{ 3.4 } & \cellcolor[rgb]{1.000000,0.920415,0.000000}\text{ 4.3 }   & \cellcolor[rgb]{0.964706,0.982422,0.000000}\text{ 4.6 }   & \cellcolor[rgb]{0.870588,0.935548,0.000000}\text{ 4.8 }   & \cellcolor[rgb]{0.745098,0.873049,0.000000}\text{ 5.1 }   & \cellcolor[rgb]{0.572549,0.787113,0.000000}\text{ 5.6 } \\
  30 & \cellcolor[rgb]{1.000000,0.892734,0.000000}\text{ 4.2 }   & \cellcolor[rgb]{0.870588,0.935548,0.000000}\text{ 4.8 }   & \cellcolor[rgb]{0.776471,0.888674,0.000000}\text{ 5.0 }   & \cellcolor[rgb]{0.713725,0.857424,0.000000}\text{ 5.2 }   & \cellcolor[rgb]{0.603922,0.802737,0.000000}\text{ 5.5 }   & \cellcolor[rgb]{0.462745,0.732426,0.000000}\text{ 5.8 } \\
\hline
\end{tabular}

\end{table}

In all these calculations we assume that the observing efficiency from Chile is $100\%$ equivalent to that from the South Pole. 
%While the forecast variant with 50\% observing efficiency from Chile has no impact on the sensitivity of the reference design, which locates all the small aperture telescopes at the South Pole, it reinforces the preference for Pole-centric configurations.
Despite long histories of CMB observations at both sites, it is still quite challenging to make a clean comparison of their observing efficiencies.
Results from the BICEP/Keck program are responsible for the leading constraints on $r$ for the last decade,
but it is still not currently possible to disentangle the role of the observing site from other factors that have contributed to the success of that program,
such as detector performance, instrument design, observing strategy, and operations management.
However, note also that even assuming equal observing efficiency at both sites (Table~\ref{tab:sigr_7yr_nodecorr_0}), there is still a difference
between $\sigma(r)$ obtained for equivalent numbers of optics tubes in Chile versus Pole, due to sky fraction, foreground complexity and delensing. In the CMB-S4 Reference Design Report we also show forecasting results that assume an observing efficiency from Chile equivalent to 50\% of South Pole efficiency. 

Note that the results where the total number of SAT optics tubes is different from the assumed number of 18 are subject to a caveat: the delensing effort is assumed to be held fixed to the reference design---one LAT at the South Pole and two in Chile. In principle, as the total effort is varied away from the reference design, one should re-optimize the fraction of delensing effort as per Figure~\ref{fig:sigr}.

\section{Conclusions}
In this paper we have presented the \textit{CMB-S4 r forecasting framework} and its evolution through three major iterations. 
To ensure forecasting realism, we have chosen to directly ground our forecasting in the end-to-end
on-sky achieved performance of Stage 2--3 experiments, rather than rely on ab initio assumptions. 
Additionally, we have checked the semi-analytic calculations with map-based simulations 
of increasing complexity, thereby creating a closed forecasting loop (presented in Figure~\ref{fig:forecasting_loop}). 
% This approach has allowed us to naturally incorporate all the effects which impact real-world experiments, 
% to iterate over all parts of the framework as new information became available, 
% and to flexibly trace the end-to-end effects of any individual change. 
This approach allows us to confidently incorporate all the effects that impact current real-world experiments, 
to flexibly trace the end-to-end effects of changes to experimental design or assumptions, 
and to iterate over the framework as more sophisticated sky and instrument models become available.

The semi-analytic calculations described in Sections~\ref{sec:fopt} and \ref{sec:optimization} indicate that for a 3\% sky fraction, $1.2\times 10^{6}$ 150\nobreakdash-GHz\nobreakdash-equivalent detector-years (or $1.8 \times 10^6$ when including marginalization over the decorrelation parameters) are required to reach the science requirement of $\sigma(r)=5\times10^{-4}$, with 30\% of this effort assigned to the delensing survey. The resulting optimal distribution of detectors across frequency bands is summarized in Table~\ref{tab:distribution}.

The map-based simulations described in Sections~\ref{sec:cdt_sims}--\ref{sec:map_results} confirm the $\sigma(r)$ results from the semi-analytic calculations. These simulations also indicate that bias in the recovered $r$ value is within $1\sigma$ for a suite of different foreground models. However, we note that foregrounds remain a serious issue that must be periodically revisited as the project progresses.
  
Additional map-based simulations indicate that the systematic bias
on $r$ can be controlled to $<1\sigma$, provided that fractional contamination levels similar to those already achieved by small aperture telescopes can be maintained.
  
Mapping the requirements from the semi-analytic calculations onto realizable instruments (see Section \ref{sec:DSR}) results in the reference design described in \cite{CMBS4-DSR} for a seven-year survey period, and in the constraints presented in Tables~\ref{tab:sigr_7yr_nodecorr_0}--\ref{tab:sigdeg_7yr_decorr_0.003}.

In Section~\ref{sec:DSR} we have also extended the semi-analytic calculations to account for realistic observation patterns and probed the dependence of $\sigma(r)$ on $r$, as a function of experiment siting, finding that the survey strategy from the South Pole is always favored in the limit of small $r$. We note that surveys with smaller footprints (such as the South Pole survey) depend more tightly on the levels of achieved delensing. Therefore, revising the delensing assumptions will be important as new studies of small-scale foregrounds become available. 

Going forward, as the CMB-S4 project matures, the collaboration will need to 
converge on increasingly specific instrumentation choices.  
To quantitatively discern between the different options, 
we intend to continue using and developing the closed-loop framework presented here. 
Among others, we anticipate adding complexity to the following directions: 
delensing treatment; foreground simulations; foreground parametrization; 
survey strategies; and instrumental systematics. 
Additionally, we expect to apply new analysis methods to our data challenges and incorporate new achieved performance levels at multiple additional frequencies from multiple sites as these data become available. The iterative nature of our framework can easily accommodate these revisions, achieving increased realism with each iteration.

\section*{Acknowledgements}
%The computations in this paper were run on the Odyssey cluster supported by the FAS Science Division Research Computing Group at Harvard University, and on the National Energy Research Scientific Computing Center (NERSC), a U.S. Department of Energy Office of Science User Facility operated under Contract No. DE-AC02-05CH11231.

%%%% Generic description %%%%
The CMB-S4 collaboration (https://cmb-s4.org/) is working to plan, construct, and operate a next-generation, multisite CMB experiment in the 2020s.  The collaboration is led by an elected Governing Board, Spokespeople, Committee Chairs, and Executive Team.  
%%%%% Project-wide funding and resources %%%%
Funding for the CMB-S4 Integrated Project Office is provided by the Department of Energy's Office of Science (project level CD-0) and by the National Science Foundation through the Mid-Scale Research Infrastructure-R1 award OPP-1935892.  
%%%% The Labs (using acknowledgment text grabbed from the web) %%%%
This research used resources of Argonne National Laboratory, a U.S. Department of Energy (DOE) Office of Science User Facility operated under Contract No. DE-AC02-06CH11357.
This document was prepared by the CMB-S4 collaboration using the resources of the Fermi National Accelerator Laboratory (Fermilab), a U.S. Department of Energy, Office of Science, HEP User Facility. Fermilab is managed by Fermi Research Alliance, LLC (FRA), acting under Contract No. DE-AC02-07CH11359.
Work at Lawrence Berkeley National Laboratory was supported by the Office of Science, Office of Basic Energy Sciences, of the U.S. Department of Energy under Contract No. DE-AC02-05CH11231.
Work at SLAC National Accelerator Laboratory is supported by the U.S. Department of Energy, Office of Science, Office of Basic Energy Sciences under Contract No. DE-AC02-76SF00515.
%%%% Other project resources %%%% 
This research used resources of the National Energy Research Scientific Computing Center, which is supported by the Office of Science of the U.S. Department of Energy under Contract No. DE-AC02-05CH11231.
%%%% Individual PI grants %%%%
In the US, work on CMB-S4 by individual investigators has been supported by the National Science Foundation (awards 1248097, 1255358, 1815887, 1835865, 1852617), the Department of Energy (awards DE-SC0009919, DE-SC0009946, DE-SC0010129), and the National Aeronautics and Space Administration (award ATP-80NSSC20K0518). 
In Australia, the Melbourne authors acknowledge support from an Australian Research Council Future Fellowship (FT150100074).
In Canada, RH is supported by the Discovery Grants program from NSERC, and acknowledges funding from CIFAR, the Sloan Foundation and the Dunlap family.
In Italy, CB acknowledges support under the ASI COSMOS and INFN INDARK programs.
In the Netherlands, DM acknowledges NWO VIDI award number 639.042.730.
In Switzerland, JC is supported by a SNSF Eccellenza Professorial Fellowship (No. 186879).
In the United Kingdom, AL, GF, JC are supported by the European Research Council under the European Union's Seventh Framework Programme (FP/2007-2013) / ERC Grant Agreement No. [616170].  AL also acknowledges STFC award ST/P000525/1.
SM is supported by the research program Innovational Research Incentives Scheme (Vernieuwingsimpuls), which is financed by the Netherlands Organization for Scientific Research through the NWO VIDI Grant No. 639.042.612-Nissanke and the Labex ILP (reference ANR-10-LABX-63) part of the Idex SUPER, received financial state aid managed by the Agence Nationale de la Recherche, as part of the programme Investissements d'avenir under the reference ANR-11-IDEX-0004-02.
%%%% Other (BB forecast specific) %%%%
Some computations in this paper were run on the Odyssey cluster, supported by the FAS Science Division Research Computing Group at Harvard University.

%------ APPENDIX ------------------%
\begin{appendix}
%\renewcommand{\thesection}{\Alph{section}}
%\section{Appendix A -- Multicomponent Theory Model}
\section{Multicomponent Theory Model}
\label{app:Appendix A}
%\numberwithin{equation}{section}
%\setcounter{equation}{0}

The forecasting framework uses a parametric model to describe the bandpower expectation values as a combination of cosmological and foreground signals. This parametrization follows the one presented in Appendix~G of \citet{Array:2018x}.

The model includes signals from the CMB (lensed-scalar and tensor contributions), Galactic dust, and Galactic synchrotron, with the possibility of spatial correlation between dust and synchrotron.
%Since these four components are uncorrelated with one another, the expectation value for cross-spectra between components is zero.
Contributions from dust and synchrotron to the $BB$ spectrum between maps at frequencies $\nu_1$ and $\nu_2$ can be written in the following form, after computing the expected spectrum as a function of $\ell$, and applying bandpower window functions are applied to calculate binned bandpower expectation values:
\begin{equation}
  \mathcal{D}_{\ell,BB}^{\nu_1 \times \nu_2} = 
  A_\mathrm{dust} \ddp f_\mathrm{d}^{\nu_1} f_\mathrm{d}^{\nu_2} \left( \frac{\ell}{80} \right)^{\ad} + 
  \As \Delta'_\mathrm{s} f_\mathrm{s}^{\nu_1} f_\mathrm{s}^{\nu_2} \left( \frac{\ell}{80} \right)^{\as} 
  + \epsilon \sqrt{A_\mathrm{dust} \As} (f_\mathrm{d}^{\nu_1} f_\mathrm{s}^{\nu_2} + f_\mathrm{s}^{\nu_1} f_\mathrm{d}^{\nu_2}) \left( \frac{\ell}{80} \right)^{(\ad + \as) / 2}.
  \label{eq:BB_model}
\end{equation}
% \begin{align}
%   \begin{split}
%     \mathcal{D}_{\ell,BB}^{\nu_1 \times \nu_2} &= 
%     A_{dust} \ddp f_\mathrm{d}^{\nu_1} f_\mathrm{d}^{\nu_2} \left( \frac{\ell}{80} \right)^{\ad} + \\
%     & + \As \Delta'_\mathrm{s} f_\mathrm{s}^{\nu_1} f_\mathrm{s}^{\nu_2} \left( \frac{\ell}{80} \right)^{\as} + \\
%     &\pm\epsilon \sqrt{A_{dust}\As} (f_\mathrm{d}^{\nu_1} f_\mathrm{s}^{\nu_2} + f_\mathrm{s}^{\nu_1} f_\mathrm{d}^{\nu_2}) \left( \frac{\ell}{80} \right)^{(\ad + \as) / 2}
%     \label{eq:BB_model}
%   \end{split}
% \end{align}
Parameters $A_\mathrm{dust}$ and $\As$ specify the dust and synchrotron power in units of $\mu\mathrm{K}_\mathrm{\mbox{\sc cmb}}^2$ at angular scale $\ell = 80$.
These are defined at pivot frequencies of \SI{353}{\GHz} for dust and \SI{23}{\GHz} for synchrotron.
The dust and synchrotron components scale as power laws in $\ell$ with slopes $\ad$ and $\as$, respectively (note that slope parameters are defined for $\mathcal{D}_\ell \equiv \clstar$).
The parameter $\epsilon$ specifies the level of spatial correlation between dust and synchrotron; this correlation coefficient is assumed to be constant across all $\ell$. If either $A_{\mathrm{dust}}$ or $A_{\mathrm{sync}}$ are negative, the contribution of the correlated component to the expectation value flips sign. Negative foreground amplitudes are technically nonphysical, but this analytic continuation becomes important when we explore entire parameter phase-spaces, for instance in the simulation re-analysis steps. 

Coefficients $f_\mathrm{d}^\nu$ and $f_\mathrm{s}^\nu$, given by equations~\ref{eq:dust_scaling} and~\ref{eq:sync_scaling}, describe the scaling of dust and synchrotron amplitude from the pivot frequencies to the actual bandpasses of the maps at frequencies $\nu_1$ and $\nu_2$. 
The SED model used for dust emission is a modified blackbody, i.e., a Planck function multiplied by a power law with emissivity spectral index $\Bd$~\citep{planckiXXII}.  We adopt a dust temperature $T_\mathrm{d} = 19.6 \mathrm{K}$.
%This simplifying assumption about the behavior of dust holds over the range of frequencies that we consider.
The SED model used for synchrotron is defined as a power law with spectral index $\Bs$ relative to a Rayleigh-Jeans spectrum.
The frequency scaling also includes the unit conversion for $\mu \mathrm{K}_\mathrm{CMB}$ from the pivot frequency to the target bandpass.

To integrate the SED and unit conversion over the bandpass of the target map, we adopt the convention used by \citet{planck2013IX}, in which our bandpass functions describe response as a function of frequency to a beam-filling source with uniform spectral radiance.
For emissivity spectral index $\beta_\mathrm{d}$ and dust temperature $T_\mathrm{d}$, the modified blackbody scaling (MBBS) from pivot frequency $\nu_\mathrm{pivot}$ to a map with bandpass $R(\nu)$ is given by
\begin{equation}
  f_\mathrm{MBBS}^{\nu} = \frac{\int{d\nu R(\nu) \nu^{3+\beta_\mathrm{d}} \left(\exp{\frac{h\nu}{kT_\mathrm{d}}} - 1\right)^{-1}}}{\nu_\mathrm{pivot}^{3+\beta_\mathrm{d}} \left(\exp{\frac{h\nu_\mathrm{pivot}}{kT_\mathrm{d}}} - 1\right)^{-1}} .
  \label{eq:greybody_scaling}
\end{equation}
For a synchrotron power-law scaling (PLS) with spectral index $\beta_\mathrm{s}$, we calculate the coefficient in a similar way,
\begin{equation}
  f_\mathrm{PLS}^{\nu} = \frac{\int{d\nu R(\nu) \nu^{2+\beta_\mathrm{s}}}}{\nu_\mathrm{pivot}^{2+\beta_\mathrm{s}}} .
  \label{eq:powerlaw_scaling}
\end{equation}
The conversion between $\mu\mathrm{K}_\mathrm{\mbox{\sc cmb}}$ units at the pivot frequency and $\mu\mathrm{K}_\mathrm{\mbox{\sc cmb}}$ units at the target map bandpass is given by
\begin{equation}
  f_\mathrm{T_{CMB}}^{\nu} = \frac{\nu_\mathrm{pivot}^4 \exp{\frac{h\nu_\mathrm{pivot}}{kT_\mathrm{CMB}}} \left(\exp{\frac{h\nu_\mathrm{pivot}}{kT_\mathrm{CMB}}}-1\right)^{-2}}{\int{d\nu R(\nu) \nu^4 \exp{\frac{h\nu}{kT_\mathrm{CMB}}} \left(\exp{\frac{h\nu}{kT_\mathrm{CMB}}}-1\right)^{-2}}} .
  \label{eq:ukcmb_conversion}
\end{equation}
Combining these factors, we obtain the scalings used in equation~\ref{eq:BB_model}, 
\begin{align}
  f_\mathrm{d}^{\nu} &= f_\mathrm{MBBS}^{\nu} \times f_\mathrm{T_{CMB}}^{\nu} \label{eq:dust_scaling} \\
  f_\mathrm{s}^{\nu} &= f_\mathrm{PLS}^{\nu} \times f_\mathrm{T_{CMB}}^{\nu} \label{eq:sync_scaling}
\end{align}

We also consider dust and synchrotron frequency decorrelation. The simplest possible model of a polarized foreground component is one with a fixed spatial pattern on the sky that scales with frequency according to a single SED. In this case, the expectation value of the cross-spectrum between any two frequencies is the geometric mean of the respective auto-spectrum expectation values. In reality, the spatial pattern might vary as a function of frequency, leading to the cross-spectra being suppressed with respect to the geometric mean of the auto-spectra~\citep{Aghanim:2017}. We refer to this phenomenon as decorrelation.

We model decorrelation in the power spectrum domain using a set of simple one-parameter models.
We define the correlation ratio of dust between two reference frequencies, 217 and 353~GHz, at pivot scale $\ell=80$ as
\begin{equation}
\dd =
\frac{\mathcal{D}_{80}(217\times 353)}{\sqrt{\mathcal{D}_{80}(217\times 217)\mathcal{D}_{80}(353\times 353)}} ,
\label{eqn:Delta}
\end{equation}
where $\mathcal{D}_{80}$ is the dust power at $\ell=80$.
Here $\dd<1$ corresponds to decorrelation. We scale to other frequency combinations using the factor suggested by \citet{Aghanim:2017},
\begin{equation}
f(\nu_1,\nu_2) = \frac{(\log (\nu_1/\nu_2))^2}{(\log(217/353))^2}.
\label{eqn:nuscale}
\end{equation}
Similarly, based on suggestions from \cite{Aghanim:2017}, we consider three possible scalings with $\ell$:
\begin{equation}
g(\ell) = \left\{ \begin{array}{ll}
                   1,             & \, \text{flat case}; \\
                   (\ell / 80),   & \, \text{linear case}; \\
                   (\ell / 80)^2, & \, \text{quadratic case.}
                   \end{array} \right.
\label{eqn:ellscale}
\end{equation}
%The $\ell$ range in which we have high signal-to-noise measurements is farily narrow, therefore this choice turns out to make little practical difference.
The scalings above can produce extreme (and non-physical)
behavior at high $\ell$ or for frequencies that are widely separated. We therefore re-map the scaled correlation coefficient using the following function:
\begin{equation}
\ddp(\nu_1,\nu_2,\ell) = \exp \left[ \log(\dd) \, f(\nu_1,\nu_2) \, g(\ell) \right] .
\label{eqn:remap}
\end{equation}
With this re-mapping, $\ddp$ remains in the
range 0 (no correlation) to 1 (perfect correlation) for all values of $f$ and $g$.
This combination of frequency scaling and non-linear re-mapping has been shown to correspond to a Gaussian spatial variation in the foreground spectral index parameter~\citep{vansyngel16}.

In a similar vein, we define the parameter $\Delta'_\mathrm{s}$, which describes decorrelation of the synchrotron pattern. %In practice, synchrotron decorrelation is centered at 1 (but is marginalized over with a flat unbounded prior), given that current
%data does offer accurate measurements of this parameter.
We do not include foreground decorrelation parameters in the dust-synchrotron correlated component.
A complete foreground model would include all correlations between dust and synchrotron foregrounds across observing frequencies, but the current data do not provide useful guidance about the form of such correlations.  

%The foreground contribution to $EE$ is similar, with $\Ad$ and $\As$ scaled by the $EE/BB$ ratios for dust and synchrotron, which are both assumed to be equal to 2 \citep{planckiLIV,krachmalnicoff18}. The model has no $EB$ contributions. We also do not include $TT/TE/TB$ spectra in the likelihood analysis, nor do we model unpolarized foregrounds.

In addition to foregrounds, we include CMB scalar and tensor contributions.
We make the simplifying assumptions that the tensor $BB$ spectrum is given by a template scaled by parameter $r$ and the $BB$ spectrum from lensed CMB is given by a template scaled by parameter $A_\mathrm{L}$.
Using CMB temperature units, the CMB contribution to the $BB$ spectrum is given by
\begin{equation}
  D_{\ell,BB}^{\nu_1\times\nu_2}=\frac{r}{0.1} D_{\ell,BB}^{\text{tensor}}+A_\mathrm{L} D_{\ell,BB}^{\text{lensing}},
\end{equation}
where $D_{\ell,BB}^{\text{tensor}}$ is the $BB$ template for a tensor signal with $r=0.1$ and $D_{\ell,BB}^{\text{lensing}}$ is the expected lensing $BB$ spectrum for $\Lambda$CDM. These are obtained using the \texttt{CAMB}\footnote{\url{https://camb.info/}} package \citep{Lewis:2011ac}.
%Similarly for $EE$:
%\begin{equation}
%    D_{l,EE}^{\nu_1\times\nu_2}=\frac{r}{0.1}D_{l,EE}^{tensor}+ D_{l,EE}^{lensing}.
%\end{equation}
%We expect that the second term dominates the $EE$ contribution unless $r$ is very large. Also, the $EE$ lensing component does not vary with $A_L$ as the lensing of B-modes into E-modes is assumed to be negligible compared to the E-mode expectation from $\Lambda$CDM.

\section{BPCM Construction and Rescaling}
\label{app:Appendix B}

The bandpower covariance matrix (BPCM) construction relies on analytic rescaling of simulations.
We calculate and store the bandpower covariance of signal and noise simulations for a particular sky and instrument model; rescaling from these covariance matrices eliminates the computational cost of creating large suites of simulations for each desired model.
In the rescaling process, we explicitly set to zero terms that have zero expectation value (under the assumption that signal and noise are uncorrelated, and physically unrelated signals are uncorrelated) to reduce Monte Carlo error in the covariance matrix that stems from the modest number (499) of simulation realizations used.
This framework is described in detail in \cite{BuzaThesis}, but we review it briefly here.

In a general case, let $m_a$, $m_b$, $m_c$, and $m_d$ denote the four maps included in our analysis, with $m_a \times m_b$ denoting the cross-spectrum between those maps.
Each map is the sum of independent signal components (CMB and foregrounds) plus a noise contribution, i.e., $m_a = \sum_i s_{ai} + n_a$.
Since the synchrotron and dust foregrounds could be spatially correlated, we choose to divide them into three mutually-independent components---uncorrelated synchrotron, uncorrelated dust, and the correlated part of synchrotron and dust.
Using the properties that these signal and noise fields are independent and have zero mean, the expectation values for many spectra can be set to zero:  $\left\langle s_{xi} \times s_{yj} \right\rangle = 0$ for $i \ne j$ (independent signal fields)
and $\left\langle s_{xi} \times n_y \right\rangle = 0 $ (signal is independent of noise). Then, a generic bandpower covariance term can be written as
\begin{align}
  \mathrm{Cov} \left( m_a \times m_b, m_c \times m_d \right) &= \left\langle ( m_a \times m_b ) ( m_c \times m_d ) \right\rangle - \left\langle m_a \times m_b \right\rangle \left\langle m_c \times m_d \right\rangle \\
  \begin{split}
    &= \sum_i \mathrm{Cov} \left( s_{ai} \times s_{bi}, s_{ci} \times s_{di} \right) \\
    &\quad + \sum_i \sum_{j \ne i} \mathrm{Cov} \left( s_{ai} \times s_{bj}, s_{ci} \times s_{dj} \right) + \sum_i \sum_{j \ne i} \mathrm{Cov} \left( s_{ai} \times s_{bj}, s_{cj} \times s_{dj} \right) \\
    &\quad + \sum_i \mathrm{Cov} \left( s_{ai} \times n_b, s_{ci} \times n_d \right) + \sum_i \mathrm{Cov} \left( s_{ai} \times n_b, n_c \times s_{di} \right) \\
    &\quad + \sum_i \mathrm{Cov} \left( n_a \times s_{bi}, s_{ci} \times n_d \right) + \sum_i \mathrm{Cov} \left( n_a \times s_{bi}, n_c \times s_{di} \right) \\
    &\quad + \mathrm{Cov} \left( n_a \times n_b, n_c \times n_d \right) ,
  \end{split}
\end{align}
where the first three terms are signal-only covariances, the next four terms are covariances between select signal--noise cross-spectra, and the last term is a noise-only covariance.
Additional terms can be set to zero if we make the further assumption that the noise is independent in each map, i.e., $\left\langle n_x \times n_y \right\rangle = 0$ for $x \ne y$.
We also set to zero the covariance between any bandpowers that are separated by more than one bin in $\ell$ (for bins with $\Delta\ell = 35$), since these correlations are very small and not well measured.

Rather than running simulations with a complicated combination of CMB and foreground signals, we calculate the above covariance terms for a simple signal model and then rescale to obtain the bandpower covariance matrix for an arbitrary model.
In practice, the signal simulations are CMB realizations for a lensed $\Lambda$CDM model.
Using $s_{x0}$ to denote the simulated signal component of map $m_x$, the rescaled bandpower covariance matrix term is given by
\begin{equation}
  \begin{split}
    \mathrm{Cov} \left( m_a \times m_b, m_c \times m_d \right) &= \mathrm{Cov} \left( s_{a0} \times s_{b0}, s_{c0} \times s_{d0} \right) \sum_i \left[ \frac{ \left\langle s_{ai} \times s_{ai} \right\rangle \left\langle s_{bi} \times s_{bi} \right\rangle \left\langle s_{ci} \times s_{ci} \right\rangle \left\langle s_{di} \times s_{di} \right\rangle }{ \left\langle s_{a0} \times s_{a0} \right\rangle \left\langle s_{b0} \times s_{b0} \right\rangle \left\langle s_{c0} \times s_{c0} \right\rangle \left\langle s_{d0} \times s_{d0} \right\rangle } \right]^{1/2} \\
    &+ \frac{1}{2} \mathrm{Cov} \left( s_{a0} \times s_{b0}, s_{c0} \times s_{d0} \right) \sum_i \sum_{j \ne i} \left[ \frac{ \left\langle s_{ai} \times s_{ai} \right\rangle \left\langle s_{bj} \times s_{bj} \right\rangle \left\langle s_{ci} \times s_{ci} \right\rangle \left\langle s_{dj} \times s_{dj} \right\rangle }{ \left\langle s_{a0} \times s_{a0} \right\rangle \left\langle s_{b0} \times s_{b0} \right\rangle \left\langle s_{c0} \times s_{c0} \right\rangle \left\langle s_{d0} \times s_{d0} \right\rangle } \right]^{1/2} \\
    &+ \frac{1}{2} \mathrm{Cov} \left( s_{a0} \times s_{b0}, s_{c0} \times s_{d0} \right) \sum_i \sum_{j \ne i} \left[ \frac{ \left\langle s_{ai} \times s_{ai} \right\rangle \left\langle s_{bj} \times s_{bj} \right\rangle \left\langle s_{cj} \times s_{cj} \right\rangle \left\langle s_{di} \times s_{di} \right\rangle }{ \left\langle s_{a0} \times s_{a0} \right\rangle \left\langle s_{b0} \times s_{b0} \right\rangle \left\langle s_{c0} \times s_{c0} \right\rangle \left\langle s_{d0} \times s_{d0} \right\rangle } \right]^{1/2} \\
    &+ \mathrm{Cov} \left( s_{a0} \times n_b, s_{c0} \times n_d \right) \sum_i \left[ \frac{ \left\langle s_{ai} \times s_{ai} \right\rangle \left\langle s_{ci} \times s_{ci} \right\rangle }{ \left\langle s_{a0} \times s_{a0} \right\rangle \left\langle s_{c0} \times s_{c0} \right\rangle } \right]^{1/2} \\
    &+ \mathrm{Cov} \left( s_{a0} \times n_b, n_c \times s_{d0} \right) \sum_i \left[ \frac{ \left\langle s_{ai} \times s_{ai} \right\rangle \left\langle s_{di} \times s_{di} \right\rangle }{ \left\langle s_{a0} \times s_{a0} \right\rangle \left\langle s_{d0} \times s_{d0} \right\rangle } \right]^{1/2} \\
    &+ \mathrm{Cov} \left( n_a \times s_{b0}, s_{c0} \times n_d \right) \sum_i \left[ \frac{ \left\langle s_{bi} \times s_{bi} \right\rangle \left\langle s_{ci} \times s_{ci} \right\rangle }{ \left\langle s_{b0} \times s_{b0} \right\rangle \left\langle s_{c0} \times s_{c0} \right\rangle } \right]^{1/2} \\
    &+ \mathrm{Cov} \left( n_a \times s_{b0}, n_c \times s_{d0} \right) \sum_i \left[ \frac{ \left\langle s_{bi} \times s_{bi} \right\rangle \left\langle s_{di} \times s_{di} \right\rangle }{ \left\langle s_{b0} \times s_{b0} \right\rangle \left\langle s_{d0} \times s_{d0} \right\rangle } \right]^{1/2} \\
    &+ \mathrm{Cov} \left( n_a \times n_b, n_c \times n_d \right) .
  \end{split}
\end{equation}
The factors of $1/2$ in the second and third terms are necessary to account for the factor of 2 difference between the variance of an auto-spectrum and the variance of the cross-spectrum between two uncorrelated fields.

An important feature of this rescaling method is that it considers signal and noise separately, rather than rescaling a combined signal-plus-noise covariance matrix.
This is important because, for the highly non-uniform hit patterns (see Figure~\ref{fig:relhits}) that result from a large field-of-view telescope making deep maps, the spatial distribution of signal and noise in the map are very different.
This leads to a significant difference in the number of signal versus noise degrees of freedom, and therefore different amounts of bandpower variance for the same overall power.
This rescaling method does not attempt to account for different degrees of freedom between the different signal types, but that is a much smaller effect.

\end{appendix}
%\vspace{0.5in}
%------ SECTION: BIBLIOGRAPHY------%
\bibliography{main}

\end{document}